\documentclass{aa}  

\usepackage{graphicx}
\usepackage[varg]{txfonts}
\usepackage[]{natbib}
\bibpunct[; ]{(}{)}{,}{a}{}{;}
\begin{document}

  \title{Rotation-activity relations and flares of M dwarfs \\ with \textit{K2} long- and short-cadence data}
  %\titlerunning{Rotation-activity relations of M dwarfs with K2 long- and short-cadence data}

   \author{St. Raetz
          \inst{1}
          \and
          B. Stelzer\inst{1,2}
          \and
          M. Damasso\inst{3}
          \and
          A. Scholz\inst{4}
%         \fnmsep\thanks{Just to show the usage
%         of the elements in the author field}
          }

   \institute{Institut f\"{u}r Astronomie und Astrophysik T\"{u}bingen (IAAT), Eberhard-Karls Universit\"{a}t T\"{u}bingen, Sand 1, D-72076 T\"{u}bingen, Germany\\
              \email{raetz@astro.uni-tuebingen.de}
         \and
             INAF - Osservatorio Astronomico di Palermo, Piazza del Parlamento 1, I-90134 Palermo, Italy
         \and
            INAF - Osservatorio Astrofisico di Torino, via Osservatorio 20, I-10025 Pino Torinese, Italy
         \and
            SUPA, School of Physics \& Astronomy, University of St Andrews, North Haugh, St. Andrews KY 16 9SS, UK
}

   \date{Received 2019; accepted }

 \abstract{Using light curves obtained by the \textit{K2} mission, we study the relation between stellar rotation and magnetic activity with special focus on stellar flares. Our sample comprises 56 bright and nearby M dwarfs observed by \textit{K2} during campaigns C0-C18 in long- and short-cadence mode. We derive rotation periods for 46 M dwarfs and measure photometric activity indicators such as amplitude of the rotational signal, standard deviation of the light curves, and the basic flare properties (flare rate, flare energy, flare duration, and flare amplitude). We found 1662 short-cadence flares, 363 of which have a long-cadence counterpart  with flare energies of up to $5.6\,\cdotp10^{34}$\,erg. The flare amplitude, duration, and frequency derived from the short-cadence light curves differ significantly from those derived from the long-cadence data. The analysis of the short-cadence light curves results in a flare rate that is 4.6 times higher than the long-cadence data.  We confirm the abrupt change in activity level in the rotation-activity relation at a critical period of $\sim$10\,d when photometric activity diagnostics are used. This change is most drastic in the flare duration and frequency for short-cadence data. Our flare studies revealed that the highest flare rates are not found among the fastest rotators and that stars with the highest flare rates do not show the most energetic flares. We found that the superflare frequency ($E\geq5\,\cdotp10^{34}$\,erg) for the fast-rotating M stars is twice higher than for solar like stars in the same period range. By fitting the cumulative FFD, we derived a power-law index of $\alpha=1.84 \pm 0.14,$ consistent with previous M dwarf studies and the value found for the Sun.}
% 5 {} token are mandatory

%  \abstract


   \maketitle
%

%________________________________________________________________

\section{Introduction}

Stellar activity and its resulting phenomena, such as stellar spots and the highly structured coronal regions, are directly linked to the existence of strong magnetic fields. These magnetic fields are believed to be generated and maintained by a dynamo that is driven by differential rotation and convection. Changes in rotation (the spindown) is in turn controlled by angular momentum losses through stellar activity. This shows that rotation and stellar activity are intimately connected. Studying the relation between stellar rotation and magnetic activity of late-type stars is essential to enhance our understanding of stellar dynamos and angular momentum evolution. 

The rotation-activity relation of late-type stars was first observed in the early 1980s \citep[e.g.,][]{1981ApJ...248..279P,1983IAUS..102..133N}. These studies used rotation periods obtained from spectroscopic measurements ($v\, \mathrm{sin}\,i$) and X-ray luminosity or chromospheric emission in the Ca II H\&K lines as diagnostics for the activity.  Other activity indicators that are frequently used are $H_{\alpha}$ \citep[e.g.,][]{2017RMxAC..49...91N} or UV emission \citep[e.g.,][]{1987ApJ...316..434S}.

Stellar rotation rates are best derived from the periodic brightness variations that are caused by cool spots on the stellar surface. Photometric observations also provide a wealth of activity diagnostics, for example, spot cycle amplitude, flare peak amplitude, and flare frequency. This means that all parameters for studying the rotation-activity relation can be obtained from the same photometric data set.

The early studies mentioned above were focused on a broad range of late-type stars. M dwarfs were underrepresented, however, especially in the slow-rotator regime where the activity level and spot amplitudes are low and rotation periods are difficult to measure. To understand the generation of magnetic fields, it is important to study M dwarfs and especially the behavior at the fully convective boundary (SpT $\sim$ M3) where theory predicts a change in dynamo mechanism \citep{2011ASPC..448..505S}. Therefore many recent publications address activity measurements of M dwarfs using various activity diagnostics \citep[e.g.,][]{2016MNRAS.463.1844S,2017ApJ...834...85N,2016Natur.535..526W}. As a result, the sample size of slowly rotating M dwarfs beyond the fully convective boundary drastically increased (see, e.g., Wright et al. 2018 and Magaudda et al. 2020,  A\&A subm.). \nocite{2018MNRAS.479.2351W}

Photometric observations of M dwarfs can best be made with space telescopes. These telescopes provide rotation periods even with low amplitudes, and their continuous monitoring capabilities enable detection of long rotation periods. \citet[][henceforth S16]{2016MNRAS.463.1844S} presented results on the rotation-activity relation of M dwarfs based on a sample of bright and nearby M dwarfs that were observed in campaigns C0-C4 of the \textit{K2} mission.

The \textit{K2} (\textit{Kepler} Two-Wheel) mission is the follow-up project of NASA's \textit{Kepler} mission and was launched in March 2009. After the failure of the second of four reaction wheels that were essential to maintain the precise pointing toward one single field, \textit{K2} observed 20 fields in a series
of sequential observing campaigns (campaign duration $\sim$80\,d). The fields were distributed around the ecliptic plane \citep{2014PASP..126..398H}. Throughout the mission, \textit{K2} observed in two cadence modes, long cadence ($\sim$30\,min data-point cadence) and short cadence ($\sim$1\,min data-point cadence). The latter was only provided for selected targets, and the long cadence was used as the default observing mode. The K2 mission was decommissioned in 2018 October after fuel exhaustion.

S16 showed that the activity level of M dwarfs changes abruptly at a critical rotation period of $\sim$10\,d when photometric activity indicators are used. The work of  S16 is based on long-cadence \textit{K2} data. The goal of the work presented here is to study the rotation-activity relation of M dwarfs using \textit{K2} short-cadence data and compare the results to those obtained from long-cadence light curves (LCs). 

While we expect that the rotation periods and the activity diagnostics related to the rotation cycle do not depend on the cadence mode, \citet{2018ApJ...859...87Y} showed that the choice of cadence will affect the parameters related to stellar flares. The goal of this paper is to compare results of long- and short-cadence data in order to address potential biases in the study of the rotation-activity relation of M dwarfs. Furthermore, we examine the flare properties in the short-cadence data and fit the flare energy distribution of our M-dwarf sample.

Our analysis gives a detailed insight into the behavior of the flare parameters with rotation period and SpT. In particular, we confirm the dramatic change in activity level at a period of $\sim$10\,d reported by S16. Our study enhances our understanding of the connection between white-light flaring and stellar rotation rate. This connection provides constraints on stellar evolution. In addition, we address differences between partly and fully convective M dwarfs, especially in their flare energy distribution.

We present our sample and  determine the stellar parameters in Sect.~\ref{sample} and Sect.~\ref{stellar_param}, respectively. We describe our data analysis in Sect.~\ref{K2_data_analysis}, the period search in Sect.~\ref{rotation}, and the flare analysis in Sect.~\ref{flare_analysis}. Our results are discussed in Sect.~\ref{results}, and we provide a summary in Sect.~\ref{summary}.

\section{ \textit{K2} M dwarf sample}
\label{sample}

The sample presented in S16 was selected from the Superblink proper motion catalog by \citet{2011AJ....142..138L}, which includes $\sim$9000 bright M dwarfs (J<10\,mag) with SpTs from K7 to M7. Our $K2$ observing project (PI Scholz) comprised all \citet{2011AJ....142..138L} M dwarfs in the \textit{K2} field of view, and was extended until the end of the mission. During 20 \textit{K2} campaigns (C0-C19), 485 LCs of 430 targets were obtained. The exploration of the rotation-activity relation by S16, however, was based only on the long-cadence LCs. Short-cadence LCs are better suited for studies of stellar flares, which is a substantial part of the work of S16. As shown by \citet{2018ApJ...859...87Y}, the basic flare properties (flare rate, flare energy, flare duration, and flare amplitude) derived from the short-cadence LCs differ significantly from those derived from the long-cadence data. For a subsample of the Superblink M dwarfs observed by \textit{K2}, short-cadence LCs are available. With the comparison of long- and short-cadence data, we can test how strongly the systematic differences of the flare properties affect results that are obtained with long-cadence LCs.

\begin{figure}
  \centering
  \includegraphics[width=0.3\textwidth,angle=270]{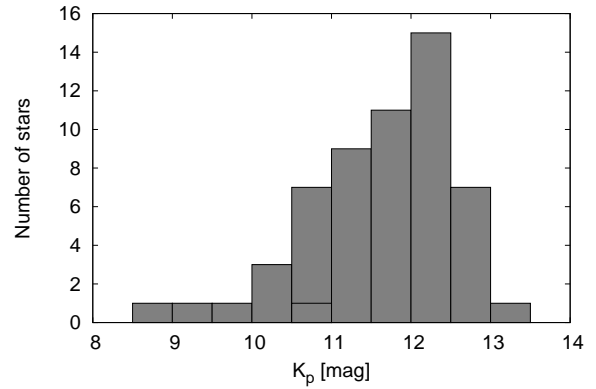}
  \caption{Distribution of the \textit{Kepler} magnitudes, $K_p$, for our \textit{K2} M-dwarf sample.}
  \label{Kp_Histogramm}
\end{figure}

In total, 87 short-cadence LCs of 79 Superblink  M dwarfs were obtained by \textit{K2} in campaigns C0 to C19. However, C19 only consists of $\text{about }$eight days of data because of fuel exhaustion, and consequently, the end of the mission. Because the observational baseline is low, we excluded C19 from our analysis. Our final sample comprising campaigns C0 to C18 includes 64 LCs of 56 targets. A histogram of the \textit{Kepler} magnitudes, $K_p$, of our targets is shown in Fig.~\ref{Kp_Histogramm}.

\section{Stellar parameters}
\label{stellar_param}

The stellar parameters (stellar mass, radius, effective temperature, and bolometric luminosity) were calculated using the empirical and semi-empirical relations of \citet[][coefficients of the relations taken from the erratum, Mann et al. 2016]{2015ApJ...804...64M},\nocite{2016ApJ...819...87M} which are based on the colors $V-J$, $J-H,$ and the absolute magnitude in the \textit{Two Micron All Sky Survey}  (2MASS) $K$-band, $M_{\mathrm{Ks}}$. The stellar magnitudes needed for these calculations were taken from the \textit{Fourth US Naval Observatory CCD Astrograph Catalog} (UCAC4) \citep{2013AJ....145...44Z}, which lists 2MASS $JHK$-photometry and the \textit{AAVSO Photometric All Sky Survey} $V$-band magnitudes \citep[APASS,][]{2014CoSka..43..518H}. The UCAC4 catalog does not give $V$-band magnitudes for four stars. In these cases, we used the magnitudes in $V$ band given in the \textit{TESS} Input Catalog \citep[TIC v8,][]{2018AJ....156..102S}. 

\begin{figure}
  \centering
  \includegraphics[width=0.3\textwidth,angle=270]{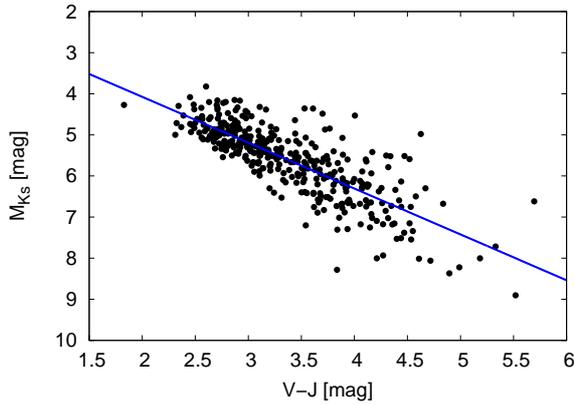}
  \caption{Absolute magnitude in the $K$ band, $M_{\mathrm{Ks}}$, vs. $V-J$ color for the 376 stars of the \textit{K2} M dwarf sample with \textit{Gaia} distance estimates given in \citet{2018AJ....156...58B}. The blue solid line shows the fit of the linear relation given in Eq.~\ref{MKs_calibration}.}
  \label{VJ_MK_Bailer-Jones_final}
\end{figure}

\begin{figure}
  \centering
  \includegraphics[width=0.3\textwidth,angle=270]{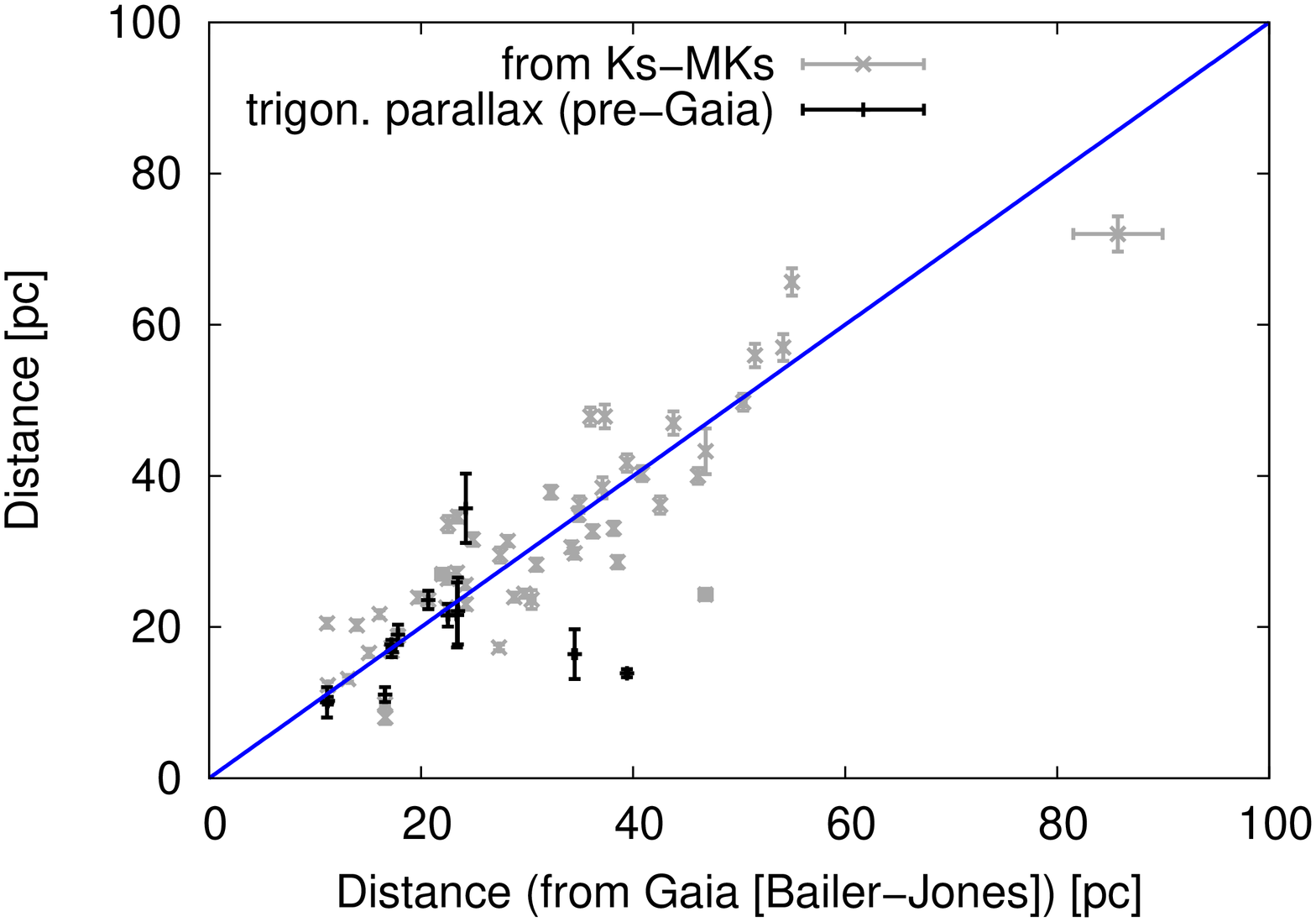}
  \caption{\textit{Gaia} distances  \citep{2018AJ....156...58B} vs. the distances determined from the trigonometric parallax given in \citet[][14 stars]{2011AJ....142..138L}, as well as those determined from the ($V-J$) - $M_{\mathrm{Ks}}$ relation for all 52 M dwarfs with \textit{Gaia} distances.}
  \label{dist_Gaia_MKs_trig}
\end{figure}

$M_{\mathrm{Ks}}$ was derived by inserting the distance and the apparent magnitude in $K$ band into the distance modulus. \citet{2018AJ....156...58B} estimated distances to 1.33 billion stars from the parallaxes provided in the \textit{Gaia} Data Release 2. Fifty-two out of our 56 targets are included in the catalog of \citet{2018AJ....156...58B}. Three of the remaining four M dwarfs do not have a \textit{Gaia} parallax, while for the fourth the parallax was found to be not reliable following the criteria by \citet[][appendix C, equations C-1 and C-2]{2018A&A...616A...2L}. Two of these four targets have trigonometric parallaxes in \citet[][from \textit{Hipparcos} or ground-based astrometry]{2011AJ....142..138L}. To determine the distances of the two remaining stars without parallaxes, we improved the empirical linear calibration from S16 between $M_{\mathrm{Ks}}$ and $V-J$ using \textit{Gaia} distances. We performed the calibration with the 376 stars from the whole \textit{K2} M dwarf sample (of 430) that have a \textit{Gaia} distance. We computed $M_{\mathrm{Ks}}$ for these stars, and then fit the ($V-J$) - $M_{\mathrm{Ks}}$ data set (see Fig.~\ref{VJ_MK_Bailer-Jones_final}) using the \begin{small}IDL\end{small}\footnote{\begin{small}IDL\end{small} is a product of the Exelis Visual Information Solutions, Inc.} routine LINFIT. In order to include the error bars, we generated 10000 data sets for which each value was randomly chosen from their error boxes defined by their $V-J$ and $M_{\mathrm{Ks}}$ uncertainties. The mean and standard deviation of the 10000 linear fits was chosen as final calibration. This relation
\begin{equation}
\label{MKs_calibration}
M_{\mathrm{Ks}}=1.844(\pm0.024)+1.116(\pm0.007)\,\cdotp(V-J)
\end{equation}
is shown together with the data in Fig.~\ref{VJ_MK_Bailer-Jones_final}.
From the $M_{\mathrm{Ks}}$ that we obtained from this relation, we calculated the distance for the two stars without parallaxes. A comparison between the \citet{2018AJ....156...58B} \textit{Gaia} distances, the distances from the \citet{2011AJ....142..138L} trigonometrical parallaxes, and the distances from our linear calibration is given in Fig.~\ref{dist_Gaia_MKs_trig} for the sample studied in this work.

To take the error bars on $V-J$, $J-H,$ and $M_{\mathrm{Ks}}$ for the calculation of the stellar properties into account, we used the same Monte Carlo approach as explained above.  We calculated a given parameter 10000 times each with a different set of $V-J$, $J-H,$ and $M_{\mathrm{Ks}}$ that were chosen randomly within their uncertainties. The final value was then determined as the mean and standard deviation of the resulting distribution. More details on this procedure are given by S16.

To determine a quiescent luminosity in the \textit{Kepler} band, $L_{\mathrm{Kp,0}}$, of our targets, we assumed that the quiescent emission corresponds to the \textit{Kepler} magnitude of the stars. We converted the $K_{\rm p}$ magnitude \citep[from the Ecliptic Plane Input Catalog, EPIC,][]{2016ksci.rept....6H} into flux using the zero-points and effective wavelengths provided at the filter profile service of the Spanish Virtual Observatory \citep[SVO,][]{2012ivoa.rept.1015R}. By applying the adopted distances, we obtained the luminosity. 

\begin{figure}
  \centering
  \includegraphics[width=0.3\textwidth,angle=270]{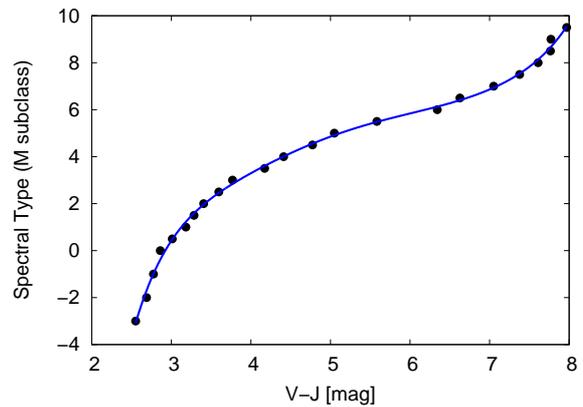}
  \caption{Relation between $V-J$ and SpT shown as subclass of M dwarfs (negative values correspond to SpTs earlier than M) based on \citet{2013ApJS..208....9P}. The blue solid line represents the seventh-order polynomial given in Eq.~\ref{SpT_calib}.}
  \label{Pecaut_Mamajek_VJ_SpT_fit}
\end{figure}

Finally, we derived a spectral type (SpT) calibration by fitting a seventh-order polynomial to the relation between $V-J$ and SpT extracted from the mean dwarf stellar color and effective temperature sequence by \citet{2013ApJS..208....9P}\footnote{We used Version 2019.3.22. The updated version of this table is available at http://www.pas.rochester.edu/$\sim$emamajek/ EEM\_dwarf\_UBVIJHK\_colors\_Teff.txt}. The update of the SpT was necessary because the $V$-band magnitudes included in the UCAC4 catalog are more precise than those used by \citet{2011AJ....142..138L} in their determination of the SpT using their own empirical relation of SpT and $V-J$. We were unable to use the SpT calibration of S16 because our sample includes targets, especially the mid- to late-M dwarfs, that are outside the valid range of their relations. The result of the fitting 
\begin{equation}
\label{SpT_calib}
SpT=A+Bx+Cx^{2}+Dx^{3}+Ex^{4}+Fx^{5}+Gx^{6}+Hx^{7},
\end{equation}
with $x\,=\,V-J$ and the coefficients given in Table~\ref{Fitparameter_SpT_calib}, is shown in Fig.~\ref{Pecaut_Mamajek_VJ_SpT_fit}.

\begin{table}
\centering
\caption{Coefficients for the SpT calibration given in Eq.~\ref{SpT_calib}.}
\label{Fitparameter_SpT_calib}
\begin{tabular}{lr}
\hline \hline
Coefficient & Value \\ \hline
$A$ & -371.5616 \\
$B$ & 481.4533 \\
$C$ & -266.4497 \\
$D$ & 81.5304 \\
$E$ & -14.8148 \\
$F$ & 1.5976 \\
$G$ & -0.0948\\
$H$ & 0.0024\\
\hline
\end{tabular}
\end{table}

\begin{figure}
  \centering
  \includegraphics[width=0.3\textwidth,angle=270]{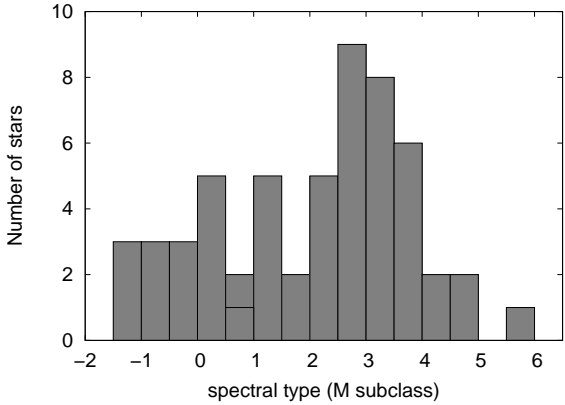}
  \caption{Distribution of the SpTs for our \textit{K2} M dwarf sample. The SpTs were obtained from the $V-J$ calibration given in Eq.~\ref{SpT_calib} and shown in Fig.~\ref{Pecaut_Mamajek_VJ_SpT_fit}. Negative values denote SpTs earlier than M.}
  \label{SpT_Histogramm}
\end{figure}

For further analysis, we divided the full sample of M dwarfs into four SpT bins, which allows to easily distinguish between partly and fully convective stars. Chabrier \& Baraffe (1997) predict the fully convective boundary to occur at M\,$\approx$\,0.35\,M$_{\odot}$. According to the tables of \citet{2013ApJS..208....9P}, this corresponds to a SpT of M3.05 (from interpolation of the mass for SpT M3 and M3.5). We therefore allocated all stars earlier than this SpT to the partially convective regime (SpT groups K7-M1 and M2-M3) and all later type stars to the fully convective regime (SpT groups M3.5-M4 and M5-M6). More detailed information about the SpT groups is given in Table~\ref{SpT_groups}. The distribution of SpTs in our target sample is shown in Fig.~\ref{SpT_Histogramm}. The full target list with the stellar parameters is given in Table~\ref{stellar_param_table} in the appendix.

\begin{table}[h]
\caption{SpT groups.}
\label{SpT_groups}
\begin{tabular}{lccc}
\hline \hline
SpT group & $V-J$ [mag] & SpT$^{a}$  & number of stars \\ 
 &  & (M subclass)  &  \\\hline
K7-M1 & $<$3.25 & $<$1.5 & 21 \\
M2-M3  & $\geq$3.25, $<$3.875 & $\geq$1.5, $<$3.05 & 18 \\
M3.5-M4 & $\geq$3.875, $<$4.70 & $\geq$3.05, $<$4.5 & 14 \\
M5-M6 & $\geq$4.70 & $\geq$4.5 & 3 \\
\hline
\end{tabular}
\\
$^{a}$ calculated with Eq.~\ref{SpT_calib} 
\end{table}

\section{\textit{K2} Data analysis}
\label{K2_data_analysis}

\begin{figure}
  \centering
  \includegraphics[width=0.8\textwidth,angle=270]{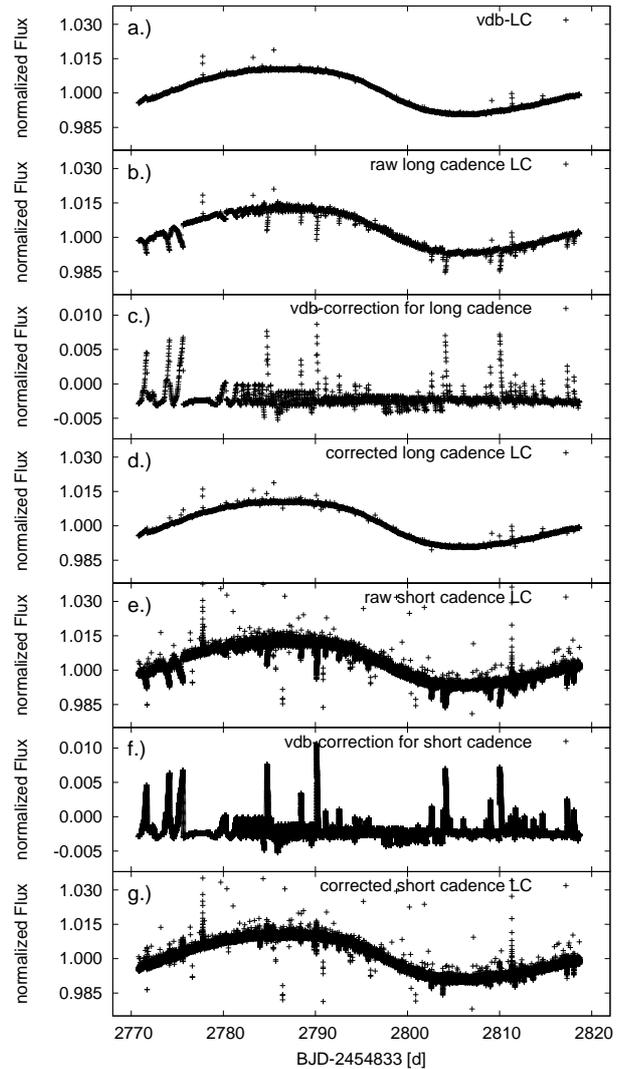}
  \caption{Light curve of EPIC\,201659529 as an example for our method for correcting the raw \textit{K2} LCs. Panel \textbf{a}  vdb-LC from which we extracted the detrending correction. \textbf{Panel b} Raw long-cadence LC extracted by us with the \textit{Kepler}GO/lightkurve code from the MAST TPF using the aperture assigned as BESTAPER in the vdb-LC. Panel \textbf{c} vdb-correction for the long-cadence data created by subtracting panel b from panel a. Panel \textbf{d} Corrected long-cadence LC. Panel \textbf{e} Raw short-cadence LC extracted by us analogously to the long-cadence LC of panel b using the aperture assigned as BESTAPER in the vdb-LC. Panel \textbf{f} vdb-correction for the short-cadence data created by subtracting panel b from panel a and interpolating to all cadences. Panel \textbf{g} Corrected short-cadence LC.}
  \label{LC_EPIC201659529_example}
\end{figure}

Unlike the original \textit{Kepler} mission, which maintained a very precise and stable pointing to its target field throughout the mission duration by using four reactions wheels, \textit{K2} balanced itself against solar radiation pressure by pointing in the spacecraft orbital plane and correcting for drifts using its thrusters. Because of this decrease in the ability to point precisely, raw \textit{K2} aperture photometry is less precise and suffers from more systematics, that is, noise due to the spacecraft pointing jitter \citep{2014PASP..126..398H}. The unstable pointing of \textit{K2} results in a movement of a star on the detector, which causes artifacts in the raw LCs that have to be corrected for. A number of groups have developed software that corrects for these instrumental effects and have made their data products publicly available, for instance, the K2 extracted light curves (K2SFF) by \citet{2014PASP..126..948V}, the LCs produced by the K2 Systematics Correction \citep[K2SC,][]{2016MNRAS.459.2408A}, and the LCs of the EPIC variability extraction and removal for exoplanet science targets pipeline \citep[EVEREST,][]{2016AJ....152..100L}.

The data reduction by \citet{2014PASP..126..948V} (vdb-LCs) provides a wealth of information (e.g., LCs for different aperture masks). However, we were unable to use the vdb-LCs directly because of two reasons: (1) \citet{2014PASP..126..948V} did not provide short-cadence LCs, and (2) they excluded all data points with a Kepler quality flag QUALITY$\neq0$ and thus also removed potentially valid points from their analysis. The 20 different quality flags assigned by the \textit{Kepler} pipeline to individual data points include data that are of poor quality, but also data that might be of lower quality or could cause problems for transit detection after applying detrending software \citep{2016ksci.rept....9T}. Removing all flagged data points by default might therefore impede the detection of real astrophysical signals or the interpretation of systematics. This concerns in particular events that might represent stellar flares, which are the topic of our work.

We developed a procedure that allows us to use the detrending by \citet{2014PASP..126..948V} (henceforth vdb-correction) and apply it to the original LCs that include points removed by \citet{2014PASP..126..948V}. First, we generated the LCs from the \textit{K2} target pixel files (TPFs) downloaded from the Barbara A. Mikulski Archive for Space Telescopes (MAST) portal. We extracted the raw LCs with the \textit{Kepler}GO/lightkurve code \citep[version: 1.0b13, August 2018;][]{2018ascl.soft12013L} using the same aperture mask as was assigned as ``best aperture'' in the \citet{2014PASP..126..948V} data.

The data points excluded by \citet{2014PASP..126..948V} based on the quality flag that we retained are described next. In all campaigns, we kept data points labeled `Impulsive outlier' (which could be real stellar flares), `Cosmic ray in collateral data', and `Thruster firing' (bits 11, 13, and 20). In later campaigns, we kept additional flagged data points because changes and errors in the onboard software resulted in additional good-quality cadences being flagged. This information is provided in the data release notes of the individual campaigns. 

Fig.~\ref{LC_EPIC201659529_example} shows one example of the method for correcting the raw LCs in long- and short-cadence mode. This method comprises the following steps: To obtain the vdb-correction, we subtracted the flux column of our extracted raw LCs from the FCOR (corrected flux) column in the vdb-LCs for all cadences that are present in both LCs. Because our extracted raw LCs in general consist of more data points than the vdb-LCs, the vdb-correction was interpolated to the missing cadences before it was added the extracted LCs. This method results in a final corrected LC that also includes all cadences with potential stellar flares. To apply the vdb-correction to short-cadence data, we extracted the LC from the short-cadence TPFs using the same aperture. Then, we interpolated the vdb-correction to all cadences of the extracted short-cadence LC. 

Fig.~\ref{LC_EPIC201659529_example} shows that our corrected long-cadence LC (Fig.~\ref{LC_EPIC201659529_example}d) is in excellent agreement with the original vdb-LC (Fig.~\ref{LC_EPIC201659529_example}a). Furthermore, this method also works well for the short-cadence data, although not all artifacts could be removed because information is missing in the vdb-correction, that is, the behavior of the LC around flux jumps caused by thruster fires is obscured by the binning to long cadence and therefore cannot be exactly reproduced by interpolation. 

\begin{figure}
  \centering
  \includegraphics[width=0.3\textwidth,angle=270]{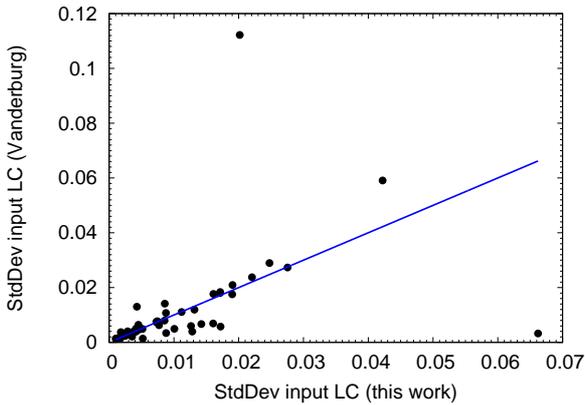}
  \caption{Comparison of the standard deviations of the original  short-cadence LCs from our work and from A. Vanderburg (priv. com.).}
  \label{Vergleich_LC_vdb}
\end{figure}

As a cross-check to determine whether our method yields LCs with comparable quality, A. Vanderburg (priv. com.) provided us with a set of short-cadence LCs for targets in our sample. The shape is almost identical for all the LCs, while the scatter varies. Fig.~\ref{Vergleich_LC_vdb} shows a comparison between the standard deviations of the original LCs produced with the two methods. Only three targets show a very different standard deviation. These deviations most likely arise because different aperture masks were used to extract the raw LCs. To test if the quality of the LCs obtained with the two methods is comparable, we ran a Student's T-test. With a probability of $\sim90\%$ (T\,=\,-0.13, p-value\,=\,0.90), the mean values of the standard deviations are equal for the two LC sets. This means that our method provides reliable LCs. In general, our method does not eliminate all systematics in every detail, but it is an easy and fast way to analyze data from the \textit{K2} mission, and it yields sufficient precision for this work.
  
In our \textit{K2} M dwarf sample we found two LCs that appeared almost identical, that is, they showed the same rotational signal, and eclipses that occurred at the same time. We identified the targets as CU\,Cnc and CV\,Cnc. They fall into the aperture mask that produced an identical LC for both stars. We therefore re-extracted the LCs from the \textit{K2} TPFs and used the individual LCs for the further analysis. The details of the analysis and the determination of the linear ephemeris of the known eclipsing binary CU\,Cnc are given in Appendix \ref{CUCnc}.

\section{Period search}
\label{rotation}

Stellar rotation rates are derived from the periodic brightness variations that are caused by cool spots on a stellar surface. Photometric observations with \textit{K2} provide rotation periods even with low amplitudes (e.g., rotational amplitudes of $\sim$900\,ppm were detected by S16). We used three methods to search for the rotation period in our M dwarfs. 

First, we computed the generalized Lomb-Scargle periodogram \citep[\begin{small}GLS\end{small};][]{2009A&A...496..577Z} for the long-cadence and for the short-cadence data. The \begin{small}GLS\end{small} implementation we used\footnote{Fortran Version v2.3.01, released 2011-09-13 by Mathias Zechmeister} can only process up to 10000 data points. We were therefore able to use \begin{small}GLS\end{small} directly only for the long-cadence data, for which a typical $K2$ campaign consists of $\sim$3000 data points. We had to bin the $\sim$95000 short-cadence data points. The binning factor was determined for each LC individually depending on the number of cadences in the input LC. We had to do this also to reduce the number of data points to a value below 10000. Although the highest used binning factor was 12, we were able to detect rotation periods down to $\sim$30\,min. This means that the binning does not compromise the detection capability for short rotation periods. 

As a second standard time-series analysis technique, we determined the autocorrelation function (ACF) of our LCs using the  \begin{small}IDL\end{small} routine A\_CORRELATE. Because evenly spaced data are required for the ACF method, we filled all data-point gaps in the LCs by interpolation. The rotation period is then given by the time lag corresponding to the first peak in the ACF times the data-point cadence. In the case of double-peaked LCs, which show two distinct sequences of peaks, which indicates that there might be two dominant spots, we chose the time lag corresponding to the first peak of the sequence of higher peaks to calculate the rotation period. We used the interpolated version of the original LC only to determine the ACF. All other period-search methods and the flare analysis (Sec.~\ref{flare_analysis}) are based on the original LC.

Finally, as a third method, we fit the LCs with a sine function using the  \begin{small}IDL\end{small} routine MPFITEXPR, which fits a user-supplied model with the Levenberg-Marquardt least-squares method to the data. We used a sinusoidal model given by  
\begin{equation}
\label{sine_fit}
f(t)=C+A\,\cdotp sin(\frac{2\pi}{P_{\mathrm{rot}}}\,\cdotp t+\phi),
\end{equation}
where $C$ is a constant, $A$ is the amplitude of the sine function, $P_{\mathrm{rot}}$ is the stellar rotation period, $t$ is the time variable, and  $\phi$ is the phase. As starting guesses for the four fitting parameters, we used 1 for all except for the rotation period,  for which we set the initial value to the result of \begin{small}GLS\end{small}. When no initial values are supplied, the procedure does not result in a reasonable fit. In general, the sine fit method yields similar results, but with a larger formal uncertainty than \begin{small}GLS,\end{small} and it allows us to constrain the rotation periods even if they exceed the \textit{K2} monitoring time baseline of 70-80\,d. 

The LCs of our sample were phase-folded with the periods of all three methods. By eye-inspection we selected the best-fitting period, which we then used in the further analysis. In the case of equally good periods from different methods, we adopted the average rotation period. Rotation periods for the targets of \textit{K2} campaigns C0-C4 (11 M dwarfs from our sample) were already derived by S16. Their values are confirmed by our analysis.

Eight stars of our sample were observed in two \textit{K2} campaigns because the target fields overlap. In these cases we also applied the period search to the combined LCs. In all except for one case, both LCs could be represented by a single rotation period. The remaining star has a long rotation period $>$\,100\,d that cannot be better constrained with two \textit{K2} campaigns separated by $\sim$2.5\,yr. For the seven stars with a reliable period estimate, we used the result from the combined period search as the final rotation period. In total, we found rotation periods for 46 stars, which is $\sim$82\% of all targets in our M-dwarf sample. Our sample therefore contains a similar but slightly higher number of rotators that show periodic variability on timescales of up to $\sim$100\,d than previous M-dwarf studies.

\section{Flare analysis}
\label{flare_analysis}

\begin{figure*}
  \centering
  \includegraphics[width=0.7\textwidth,angle=270]{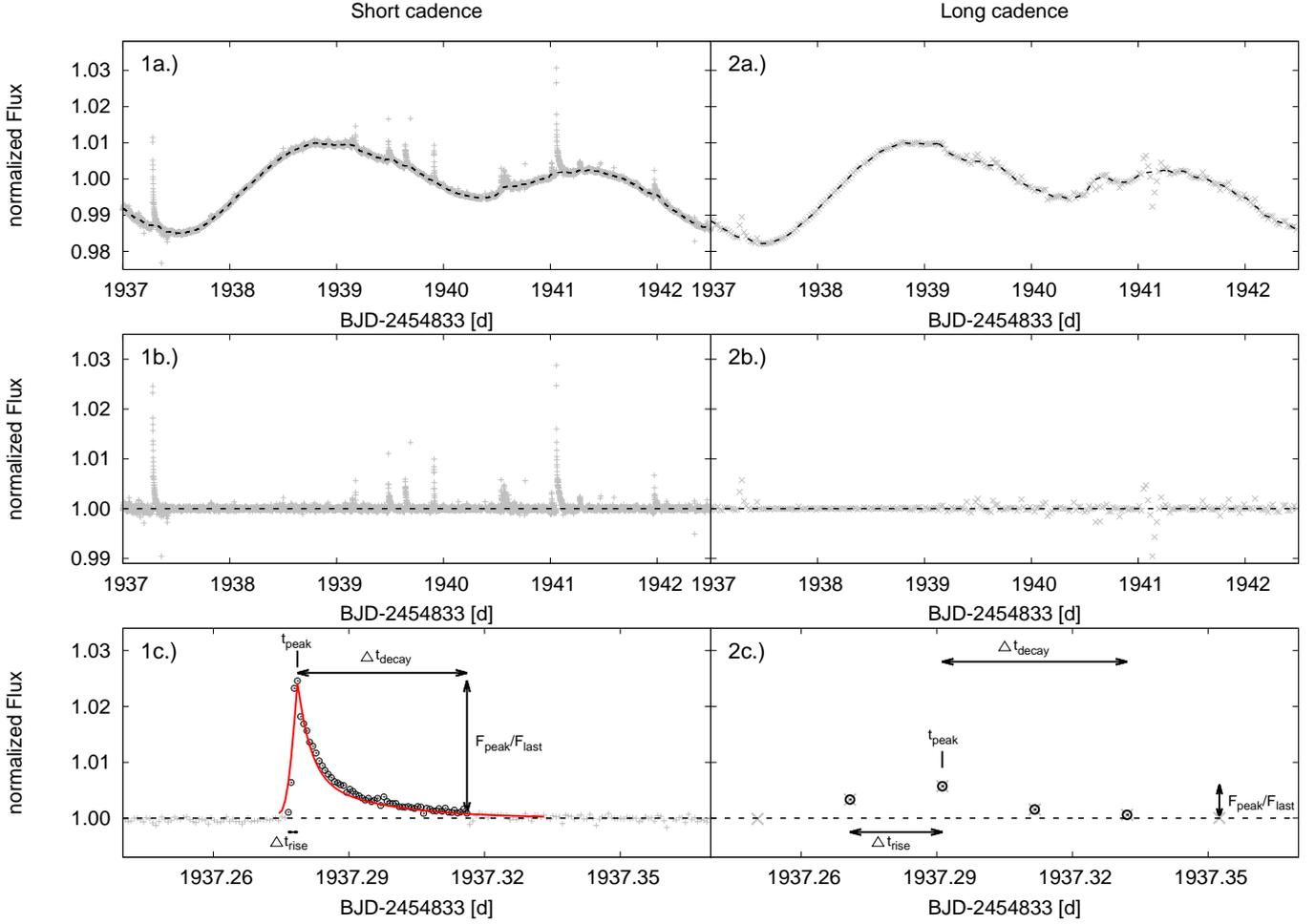}
  \caption{Example for our flare detection and validation procedure. The left and right plots show the short- and long-cadence data, respectively. \textbf{Top}: Observed and smoothed original (black dashed) LC. \textbf{Middle:} Flattened LC (without rotational signal). \textbf{Bottom:} Zoom on the first flare candidate from the panels above. All relevant parameters for the flare validation are also shown: $t_{\mathrm{peak}}$ is the time of the flare maximum, $\Delta t_{\mathrm{rise}}$ is the duration of the rise phase,  $\Delta t_{\mathrm{decay}}$ is the duration of the decay phase,  $F_{\mathrm{peak}}/F_{\mathrm{last}}$ is the flux ratio between the flare maximum and the last flare point (for long-cadence data, ${F_{\mathrm{last}}}$ is defined as the flux of the first point after the flare). The red solid line is the best-fitting flare template following \citet{2014ApJ...797..122D}.}
  \label{Epic202059229_example_FlareDet}
\end{figure*}

To perform a detailed flare analysis and conduct a diagnostic comparison between the properties of flares seen in long- and short-cadence data, we developed an algorithm that includes the detection and validation of flares. The procedure yields all relevant flare properties for all validated flares, that is, the times of flare start ($t_{\mathrm{first}}$) and flare peak ($t_{\mathrm{peak}}$), the duration ($\Delta t_{\mathrm{Flare}}$), the peak flare amplitude ($A_{\mathrm{peak}}$), the number of flare points, and the equivalent duration ($ED$). The latter is computed as the integral under the flare. Because the LCs are given in relative flux, the $ED$ is given in seconds. Following \citet{2016ApJ...829...23D}, we calculated a relative flare energy, $E_{\rm F}$, by multiplying the $ED$ by the quiescent stellar luminosity ($L_{\mathrm{Kp,0}}$, see Sect. \ref{stellar_param}). An example demonstrating our flare detection and validation procedure for the long- and short-cadence data is shown in Fig.~\ref{Epic202059229_example_FlareDet}.

\subsection{Flare detection}
\label{flare_detection}

The algorithm for the detection of stellar flares is based on the routine developed by S16. For the flare detection, we first created a smoothed LC. This was done with an iterative process of boxcar smoothing. The original and smoothed LC are shown in the top panel in Fig.~\ref{Epic202059229_example_FlareDet}. When an LC showed data-point gaps of $>$4\,h, we split the LC into segments of continuous observations and treated each segment individually. In the second step we subtracted the smoothed from the original LCs to remove the rotational signal (flattened LC, middle panel in Fig.~\ref{Epic202059229_example_FlareDet}). Then we removed all points that deviated by more than 2$\sigma$ from the subtracted LC. Here we chose a threshold of 2$\sigma$ to obtain a smoothed LC that was as clean as possible. These three steps were repeated three times with a successively smaller width of the boxcar. The optimal boxcar width was chosen with respect to the rotation period of the M dwarf, and this period was determined from the original LC (see Sect. \ref{rotation}) in such a way that the largest boxcar width was smaller than one-third of the rotation period for the fastest rotators. The smoothed LC was then interpolated to all data points of the original LC. Finally, all points of the original LC that lay 3$\sigma$ above the final smoothed LC were flagged. 

To avoid the false identification of spurious events as flares, we only assigned groups of consecutively flagged data points as potential flares. In particular, we selected all groups of at least two and five consecutive upward outlying points for the long- and short-cadence data, respectively. As a result of our flare detection procedure, all potential flares are significant with $>$3$\sigma$ (the actual flare significance is even higher because at least two and five points for long and short cadence, respectively, are required to lie above 3$\sigma$) and last at least $\sim$30\,min in long cadence (time between two points with a cadence of 29.4\,min) and $\sim$4\,min in short cadence (time between the first and the fifth point, each with a cadence of 58.9\,s).

The subtraction of the smoothed LC from the original LC and the removal of all 3$\sigma$ outliers results in a cleaned and flattened LC with a standard deviation, $S_{\mathrm{flat}}$. In the case of several LC segments, the full LC of one campaign could have several values of $S_{\mathrm{flat}}$. The average $S_{\mathrm{flat}}$ is then the mean of the $S_{\mathrm{flat}}$ values for the different LC segments. 

These criteria for our flare detection procedure (flare detection threshold and the number of consecutive outliers) were selected by testing various combinations. Two consecutive points are the minimum requirement for our flare validation process. For the long-cadence LCs, we decided to use this minimum because the data-point cadence already implies a high minimum detectable flare duration and we did not wish to increase it further. To quantify the probability of obtaining spurious detections at our chosen detection criteria, we simulated a normalized flattenend LC by adding Gaussian noise (white noise) together with a $1/f^{2}$ component (red noise) to a flux value of 1 for all cadences of an observed \textit{K2} LC. The standard deviation of the simulated LC was set to the $S_{\mathrm{flat}}$ value of the same \textit{K2} LC. The red noise accounts for residual systematics that could not be removed in the data analysis and flattening process. In this simulated LC, we counted the number of consecutive upward outliers. We repeated this test 10000 times. For a typical long-cadence LC with $\sim$3000 data points, we found a probability of $\sim 3\cdotp10^{-4}$\% of a spurious flare detection (two consecutive points above 3$\sigma$). However, these are flare candidates, and they still have to pass the subsequent flare validation process.

Because our validation process of short-cadence flares includes the fitting of a flare template (see Sect.~\ref{flare_validation}), we chose the number of consecutive outliers to be $>$3 to ensure a reasonable fit\footnote{For the same reason, we did not fit a flare template to the long-cadence flare candidates.}. The simulations for the short-cadence LCs resulted in a spurious detection probability of $\sim 4\cdotp10^{-7}$\% for four consecutive points above 3$\sigma$, while we were unable to detect any group of five points above 3$\sigma$. Our simulations might underestimate the probabilities in the short-cadence case because as shown in Sect.~\ref{K2_data_analysis}, we were unable to remove all systematics in every detail. Our threshold of five consecutive outliers for a flare candidate is therefore justified because it minimizes spurious detections.

\subsection{Flare validation}
\label{flare_validation}

To validate potential flares as confirmed flares, several criteria have to be fulfilled. The steps of flare validation are depicted in the bottom panel of Fig.~\ref{Epic202059229_example_FlareDet}. In each group of consecutive outlying points, we determined the flare amplitude, $A_{\mathrm{peak}}$, by subtracting the continuum flux level given by the smoothed and interpolated LC at the time of the highest outlier from the corresponding flux. The flare duration was calculated as the time between the first and last flare point $\Delta t_{\mathrm{Flare}}=t_{\mathrm{last}}-t_{\mathrm{first}}$, where $t_{\mathrm{last}}$ is defined as the last continuous point after each group of consecutive outlying points that is higher than 1$\sigma$ above the smoothed and interpolated LC.

We first removed potential flares that occurred right before or after a data-point gap or within $\pm$0.005\,d around a thruster-firing event of the spacecraft. The times of the thruster fires were obtained from the quality flags in the LC. The sudden movement of the telescope might cause flux jumps and outliers that might mimic a flare. These deviating points are usually not the points where the thruster fires occurred (the measurements with the thruster-fire quality flag), but the data points following this event. On the other hand, real flares might be removed in that step. However, the thruster fires will affect the shape of a flare, and this might introduce a systematic bias in the interpretation of the results if we did not remove them from our analysis.

%As another criterion we checked if the ratio between the flare amplitude and the standard deviation of the flattened and cleaned LC (rotational signal and outliers removed) is greater than 3, which should be the case by default as the 3$\sigma$ threshold was required for detection of potential flares. 

As another criterion, we required the flux ratio between the flare maximum and the last flare point $\frac{F_{\mathrm{peak}}}{F_{\mathrm{last}}}$ for short-cadence data to be $\geq$2. For the long-cadence data, the same flux ratio is required for the flare maximum, and the first point after the flare, that is, ${F_{\mathrm{last}}}$ is to be replaced by the flux of the first point after the flare. This criterion was chosen to allow for two-point flare candidates, which have a flux ratio between the first and the last point that is lower than 2. Without this replacement, such flares would not pass the validation although they might be true flares. To have a further criterion for a proper flare shape, the decay phase $\Delta t_{\mathrm{decay}}$ should be longer than the rise phase $\Delta t_{\mathrm{rise,}}$ and the flare maximum was not allowed to be the last point of the potential flare. Finally, of the potential long-cadence flares, only those with a short-cadence counterpart were considered a confirmed flare.

The short-cadence data allow studying the morphology  of a flare in more detail. Therefore it is possible to use the flare shape as an additional criterion for the flare validation. We used the empirical flare template suggested by \citet{2014ApJ...797..122D}, which was constructed using 885 classical (only one peak) flares observed on the highly active mid-M dwarf GJ\,1243. This flare template consists of a fourth-order polynomial in the rise phase and a sum of two exponential functions in the decay phase. The flare template was fit to each potential flare with two independent parameters: the flare start time, and the flare duration. To shorten the computing time for each individual flare candidate, we imposed the following fitting limits: (1) The flare start time was allowed to vary by about three data points around the first detected flare point (first 3$\sigma$ outlier of a potential flare). (2) As lower and upper limit for the flare duration, we chose one and six times the time between the first and last point of the potential flare, respectively. (3) We used the relation between flare amplitude and flare duration found by \citet[][their Fig.~10]{2014ApJ...797..121H} for a sample of 833 flares from two months of \textit{Kepler} observations of GJ\,1243. Instead of fitting all amplitudes, we thus chose those that were connected to the flare duration through this relation. The scatter of this relation of about one order of magnitude was included in our fitting process. The potential short-cadence flare was finally confirmed as flare if the flare template fit the data better (lower rms) than a linear fit through the same flare points.

In total, we found 1662 validated flares in the short cadence and 363 validated flares in the long cadence of our 56 M dwarfs. Of these, 1644 and 356 for short and long cadence, respectively, are on the 46 stars with a measurable rotation period. When we considered long-cadence data alone, we found 859 flares. This means that only 42\% of the long-cadence flares have a short-cadence counterpart. 

\section{Results}
\label{results}

\subsection{Flare amplitude, duration, and energy}
\label{sect:comp_lc_sc_flares}

\begin{figure}
  \centering
  \includegraphics[width=0.3\textwidth,angle=270]{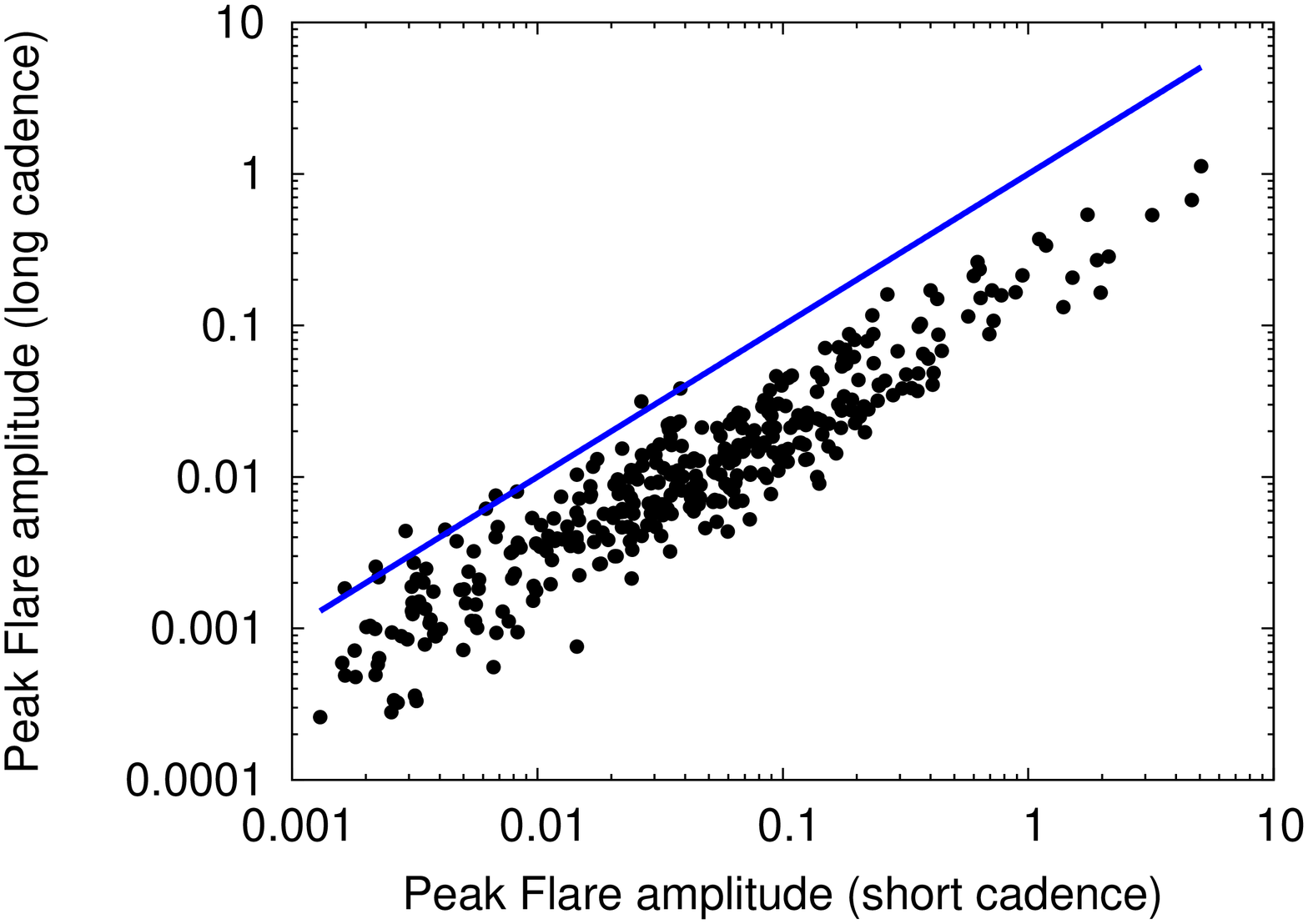}
  \includegraphics[width=0.3\textwidth,angle=270]{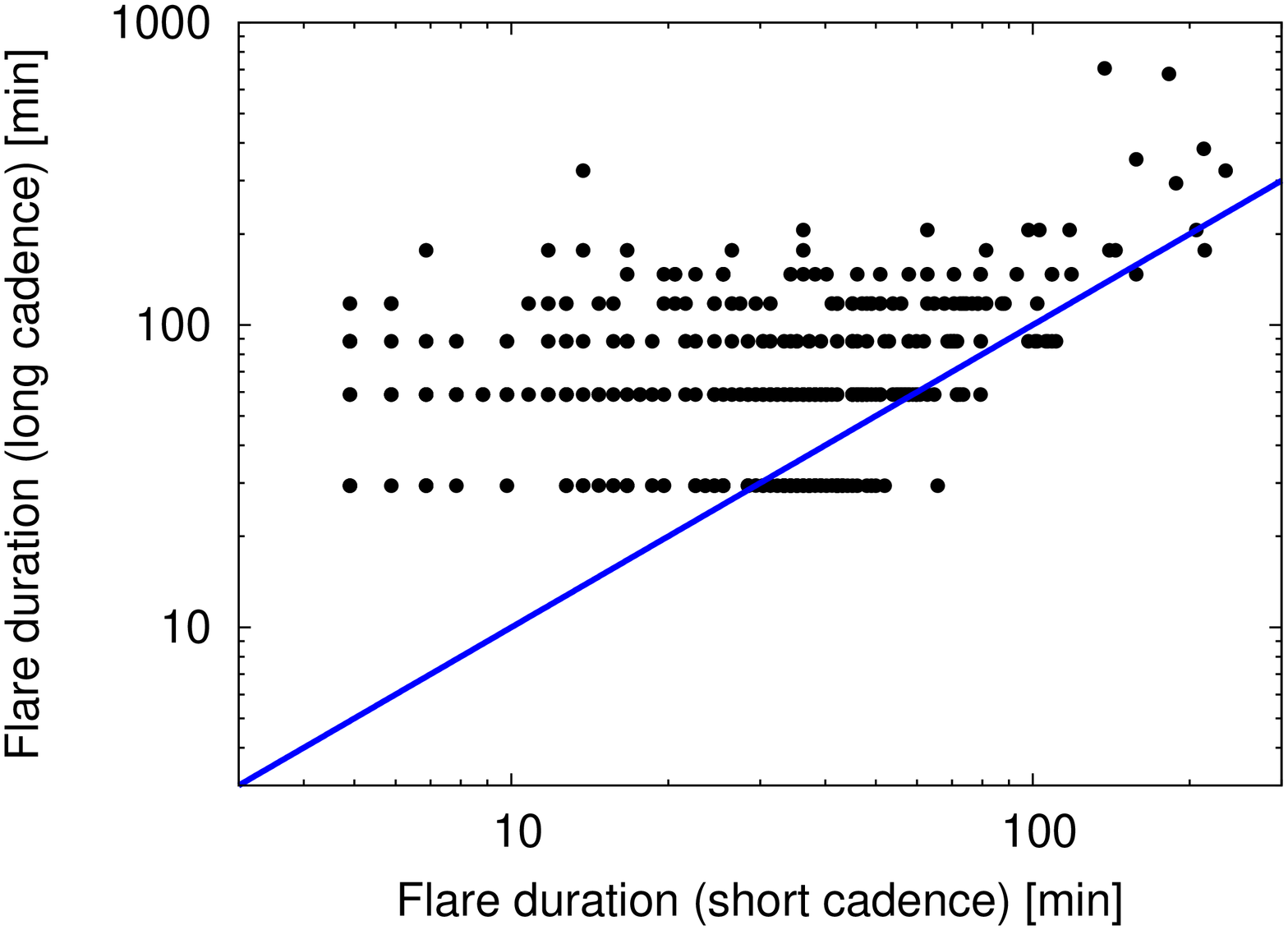}
  \includegraphics[width=0.3\textwidth,angle=270]{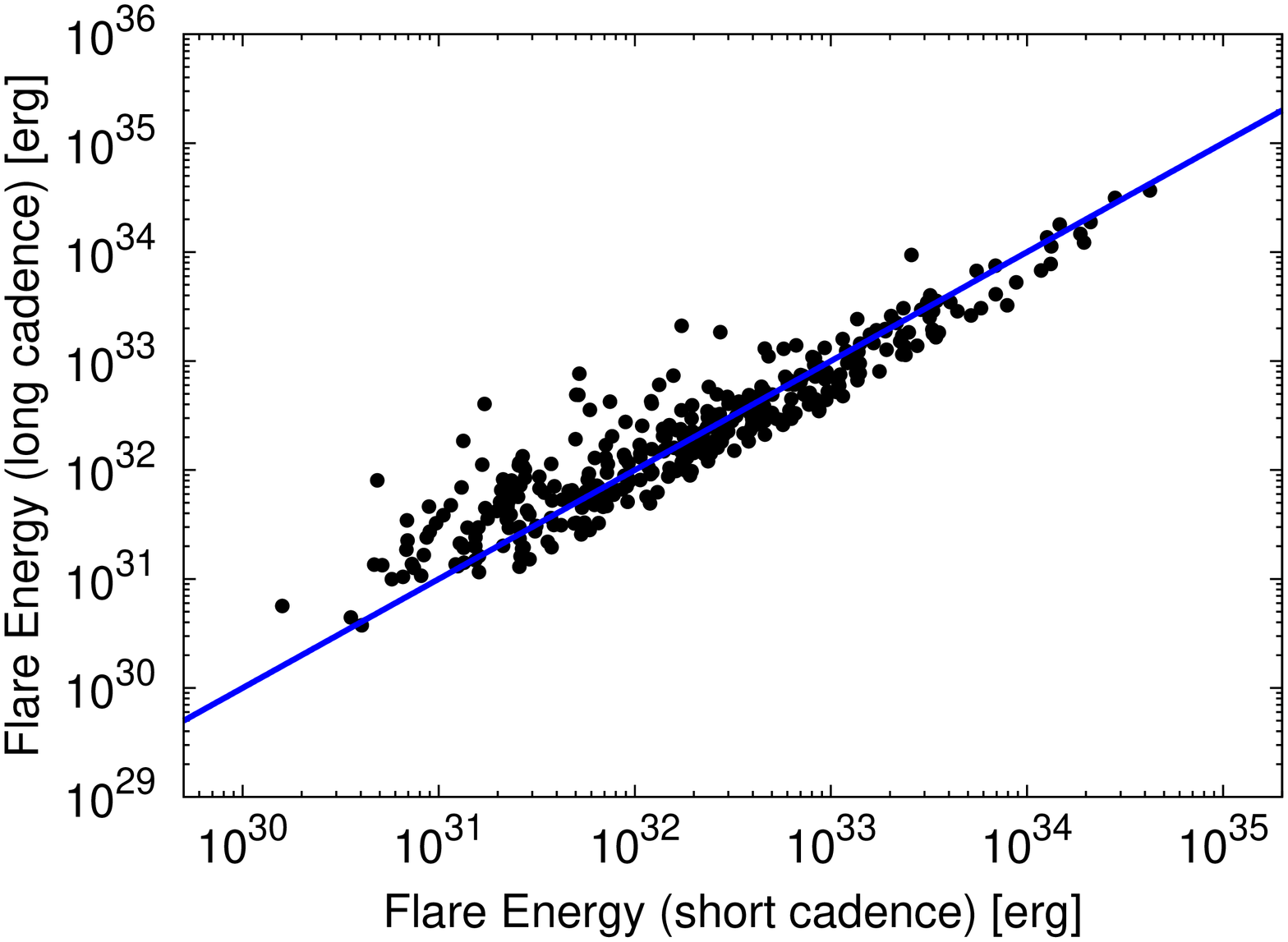}
  \caption{Comparison of the amplitude, duration, and energy between short-cadence and long-cadence flares. The blue solid line shows where long-cadence equal short-cadence values. The underestimated amplitude and overestimated duration of the long-cadence flares appear to be compensated for in the energy.}
  \label{comparison_SC_LC}
\end{figure}

\citet{2018ApJ...859...87Y} carried out a flare analysis for 5140 targets of the main \textit{Kepler} mission that have short-cadence data. The autors only used data of observing windows where long- and short-cadence data overlapped. In their comparison of the basic properties of long- and short-cadence flares they found that energies and amplitudes for long-cadence flares are underestimated and the flare durations are overestimated. Their work, however, was based on the entire \textit{Kepler} mission (Q1-Q17) and did not distinguish between different types of stars and SpTs. 

In the analysis of our \textit{K2} M dwarf sample, we found 363 long-cadence flares that have a short-cadence counterpart (358 for stars with a reliable period estimate). The comparison between long-and short-cadence flare energies, amplitudes, and durations for this subsample of flares is shown in Fig.~\ref{comparison_SC_LC}. We found that the flare amplitudes in long-cadence LCs are underestimated by on average $30\pm20\%,$ which confirms the finding by \citet{2018ApJ...859...87Y}. The long-cadence durations are overestimated by on average $60\pm40\%,$ which is a smaller factor than was found by \citet{2018ApJ...859...87Y}. The strong artificial quantization of the long-cadence flare duration shown in Fig.~\ref{comparison_SC_LC} is a consequence of the $\sim$30\,min observation cadence, that is, all long-cadence flares with 2 data points by definition have a duration of $\sim$30\,min, those with 3 points have $\sim$60\,min, those with 4 points have $\sim$90\,min, etc. The longest flare duration detected in long cadence was $\sim$700\,min, which corresponds to 24 flare points.

The bottom panel of Fig.~\ref{comparison_SC_LC} shows that the underestimated amplitude and the overestimated duration of the long-cadence flares almost compensate for each other in the flare energy. As explained in Sect.~\ref{flare_analysis}, the flare energy is proportional to the equivalent duration $ED,$ which is the integral under the flare. The longer duration and the lower amplitude of the long-cadence flares result in the same $ED$ as a short-cadence flare with higher amplitude and shorter duration. The energy of the short-cadence flares is indeed almost identical (median value of the ratio between short- and long-cadence flares is 0.99) to the long-cadence flare energy.

Our results for the flare amplitudes, durations, and energies were confirmed by a T-test. While the mean values of the durations and amplitudes of the long- and short-cadence flares are significantly different (T\,=\,-5.01, p-value\,=\,$6.95\cdotp10^{-7}$ and T\,=\,10.68, p-value\,=\,$7.44\cdotp10^{-25}$ for amplitudes and durations, respectively), we found that with a probability of $\sim61\%$ (T\,=\,-0.51, p-value\,=\,0.61), the mean values of the energy are equal for long- and short-cadence flares.

\subsection{Rotation-activity relation}

S16 studied the relation between stellar rotation and photometric activity indicators, that is, the amplitude of the rotation signal, the standard deviation of the LCs, the peak flare amplitude, and the flare frequency in long-cadence \textit{K2} LCs of Superblink M dwarfs. They found a bimodal distribution of these parameters with an abrupt change in activity level at a rotation period of $\sim$10\,d. 
%\textbf{A bimodal behavior was also seen by \citet{2015ApJ...812....3W} in the fraction of their sample M dwarfs with detectable H$_{\alpha}$ emission. Fig.~5 of \citet{2015ApJ...812....3W} shows a strong decrease in activity fraction with increasing rotation period. Furthermore, \citet{2015ApJ...812....3W} measured a decrease in strength of magnetic activity with incresing rotation period. Hence, the lower fraction of active M dwarfs with a lower level of activity for the slowest rotators is reflected by the behavior of the photometric activity indicators.}

We determined the same parameters of the photometric activity as S16 for our sample of 56 M dwarfs with short-cadence \textit{K2} LCs. First, we only consider results from the analysis of the short-cadence LCs. Subsequently, we also compare the long- and short-cadence data of our \textit{K2} M dwarf sample. The rotation periods and measured photometric activity indicators for our targets are given in Table~\ref{rot_flares_table} in the appendix.

\subsubsection{Activity diagnostics related to the rotation period}
\label{paragr:act_rot}

\begin{table*}[]
\centering
\caption{Results of the T-tests for the amplitude of rotation modulation and standard deviation of the full LC of slow and fast rotators. The top half of the table shows the full sample. The bottom half includes only stars with values higher than the value that corresponds to the average $S_{\mathrm{flat}}$ of the fast-rotating M dwarfs.}
\label{T-test-results}
\begin{tabular}{lcccc}
\hline \hline
Parameter & mean value ($P_{\rm rot}<10$\,d)  & mean value ($P_{\rm rot}\geq10$\,d) & T-statistic &  p-value  \\\hline
$R_{\mathrm{per}}$ & $2.2\pm1.1\%$ & $0.8\pm0.6\%$ & 5.1 & $6.2\cdotp10^{-6}$ \\
$S_{\mathrm{ph}}$ & $18531\pm9466$\,ppm & $4518\pm3867$\,ppm & 7.2 & $5.8\cdotp10^{-9}$ \\\hline
$R_{\mathrm{per},>0.52\%}$ & $2.2\pm1.1\%$ & $1.2\pm0.6\%$ & 3.7 & $8.5\cdotp10^{-4}$ \\
$S_{\mathrm{ph},>2600\,\rm ppm}$ & $18531\pm9466$\,ppm & $5378\pm4088$\,ppm & 6.0 & $6.8\cdotp10^{-7}$\\\hline
\hline
\end{tabular}
\end{table*}

\begin{figure}
  \centering
  \includegraphics[width=0.3\textwidth,angle=270]{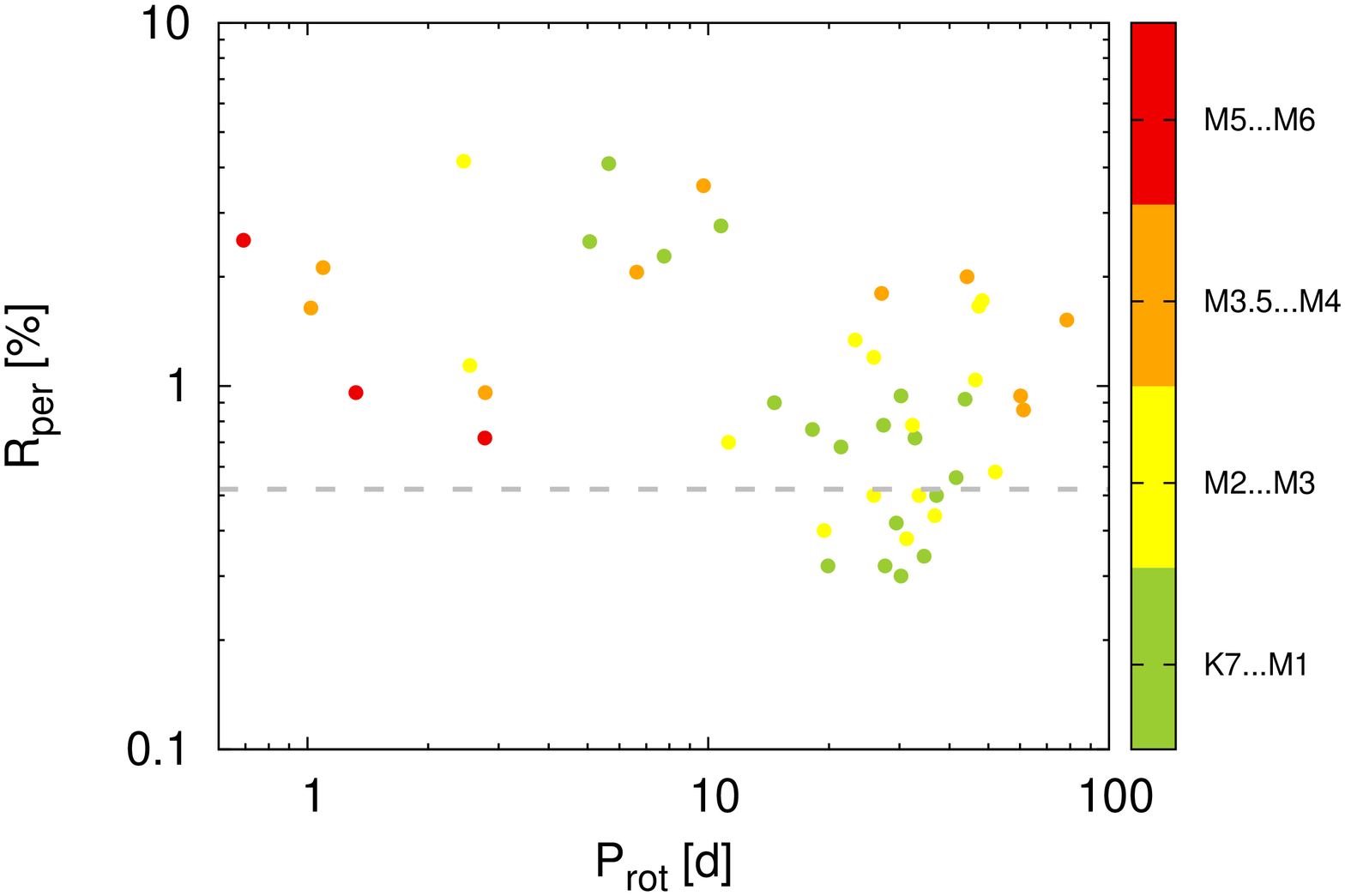}
  \includegraphics[width=0.3\textwidth,angle=270]{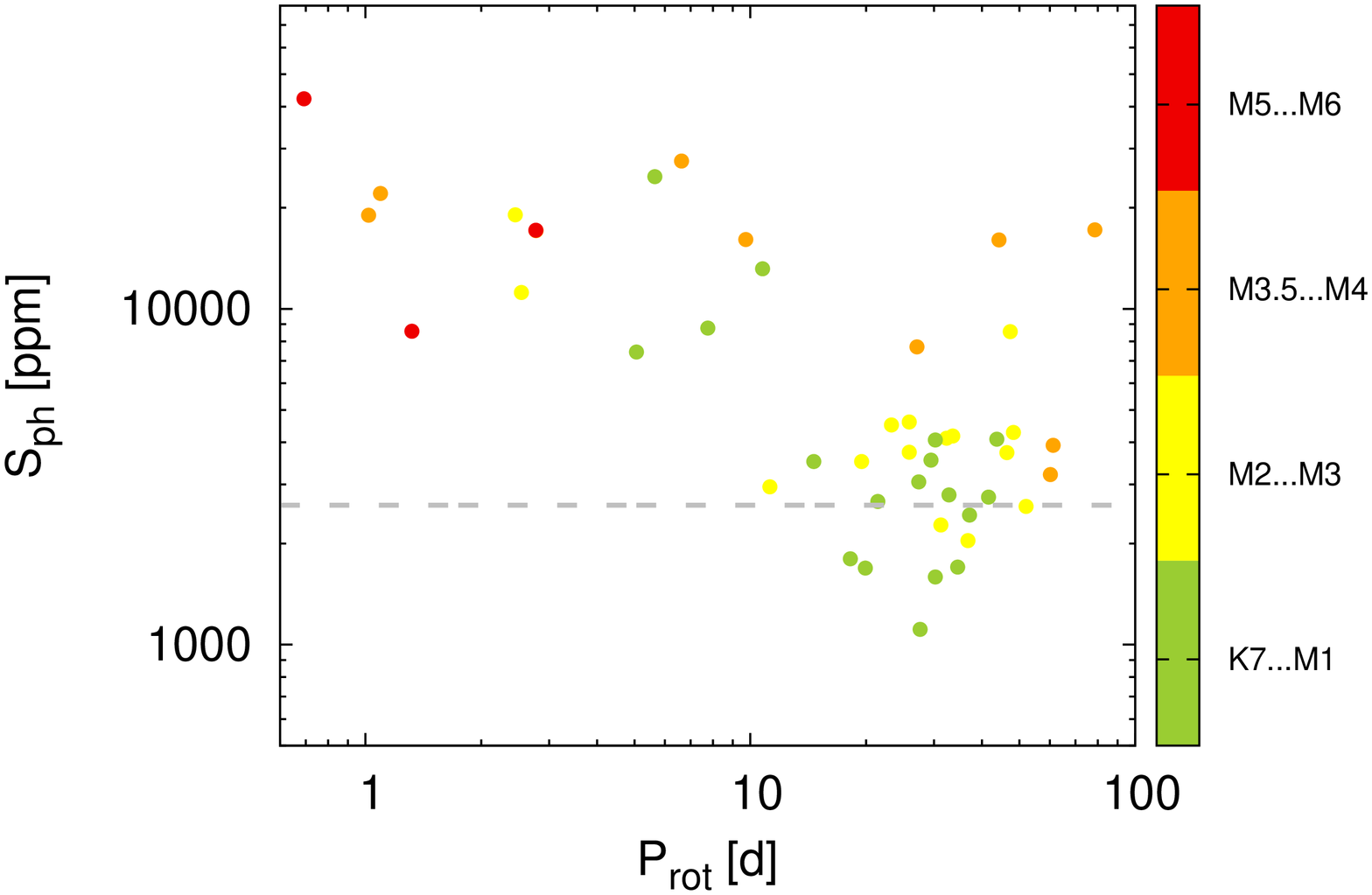}
  \caption{Full amplitude of the rotation signal given in percent ($R_{\mathrm{per}}$, top) and the standard deviation of the full LC ($S_{\mathrm{ph}}$, bottom) vs. the rotation period for the short-cadence \textit{K2} LCs of our sample targets. The dashed gray line corresponds to the average $S_{\mathrm{flat}}$ of the fast-rotating M dwarfs. All values above this line were considered for our statistical analysis to be unaffected by a detection bias (see text).}
  \label{period_Rper_Sph}
\end{figure}

Fig.~\ref{period_Rper_Sph} shows our values from the short-cadence data for the full amplitude of the rotation signal given in percent ($R_{\mathrm{per}}$) as defined by \citet{2013MNRAS.432.1203M} and the standard deviation of the full LC \citep[$S_{\mathrm{ph}}$,][]{2014JSWSC...4A..15M}, which is dominated by the rotation modulation because the flares only contribute a few points compared to the total number of points in the LC. Both parameters show the same behavior with $P_{\rm rot}$ as was reported in the long-cadence data by S16. For a more quantitative analysis, we grouped the targets into fast rotators ($P_{\rm rot}<10$\,d) and slow rotators ($P_{\rm rot}\geq10$\,d), determined the mean and standard deviation of $R_{\mathrm{per}}$ and $S_{\mathrm{ph}}$ for each group, and performed a Student's T-test.  

The T-statistic T\,=\,5.1 and the p-value\,=\,$6.2\cdotp10^{-6}$ of $R_{\mathrm{per}}$ indicate that the two groups have significantly different means ($R_{\mathrm{per}}(P_{\rm rot}<10$\,d)\,=\,$2.2\pm1.1\%$, $R_{\mathrm{per}}(P_{\rm rot}\geq10$\,d)\,=\,$0.8\pm0.6\%$). The T-test for $S_{\mathrm{ph}}$ yielded an even more convincing result. We found that the mean of the fast rotators, $S_{\mathrm{ph}}(P_{\rm rot}<10$\,d)\,=\,$18531\pm9466$\,ppm, and the mean of the slow rotators, $S_{\mathrm{ph}}(P_{\rm rot}\geq10$\,d)\,=\,$4518\pm3867$\,ppm, is significantly different (T\,=\,7.2, p-value\,=\,$5.8\cdotp10^{-9}$). To rule out the possibility of a detection bias, we repeated the test with only those data that lie above a given detection threshold. Specifically, we considered all measurements above 0.0026 ($R_{\mathrm{per}}>0.52\%$ for full amplitudes, $S_{\mathrm{ph}}>2600$\,ppm; this threshold is shown in Fig.~\ref{period_Rper_Sph} with horizontal line), which corresponds to the average $S_{\mathrm{flat}}$ of the fast-rotating M dwarfs. This reduced the sample of the slow rotators by 11 and 8 for $R_{\mathrm{per}}$ and $S_{\mathrm{ph}}$, respectively. By definition, all fast rotators are already above this threshold so that the mean values for this group do not change. The mean values of $R_{\mathrm{per}}(P_{\rm rot}\geq10$\,d)\,=\,$1.2\pm0.6\%$ and $S_{\mathrm{ph}}(P_{\rm rot}<10$\,d)\,=\,$5378\pm4088$\,ppm for the slow rotators are still significantly different from those of the fast-rotator groups (T\,=\,3.7, p-value\,=\,$8.5\cdotp10^{-4}$ and T\,=\,6.0, p-value\,=\,$6.8\cdotp10^{-7}$ for $R_{\mathrm{per}}$ and $S_{\mathrm{ph}}$, respectively). All results of the T-tests are also summarized in Table~\ref{T-test-results}. We thus verified that slow- and fast-rotating M dwarfs have systematically different amplitudes of the rotation modulation.

For heavily spotted super active stars, the spot amplitudes decrease for an increasing level of activity. This anticorrelation requires spots that cover $>$50\% of the stellar surface \citep[e.g.,][]{2017ApJ...834...85N}. We estimated the spot coverage of our targets using the relation given in \citet{2019ApJ...876...58N} and found a maximum spot filling factor of $\sim$9\%. This shows that the amplitude of the spot modulation is a reliable activity indicator for our sample.

\subsubsection{Flares}
\label{paragr:act_flares}

\paragraph{Detection sensitivity:}

\begin{figure}
  \centering
  \includegraphics[width=0.3\textwidth,angle=270]{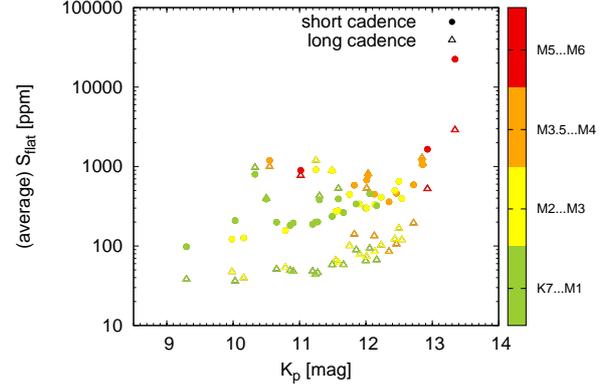}
  \caption{Average standard deviation of the flattened and cleaned LC, $S_{\rm flat}$, vs. $K_{p}$, the magnitude in the \textit{Kepler} band. Open triangles represent long-cadence LCs, and filled circles show short-cadence LCs.}
  \label{Kp_SFlat_SC}
\end{figure}

The detection of flares relies on the identification of upward outliers in the LC, which in turn depends on the noise level of the stars. We quantify this noise as the standard deviation of the LC after the rotational modulation and the outliers, $S_{\rm flat}$ (see Sect.~\ref{flare_detection}), are removed. This parameter depends on the apparent brightness of the star: fainter stars display a higher noise level. In Fig.~\ref{Kp_SFlat_SC} we show the relation between $S_{\rm flat}$ and $K_{\rm p}$ for our sample, displaying $S_{\rm flat}$ values from the short- and long-cadence LCs. For LCs that consist of more than one segment (see Sect.~\ref{flare_detection}), $S_{\rm flat}$ is the average of the values for all segments. The increase in $S_{\rm flat}$ for fainter stars is clearly visible. The $S_{\rm flat}$ in the short-cadence data is systematically higher than in the long-cadence data by factor 3.6 on average. Reasons for this discrepancy might be instrumental noise, which is averaged out in the binning process from short to long cadence, as well as an additional unresolved astrophysical signal that we explain in more detail below.

Our flare detection algorithm requires that each data point is at least 3$\sigma$ above $S_{\rm flat}$. The range of detection thresholds covered by our sample stars therefore needs to be considered when we interpret the activity diagnostics that we extract from the LCs. 

%Clearly, the increase of $S_{\rm flat}$ for fainter stars is seen but for a given $K_{\rm p}$ some scatter is superposed

\paragraph{Activity diagnostics related to flares:}

\begin{figure}
  \centering
  \includegraphics[width=0.3\textwidth,angle=270]{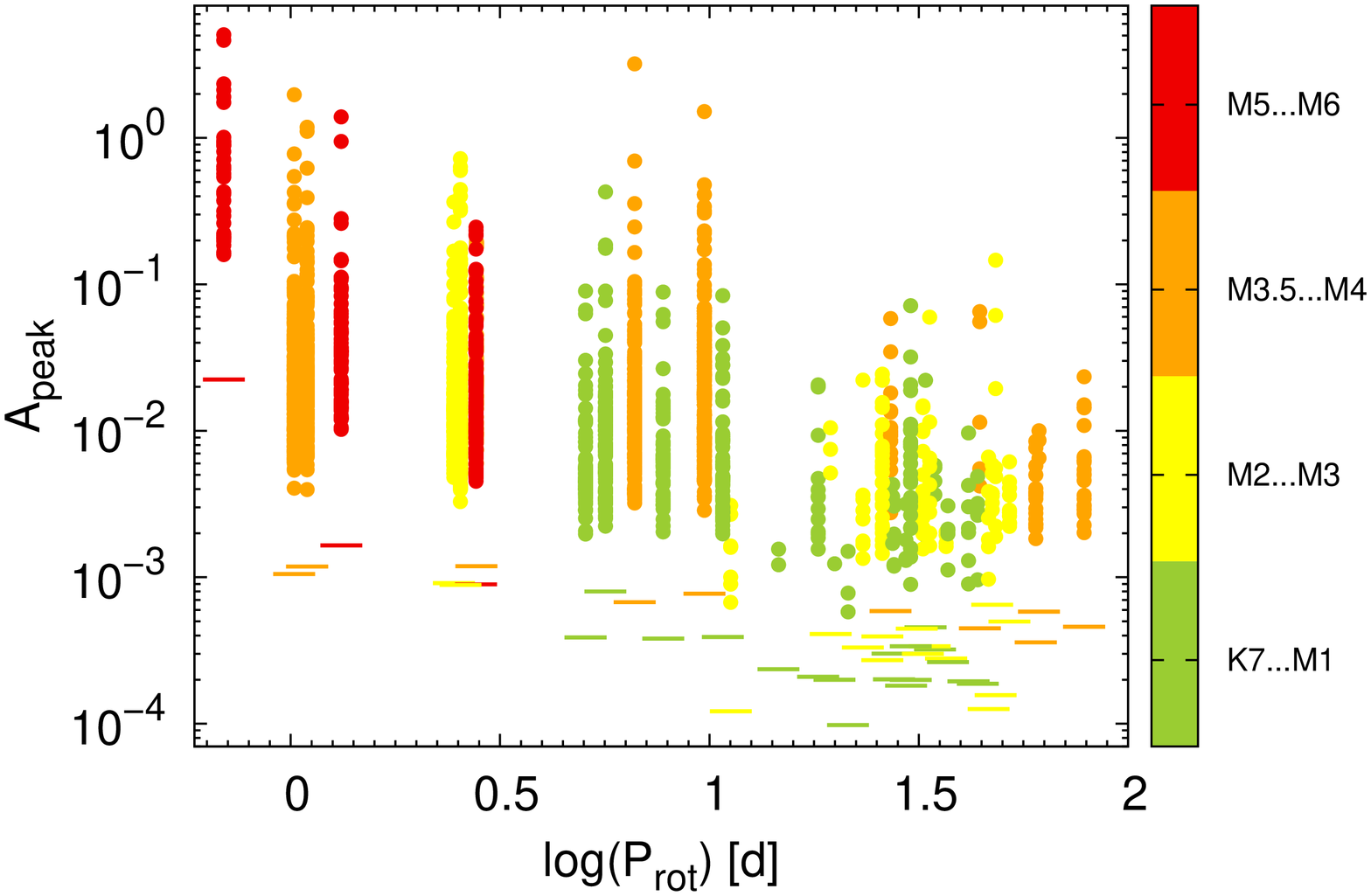}
  \includegraphics[width=0.3\textwidth,angle=270]{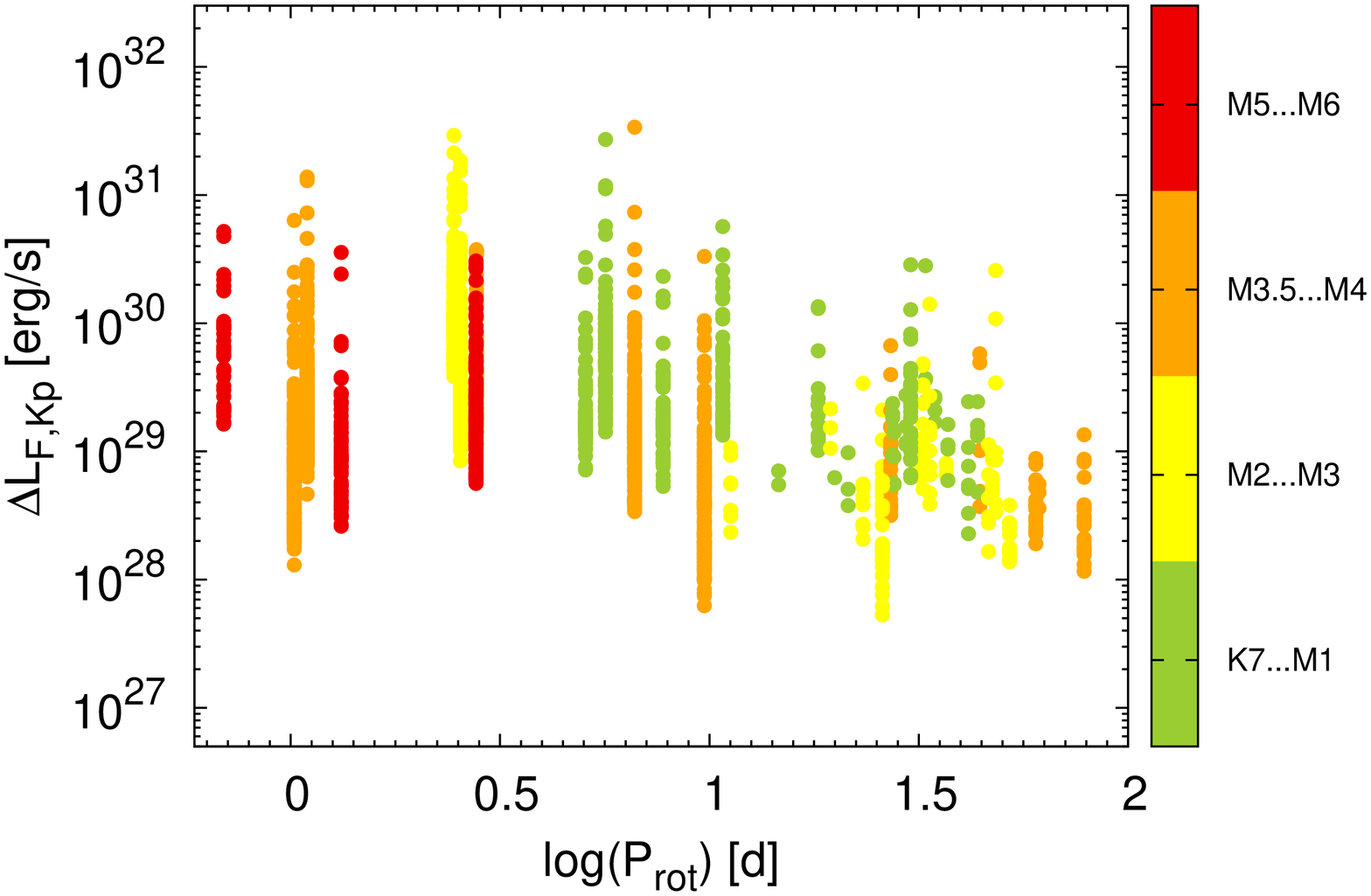} 
  \includegraphics[width=0.3\textwidth,angle=270]{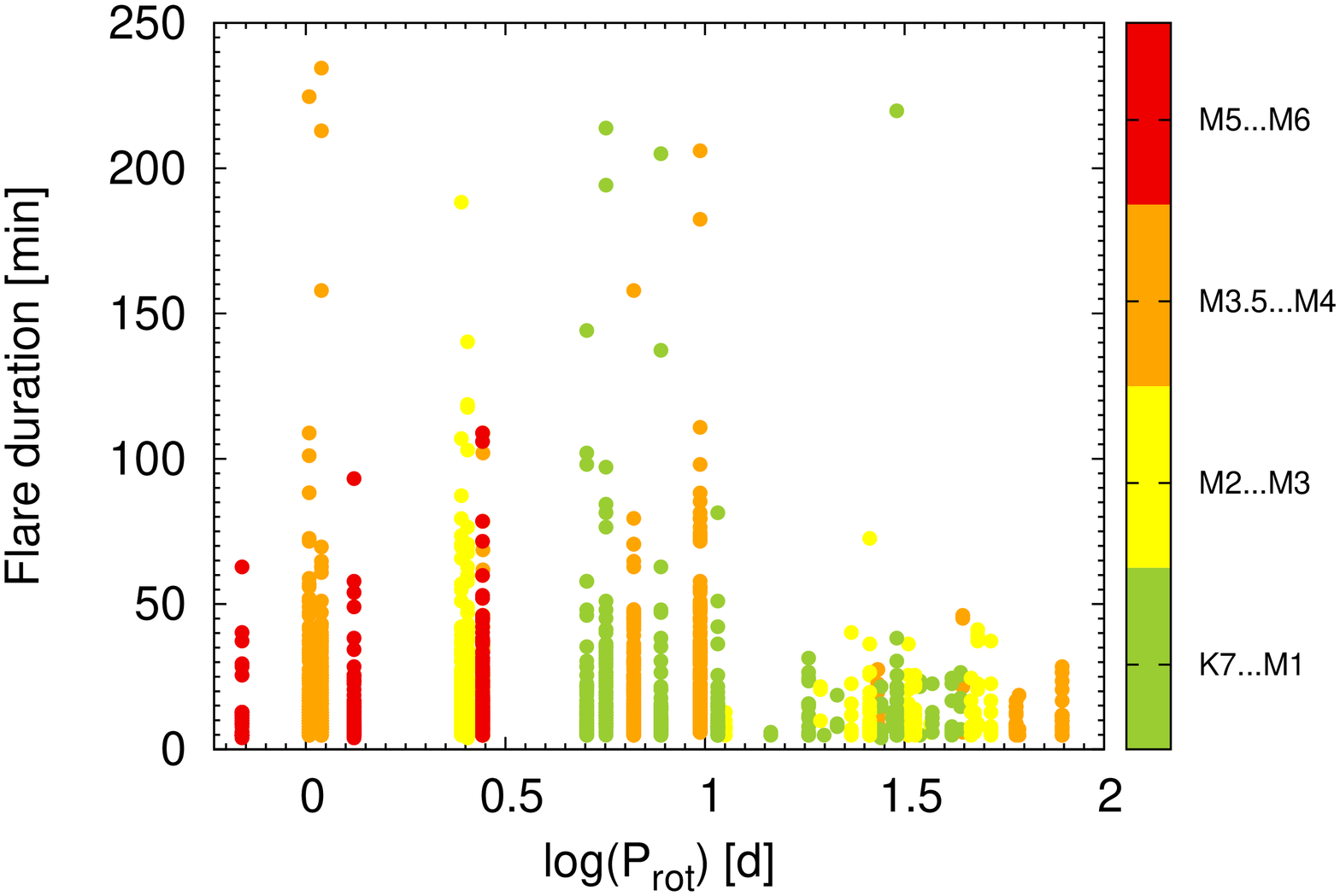} 
  \caption{Relation between the stellar rotation period and the flare parameters: relative amplitude (top), absolute amplitude (middle), and duration (bottom) for all flares detected in the short-cadence data. The small horizontal bars in the top panel represent the standard deviation of the cleaned and flattened LCs ($S_{\mathrm{flat}}$), which is a measure of our detection threshold.}
  \label{period_A_L_A_D_SC}
\end{figure}

\begin{figure}
  \centering
  \includegraphics[width=0.3\textwidth,angle=270]{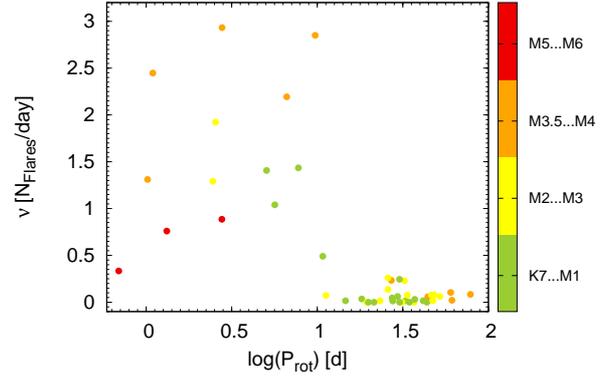}
  \caption{Short-cadence flare frequency vs. rotation period for all 46 stars with measurable rotation periods.}
  \label{period_Flare_Frequency}
\end{figure}

In addition to $R_{\mathrm{per}}$ and $S_{\mathrm{ph}}$ ,  the flare parameters also show a different behavior for fast and slow rotators. The top and bottom panels of Fig.~\ref{period_A_L_A_D_SC} show the rotation-activity relation in terms of relative flare amplitude ($A_{\mathrm{peak}}$), and the flare duration ($\Delta t_{\mathrm{Flare}}$) for all short-cadence flares. The peak flare amplitude seems to lack low-amplitude and high-amplitude flares for the fast and slow rotators, respectively. The standard deviation of the cleaned and flattened LCs ($S_{\mathrm{flat}}$) are given as the small horizontal bars in the top panel of Fig.~\ref{period_A_L_A_D_SC} as an indication of our detection threshold of 3$S_{\mathrm{flat}}$. This means that we may have missed low-amplitude flares for the fast rotators, but the dearth of high-amplitude flares for the slow rotators is real. 

Because our sample contains a range of different $K_p$ magnitudes from 8.9 - 13.3\,mag with different detection sensitivities, the comparison of relative amplitudes might be biased. To avoid this problem, we converted the relative values $A_{\mathrm{peak}}$ into absolute values in terms of luminosities $\Delta L_{\mathrm{F,Kp}} =L_{\mathrm{Kp,0}}\,\cdotp A_{\mathrm{peak}}$, where  $L_{\mathrm{Kp,0}}$ is the quiescent luminosity that can be associated with the flattened LCs. The middle panel of Fig.~\ref{period_A_L_A_D_SC} shows $\Delta L_{\mathrm{F,Kp}}$ over the rotation period. The overall bimodal shape of two activity levels can still be recognized, although not as strongly as for the relative amplitudes. The two plots clearly show that the slow rotators do not have very high amplitude flares. This finding confirms the results of S16, which was based on long-cadence LCs.

\begin{figure}
  \centering
  \includegraphics[width=0.3\textwidth,angle=270]{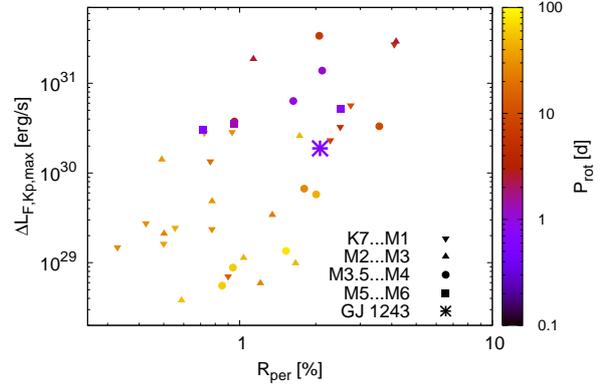}
  \caption{Correlation between the amplitude of spot modulation and the maximum of the absolute flare amplitude for each star. The highly active M4 star, GJ\,1243 \citep{2014ApJ...797..121H}, is given as a comparison to our sample.}
  \label{Rper_maxL_Apeak_SC}
\end{figure}

\citet{2019ApJ...876...58N} found a correlation between amplitude of spot modulation and the occurrence rate of superflares in solar-like stars. They concluded that the majority of these strongest flares occur on stars with large spots. As shown in Fig.~\ref{Rper_maxL_Apeak_SC}, we found that the flare amplitude $\Delta L_{\mathrm{F,Kp}}$ is correlated with the spot amplitude, which was shown to be a reliable activity indicator in Sect.~\ref{paragr:act_rot}, for the stars in our sample. Fig.~\ref{Rper_maxL_Apeak_SC} also shows the fast-rotating ($P_{\rm rot} = 0.59\,\rm d$) highly active M4 star GJ\,1243 \citep[values extracted from][]{2014ApJ...797..121H}. GJ\,1243, which falls in our third SpT group (M3.5-M4), follows the same correlation as the stars in our \textit{K2} M dwarf sample.

For the fast rotators we found flares with durations ranging from 4 - 230\,min (4\,min is the default minimum duration set by the detection algorithm, see Sect.~\ref{flare_analysis}), while there are hardly any flares with durations longer than 50\,min in the slow-rotator regime. Hence, the short cadence data allows us here to show for the first time that the same dichotomy between fast and slow rotators is present in the duration of flares: the flares on stars with $P_{\rm rot} > 10$\,d are systematically shorter than the flares on faster rotators. The bottom panel of Fig.~\ref{period_A_L_A_D_SC} also shows that across the full range of rotation periods, the durations of most of the flares that were detected in short-cadence data are about $1$\,hr or shorter. These short events are the reason that the flare rate in long-cadence data is underestimated, as we explain in more detail in Sect.~\ref{comp_SC_LC_rotation}. 

%Finally, the flare energy, which is shown in the bottom plot in Fig.~\ref{period_L_Flare_Amplitude_Energy_SC}, is very similar to the amplitude $\Delta L_{\mathrm{F,Kp}}$. The slow rotators do not flare with very high energies which is another confirmation of the previous findings as the detected flares for this target group have short durations and low amplitudes.

By dividing the number of flares in each LC by the duration of the corresponding observing campaign, we derived the flare frequency $\nu$[$N_{\mathrm{flares}}/d$]. Fig.~\ref{period_Flare_Frequency} shows the $\nu$[$N_{\mathrm{flares}}/d$] over $P_{\rm rot}$ for all 46 stars with measurable rotation periods. Clearly, the behavior for fast and slow rotators is bimodal. We found values ranging from 0.04 to 2.9 $N_{\mathrm{flares}}/d$ and from 0 to 0.65 $N_{\mathrm{flares}}/d$ for the fast and slow rotators, respectively. 

\begin{figure}
  \centering
  \includegraphics[width=0.3\textwidth,angle=270]{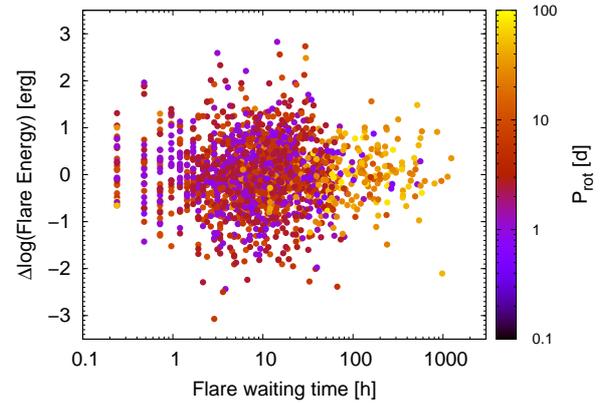}
  \caption{Difference in flare energy between two flares over the waiting time for all short-cadence flares. No correlation is found, i.e., the flare energy is not higher when the waiting time was longer.}
  \label{Flare_waiting_time_Energy}
\end{figure}

The time between two flares in a single LC is usually called waiting time. In Fig.~\ref{Flare_waiting_time_Energy} we plot the difference between the energy of two consecutive flares over the waiting time for all 1644 flares on stars with measured rotation period. We did not find any correlation between waiting time and energy (Pearson correlation coefficient: 0.003). This confirms the results of \citet{2014ApJ...797..121H} for GJ\,1243. The slowest rotators tend to have the longest waiting time. This is a result of their low flare frequency, as discussed above.

To summarize, all flare parameters show a change in activity level at a period of $\sim 10$\,d. This change is most drastic in the flare duration (bottom panel Fig.~\ref{period_A_L_A_D_SC}) and the flare frequency (Fig.~\ref{period_Flare_Frequency}). The same behavior has previously been noted for the flare frequency in the study of S16 with long-cadence data, but it is presented here for the first time for the flare duration. We do not see any dependence on SpT. Figures~\ref{period_A_L_A_D_SC} and \ref{period_Flare_Frequency} show that the change in activity is present in early- and late-M dwarfs. Only for the latest SpT bin (SpT group M5-M6) are we unable to draw any conclusion because this group only includes three stars that are all fast rotators. 

\paragraph{Activity diagnostics hidden in the noise:}

\begin{figure}
  \centering
  \includegraphics[width=0.3\textwidth,angle=270]{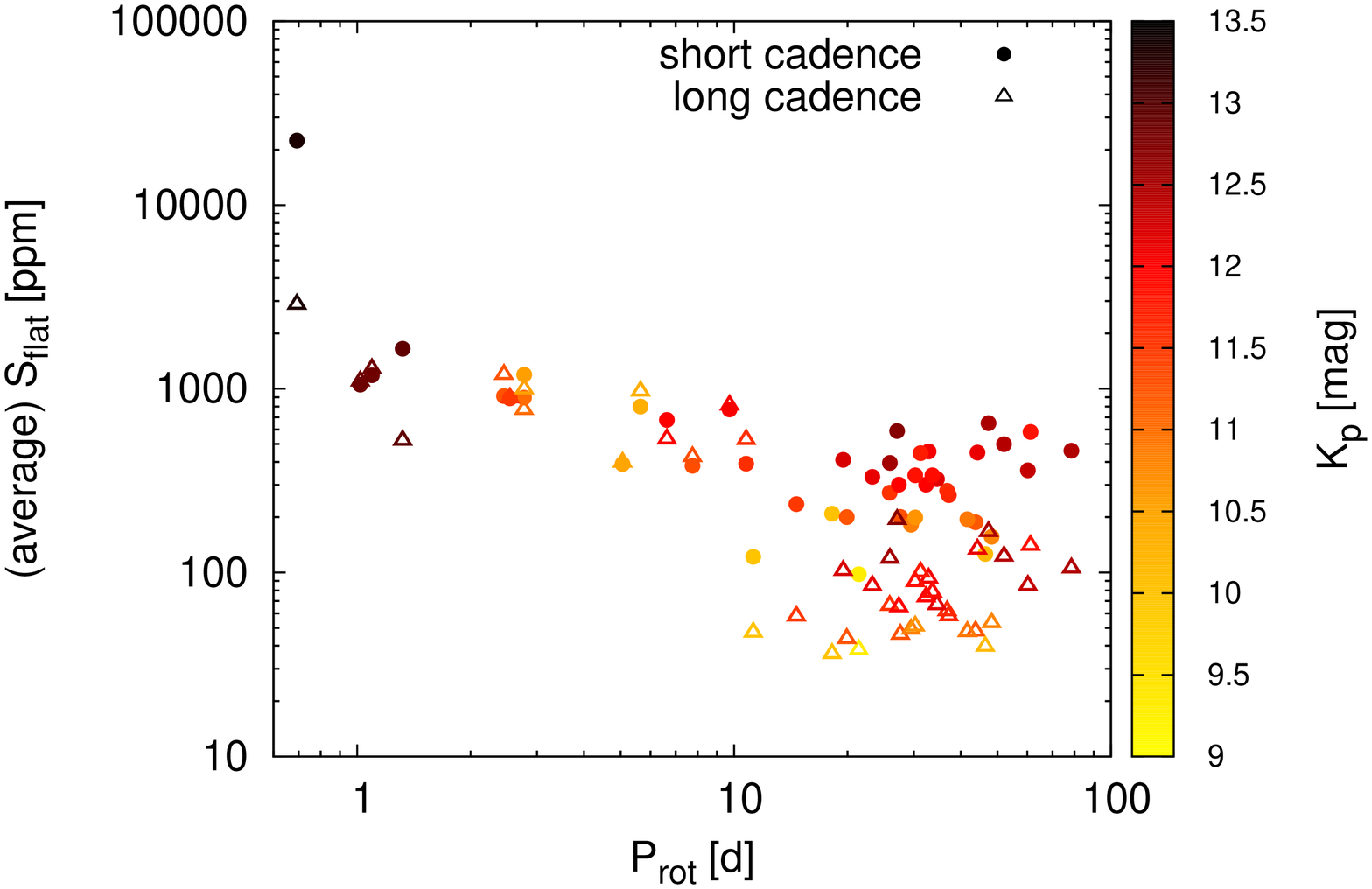}
  \caption{Average standard deviation of the flattened and cleaned LC vs. rotation period for all 46 stars with measurable rotation periods. Open triangles show long-cadence LCs, and filled circles represent short-cadence LCs.}
  \label{period_SFlat_SC}
\end{figure}

The bimodality in terms of rotation period for all activity diagnostics discussed above was found by S16 to also be present in the standard deviation of the flattened and cleaned LC, $S_{\rm flat}$, for long-cadence LCs. This was interpreted to mean that additional unresolved astrophysical processes were hidden in the noise level. We examined the relation between $S_{\rm flat}$ and $P_{\rm rot}$ here for the first time for short-cadence data (see Fig.~\ref{period_SFlat_SC}). We see a trend of increasing $S_{\rm flat}$ for faster rotators. Compared to long-cadence data for the same stars, the bimodality is much less evident in the short-cadence data. This is a result of the upward shift of the $S_{\rm flat}$ values in short-cadence with respect to long-cadence data that we described in the discussion of the detection sensitivity (see Fig.~\ref{Kp_SFlat_SC}). Interestingly, Fig.~\ref{period_SFlat_SC} shows that only short-cadence slow rotators are affected by the increased $S_{\rm flat}$ values, while in the fast-rotator regime, $S_{\rm flat}$ is very similar for long- and short-cadence LCs. In relation to Fig.~\ref{Kp_SFlat_SC}, we also noted a spread of $S_{\rm flat}$ for given $K_{\rm p}$ that cannot be due to a sensitivity issue. Fig.~\ref{period_SFlat_SC} shows us that this spread originates in the different rotation periods of the stars combined with the empirical dividing line of strong activity versus weak activity at $P_{\rm rot}\sim 10$\,d. To rule out that the difference in $S_{\rm flat}$ between fast and slow rotators is a result of the use of different boxcar widths in the flare-finding procedure, we selected a few fast and slow rotators from our sample and derived $S_{\rm flat}$ with various combinations of boxcar widths. For the more than 4000 tested combinations, which included the combinations that were used, we determined the range of $S_{\rm flat}$ values for a given star. We find the uncertainties due to our analysis method to be about a few dozen ppm for the slow rotators and a few hundred ppm for the fast rotators. These values are too low, and the choice of the boxcar width is therefore not responsible for the observed difference of $S_{\rm flat}$ between the fast- and slow-rotator regime ($\text{approximately}$ one order of magnitude in the long-cadence data, see Fig.~\ref{period_SFlat_SC}). We thus confirm the conclusion by S16 that intrinsic stellar properties are hidden in the standard deviation of the LC after rotation and flares are removed. In the same vein, we can interpret the fact that for a given star, $S_{\rm flat}$ is higher in short-cadence LCs than in long-cadence data: regardless of the astrophysical feature that is contained in $S_{\rm flat}$, for example, unresolved mini-flares or mini-spots, it has a stronger influence on short-cadence data than on long-cadence data because these very short and low-amplitude events are smeared out even more for long-cadence than for short-cadence data (see Sect.~\ref{sect:comp_lc_sc_flares}). This means that the long cadence data have a higher sensitivity for the bimodality in $S_{\rm flat}$ versus $P_{\rm rot}$.

\subsubsection{Comparison between short and long cadence}
\label{comp_SC_LC_rotation}

\begin{figure}
  \centering
  \includegraphics[width=0.3\textwidth,angle=270]{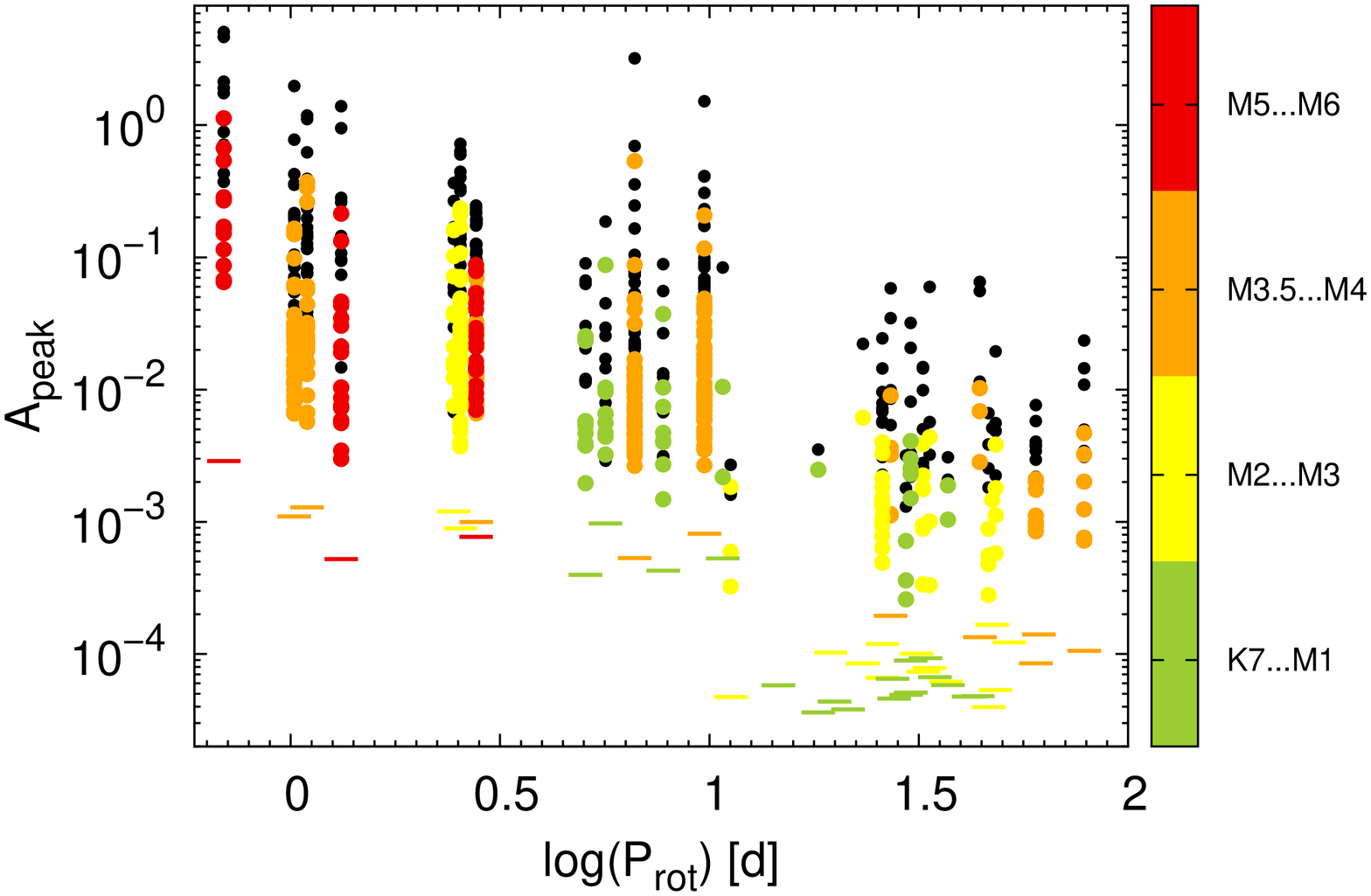} 
  \includegraphics[width=0.3\textwidth,angle=270]{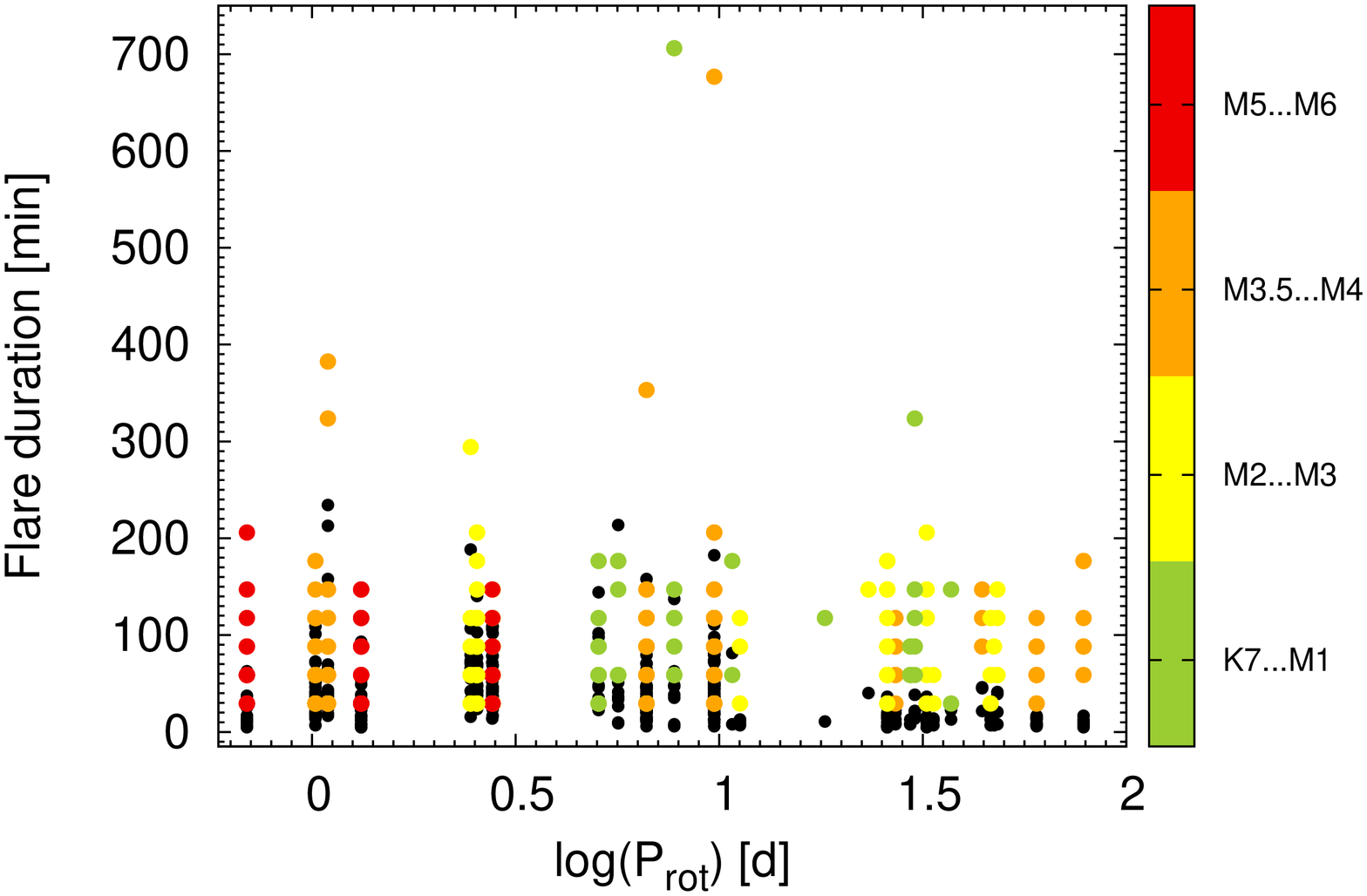}
  \includegraphics[width=0.3\textwidth,angle=270]{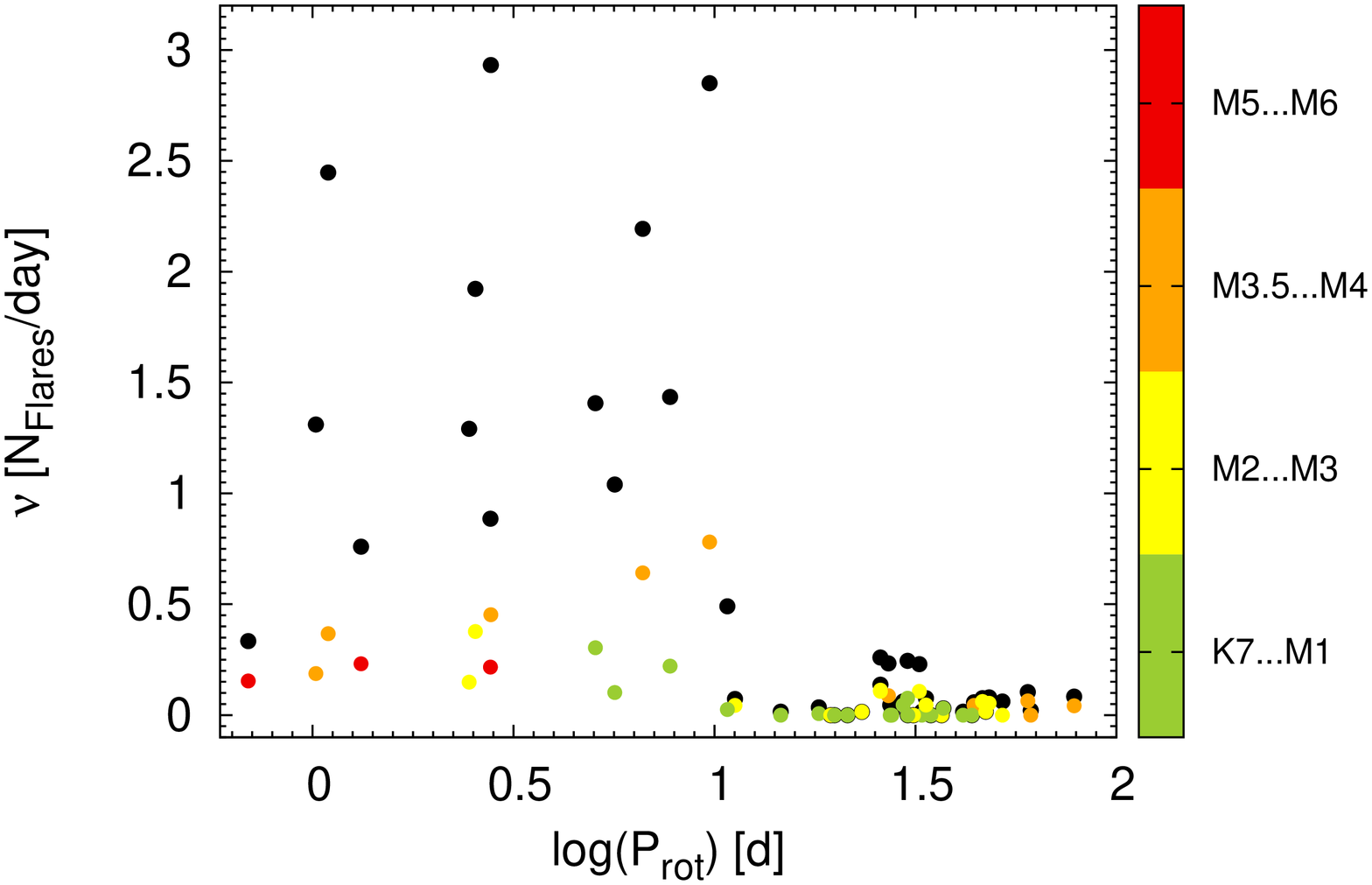}
  \caption{Flare amplitude (top), flare duration (middle), and flare frequency (bottom) vs. rotation period for all targets with measurable rotation periods that show flares in long cadence with a short-cadence counterpart. Black symbols denote short-cadence flares (see Fig.~\ref{period_A_L_A_D_SC} top and middle panel, and Fig.~\ref{period_Flare_Frequency}). Short-cadence flares without a long-cadence counterpart are not considered. The small horizontal bars in the top panel represent $S_{\rm flat}$ for the long-cadence LCs.}
  \label{period_Flare_A_D_LC_SC}
\end{figure}

In Sect.~\ref{paragr:act_flares} we studied the relation between the rotation period and the main flare parameters (amplitude and duration) for the short-cadence data. To determine whether the data-point cadence introduced biases in studies that are based on long-cadence LCs, we compared our short-cadence results to those obtained using the long-cadence LCs of the same stars.

Fig.~\ref{period_Flare_A_D_LC_SC} shows the final comparison. The short-cadence data shown are a subsample of Fig.~\ref{period_A_L_A_D_SC} (top and bottom) because short-cadence flares without a long-cadence counterpart are not considered in Fig.~\ref{period_Flare_A_D_LC_SC}. As shown in Sect.~\ref{sect:comp_lc_sc_flares}, we again see that the flare amplitudes are underestimated in long-cadence data, but the durations are overestimated. 

For the flare amplitude (top panel of Fig.~\ref{period_Flare_A_D_LC_SC}) the change in behavior at $P_{\rm rot}\sim 10$\,d is very similar in long-cadence to the short-cadence data. Although the values of the flare amplitudes are lower in long-cadence than in short-cadence data, the difference in the mean values of $A_{\rm peak}$ of the largest flare for each star (upper envelope) between fast and slow rotators is the same within the error bars ($\sim$1.75 orders of magnitude). For the lower envelope, however, the change in activity level at $P_{\rm rot}\sim 10$\,d is more obvious for the long-cadence than for the short-cadence data. This is a consequence of the detection sensitivity that we discussed in the previous section. Because $S_{\rm flat}$ (given as the small bars in the top panel of Fig.~\ref{period_Flare_A_D_LC_SC}) was found to have higher sensitivity for the bimodality in the long-cadence data (see Fig.~\ref{period_SFlat_SC}), we found the same trend in the flare amplitudes.

Interestingly, the bimodality of the activity levels is only barely visible for the durations of long-cadence flares. The middle panel of Fig.~\ref{comparison_SC_LC} shows that the durations are more overestimated for shorter flares. This is a consequence of the data-point cadence of $\sim$30\,min, which causes a strong artificial quantization.

\begin{figure}
  \centering
  \includegraphics[width=0.3\textwidth,angle=270]{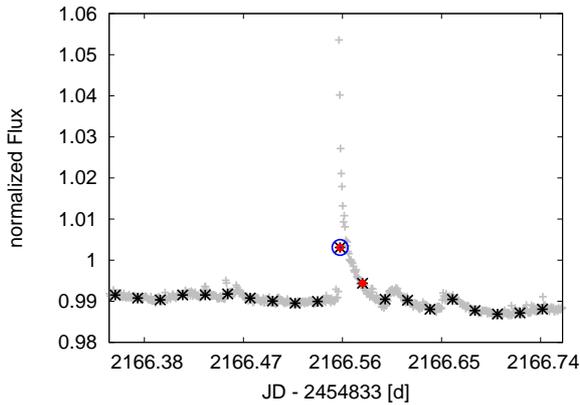}
  \caption{Example for a two-point flare in our long-cadence LCs. One point is flagged as ``impulsive outlier''. This flare was not detected by S16 because they removed impulsive outliers. The black points give our long-cadence LC. Red marks the points that were found as flare candidates by our detection algorithm. The blue circle shows the point that is flagged as an impulsive outlier. The short-cadence LC of the flare is shown in gray for comparison.}
  \label{Epic20626336_LC_Flare78}
\end{figure}

In the fast-rotator regime, our study of the short-cadence data obtained far higher flare rates, up to 2.9 $N_{\mathrm{flares}}/d$ versus values $< 0.2\,N_{\mathrm{flares}}/d$ detected by S16 for the same range of rotation periods in long-cadence data. In general, the flare frequency is significantly lower in the long-cadence data (see Fig.~\ref{period_Flare_A_D_LC_SC}, bottom). For the fast-rotator regime from our long-cadence data, we found an average flare frequency of $0.31 N_{\mathrm{flares}}/d,$ which is higher than the frequency reported by S16 ($0.11 N_{\mathrm{flares}}/d$). The two main reasons for the disagreement are differences in the flare validation process and the use of different LCs. S16 used the vdb-LCs, from which data points flagged as ``impulsive outliers'' and, hence, real flares, were removed. An example of a flare consisting of two points in long-cadence data of which one point was flagged as  impulsive outlier is shown in Fig.~\ref{Epic20626336_LC_Flare78}. This flare was consequently not detected by  S16. 

By comparing our long- and short-cadence data, we found for the fast rotator regime a flare rate that is $\text{about five}$ times higher for the short-cadence data and $\text{about three}$ times higher in the slow-rotator regime. In the whole sample, the analysis of the short-cadence LCs results in a 4.6 times higher flare rate than the analysis of the long-cadence data. This value is consistent with the true flare rate of \citet[][$\text{approximately }$4.1 times higher flare rate in short cadence, see their Table~4]{2018ApJ...859...87Y}. 

Almost all flares detected in slow rotators in the short-cadence data (right part of Fig.~\ref{period_Flare_Frequency}) have durations $< 1$\,hr. In the long-cadence data these events would either have gone unnoticed or, as we showed in Sect.\ref{sect:comp_lc_sc_flares}, would be detected with a longer duration. In our long-cadence LCs we find the flare frequency on stars with $P_{\rm rot} > 10$\,d to be near to zero (average flare rate $0.02 N_{\mathrm{flares}}/d$). Only 33\,\% of the short-cadence flares are also detected in long-cadence data.

\subsection{Relation of flare duration to amplitude}

\begin{figure}
  \centering
  \includegraphics[width=0.3\textwidth,angle=270]{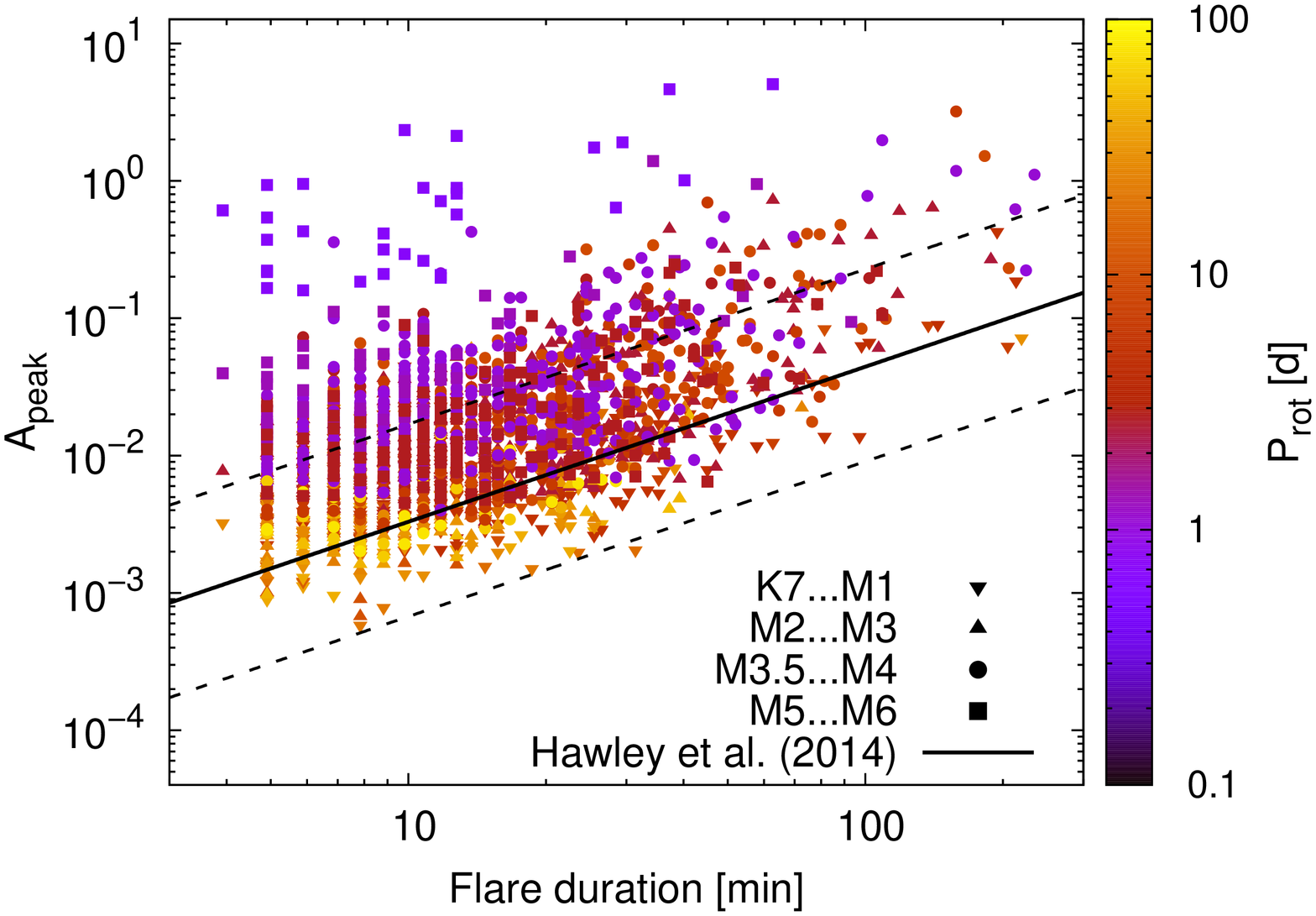}
  \includegraphics[width=0.3\textwidth,angle=270]{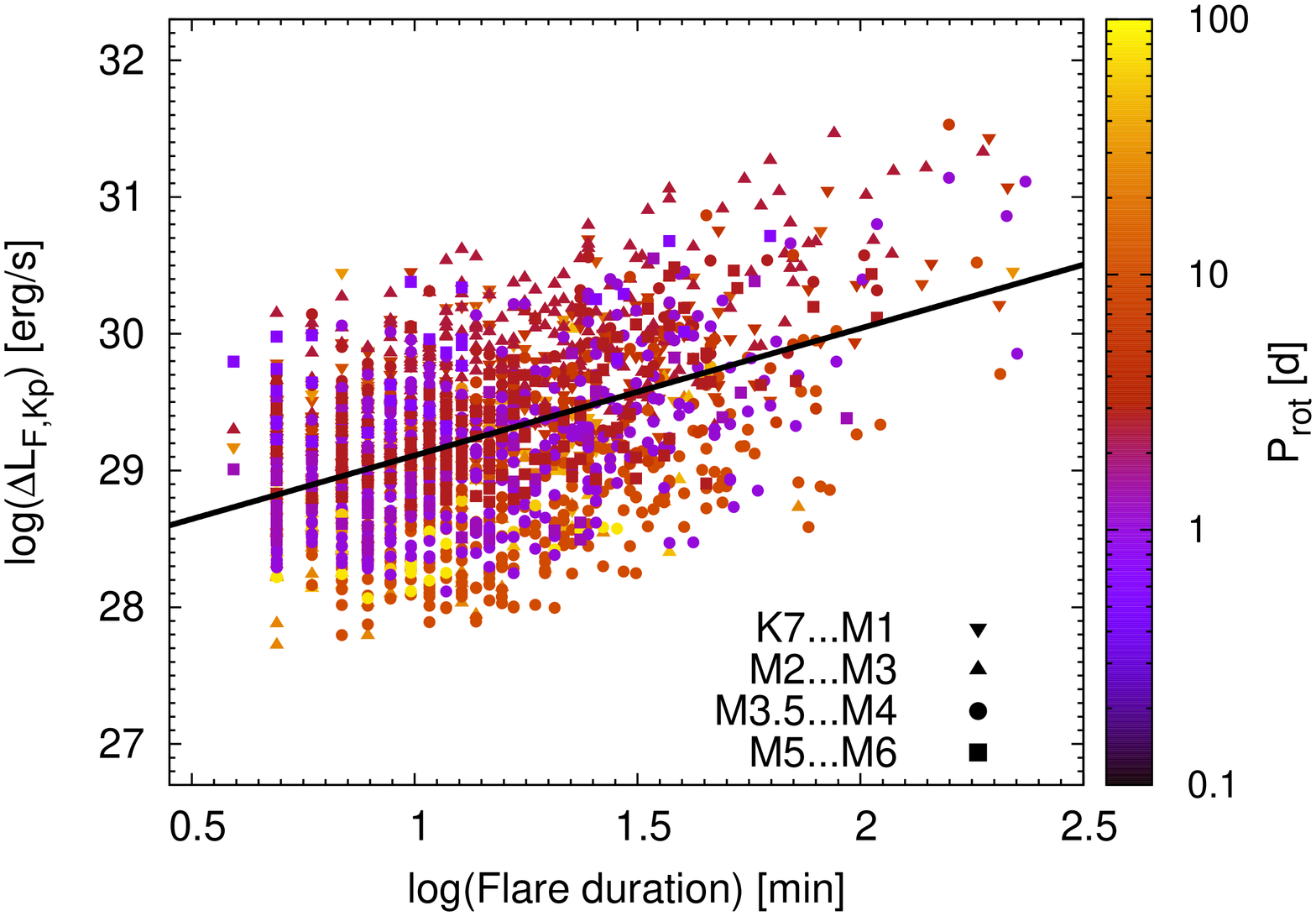}
  \caption{\textbf{Top:} Relation between flare duration and relative amplitude for all validated short-cadence flares on stars with measured rotation periods. The solid black line represents the duration - amplitude relation from \citet{2014ApJ...797..121H}, which  we extracted from their Fig.~10. The dashed lines give the lower and upper envelope of their relation. \textbf{ Bottom:} Relation between flare duration and absolute peak flare amplitude. Our linear relation (Eq.~\ref{duration-amplitude-relation}) is given as the black solid line.}
  \label{Flare_Duration_Amplitude_SC}
\end{figure}

%\textbf{the fast rotating ($\mathbf{P_{\rm rot} = 0.59\,\rm d}$) M4 star GJ\,1243}
\citet{2014ApJ...797..121H} stated that if the LC shapes of all classical flares were identical, the flare parameters (amplitudes, durations, and energies) would be highly correlated. Their measurements of the flare parameters for a sample of 833 flares from two months of \textit{Kepler} observations of GJ\,1243 indeed show a strong correlation \citep[Fig.~10 in ][]{2014ApJ...797..121H}. To shorten computing time, when we fit a flare template in Sect.~\ref{flare_validation}, we used as an input the relation between amplitude and duration that we extracted from the right panel of Fig.~10 of \citet{2014ApJ...797..121H}. Their relation, however, is based on a single M dwarf. To verify that this relation is universally valid, we measured the amplitudes and durations of all 1644 flares we found in the short-cadence data of the 46 targets with a measurable rotation period. The result is shown in Fig.~\ref{Flare_Duration_Amplitude_SC} (top) together with the relation from \citet{2014ApJ...797..121H}, which we extracted from their plot. The dashed lines give the scatter of approximately one order of magnitude in the Hawley data set. In general, we confirm the trend that flares with longer durations have higher amplitudes. Interestingly, however, many flares in our sample lie above the upper boundary of this relation. In addition, we found a dependence on SpT and rotation period. In Table~\ref{Amplitude_scatter} we summarize the scatter of the duration-amplitude relation in terms of the difference between lowest and highest amplitude for a given flare duration (corresponding to the difference between lower and upper envelope of the duration-amplitude relation), $\Delta \rm log(A_{\mathrm{peak}})$, for different groups of stars.

\begin{table}[h]
\caption{Scatter (difference between lower and upper envelope) of the duration-amplitude relation (Fig.~\ref{Flare_Duration_Amplitude_SC}, top panel), $\Delta \rm log(A_{\mathrm{peak}})$, for different groups of flares.}
\label{Amplitude_scatter}
\begin{tabular}{lc}
\hline \hline
Flare group & $\Delta \rm log(A_{\mathrm{peak}})$  \\ \hline
GJ\,1243$^{\ast}$ & 1.4 \\
all & 3.5 \\\hline
K7...M1 &  1.8 \\
M2...M3  & 2.1  \\
M3.5...M4 & 2.4 \\
M5...M6 & 2.9 \\\hline
$P_{\rm rot} < 5$\,d & 3.0 \\
$5\,\rm d < P_{\rm rot} < 10$\,d & 2.2 \\
$P_{\rm rot} > 10$\,d & 1.9 \\\hline
\hline
\end{tabular}
\\
$^{\ast}$extracted from Fig.~10 of \citet{2014ApJ...797..121H}.  
\end{table}

No flares of very early M dwarfs (SpT group K7-M1) lie above the upper boundary of the \citet{2014ApJ...797..121H} relation, while flares of the late-M dwarfs (SpT group M5-M6) are rarely seen close to the lower boundary. This lower boundary, however, is defined by our detection threshold and it is therefore obvious that we cannot detect low-amplitude flares on the fainter stars, and the latest SpT bin includes the faintest targets. We found that the maximum relative flare amplitude increases for cooler SpT. This is confirmed by the increase in $\Delta \rm log(A_{\mathrm{peak}})$  with the SpT (see Table~\ref{Amplitude_scatter}). 

A similar statement can be made for the rotation periods. No flares of slow rotators lie above the upper boundary of the \citet{2014ApJ...797..121H} relation. In general, the longer the rotation period of a star, the lower the maximum amplitudes for a given flare duration. This means that $\Delta \rm log(A_{\mathrm{peak}})$ decreases for slower rotation. The flares that lie more than one order of magnitude above the Hawley relation belong to the very fast rotating latest M dwarfs (SpT group M5-M6).

Because the relative amplitudes are biased by the range of different $K_p$ magnitudes, as mentioned in Sect.~\ref{paragr:act_flares}, we also plotted the duration-amplitude relation with the flare amplitude expressed in luminosity. This is shown in Fig.\ref{Flare_Duration_Amplitude_SC} (bottom). The removal of the bias results in a tighter relation. The scatter $\Delta \rm log(\Delta L_{\mathrm{F,Kp}})$ for the full sample is reduced to 2.9 orders of magnitude with respect to 3.5 (see Table~\ref{Amplitude_scatter}). A linear fit  shown as black line in the bottom panel of Fig.~\ref{Flare_Duration_Amplitude_SC} results in a relation of duration to absolute amplitude given by 
\begin{equation}
\label{duration-amplitude-relation}
\rm log(\Delta L_{\mathrm{F,Kp}})=(28.18 (\pm0.05) + \rm log(\Delta t_{\mathrm{Flare}})\mathrm{[min]}\cdot 0.93 (\pm0.04)).
\end{equation}

We found that the dependence on rotation period of the absolute amplitudes almost disappears when the difference in the intercepts of the linear fits for slow and fast rotators is lower than 3\,$\sigma$. We kept the slope fixed on the value given in Eq.~\ref{duration-amplitude-relation}. The dependence on the SpT, however, is still valid. The linear fits of the partly convective (SpT groups K7-M1 and M2-M3) and fully convective (SpT groups M3.5-M4 and M5-M6) results in a difference in the intercepts on a 12\,$\sigma$ level. Our results confirm the findings of \citet{2009AJ....138..633K} that higher mass M dwarfs tend to have more luminous flares than the cooler less massive M dwarfs.

\begin{figure}
  \centering
  \includegraphics[width=0.3\textwidth,angle=270]{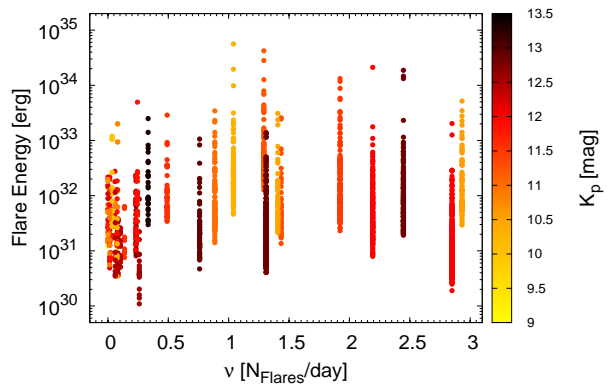}
  \caption{Number of short-cadence flares vs. flare energy for all flares. One flare number corresponds to one target. The energy is given for all flares of that star. }
  \label{Flare_Number_energy_SC}
\end{figure}

When we compared our measurements with the flares of GJ\,1243 studied by \citet{2014ApJ...797..121H}, a fast-rotating ($P_{\rm rot} = 0.59\,\rm d$) M4 star with $K_{\rm p} = 12.7$\,mag, we found that our flare amplitudes seem to be generally higher. The reason for the detection of flares with lower amplitudes on GJ\,1243 is that the precision of the main \textit{Kepler} mission is higher than that of \textit{K2}. The lack of high-amplitude flares on GJ\,1243, however, is an interesting finding for such a highly active star. For the star that is most similar to GJ\,1243 in our sample (EPIC\,212611828, SpT M3.2, $K_{\rm p} = 12.8$\,mag, $P_{\rm rot} = 1.09\,\rm d$), we found 160 flares in K2 campaign C17 (campaign duration 65.4\,d), $\text{about }$44\% of which have a higher relative amplitude than any of the flares on GJ\,1243 in the first two months of \textit{Kepler} observations. When we converted the relative amplitudes into absolute values, we found that $\text{about }$4.4\% of the flares of EPIC\,212611828 are more luminous than the strongest flare of GJ\,1243. The number of superflares (flare energy $> 10^{33}$\,erg) in this flare sample for GJ\,1243 is close to zero (between 0.1 - 0.4\% superflares according to the FFDs of Hawley et al. 2014 and Davenport 2016), while we found that $\text{about }$3\% of the flares of EPIC\,212611828 have energies in the superflare regime.  The number of flares on GJ\,1243 in the same observing time, however, is more than five times higher than for EPIC\,212611828. The energy is released in more smaller amplitude flares, while for EPIC\,212611828, the number of flares is lower, but they have a higher energy. When we plot the flare energy over the flare rate for our  M-dwarf sample (Fig.~\ref{Flare_Number_energy_SC}), we find an increase in energy of the strongest flare with flare rate up to $\text{about}\,1.25\,N_{\mathrm{flares}}/d$. Above this value, the energy of the strongest flare decreases with flare rate. Although the highest flare rate in our sample is $\text{about}$\,6.5 times lower than the rate for GJ\,1243, we found the most energetic flares not on the star with the highest flare rate.

%\begin{figure}
%  \centering
%  \includegraphics[width=0.3\textwidth,angle=270]{Ratio_ProtDur_Amplitude_SC.ps}
%  \caption{Peak flare amplitude as a function of the ratio between rotation period and flare amplitude for all validated short cadence flares.}
%  \label{Ratio_ProtDur_Amplitude}
%\end{figure}

\subsection{Energy distribution of the flare number}
\label{Sect_FFDs}

\begin{figure}
  \centering
  \includegraphics[width=0.3\textwidth,angle=270]{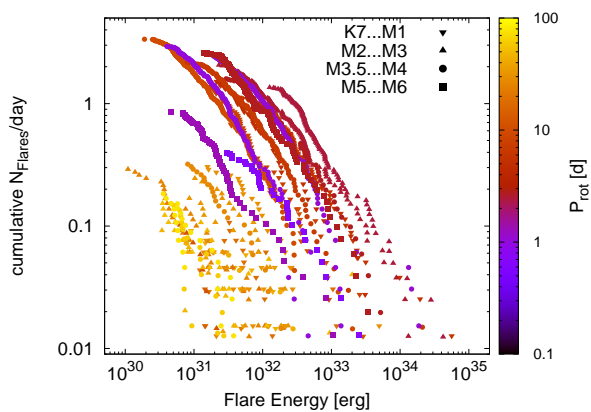}
  \caption{Cumulative flare energy frequency distribution for all targets with measurable rotation periods that show flares in short-cadence LCs. The rotation period of the stars is color-coded. In Fig.~\ref{FFD_SpT_groups} in the appendix we show a separate plot for the FFDs of each SpT group.}
  
  \label{kumulative_Verteilung_Flare_Energy_all}
\end{figure}

\begin{figure}
  \centering
  \includegraphics[width=0.3\textwidth,angle=270]{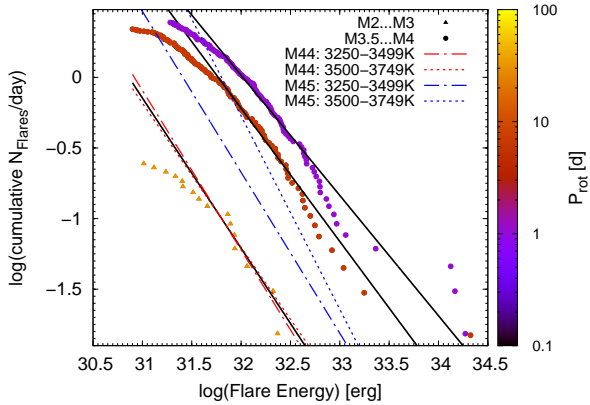}
  \caption{Cumulative flare frequency diagram for three selected targets from our sample with different rotation periods. The black solid lines give the power-law fit including all flares above a given energy detection threshold (see text in Sect.~\ref{Sect_FFDs}). We also show the power-law fits of \citet[][from their Table~4]{2019A&A...622A.133I} for Praesepe (M\,44,  630\,Myr, red) and the Pleiades (M\,45, 125\,Myr, blue) in the same temperature range of our selected targets.}
  \label{kumulative_Verteilung_Flare_Energy_selected}
\end{figure}

Because flares span many orders of magnitude in energy, flare rates are usually described by the cumulative flare frequency distribution \citep[FFD; e.g.,][]{1976ApJS...30...85L,2016ApJ...829...23D}. The FFD yields information about the flare frequency above a certain energy and is therefore more detailed than the total flare frequency  $\nu$[$N_{\mathrm{flares}}/d$] that we discussed in Sect.~\ref{paragr:act_flares} and Sect.~\ref{comp_SC_LC_rotation}. The FFDs of our targets are shown in Fig.~\ref{kumulative_Verteilung_Flare_Energy_all}. An alternative version of this plot with an individual panel for each SpT group is given in Fig.~\ref{FFD_SpT_groups} in the appendix.

As we showed in Fig.~\ref{period_Flare_Frequency} and explained in Sect.~\ref{paragr:act_flares}, slow rotators ($P_{\rm rot}>10$\,d) have a significantly lower flare frequency than fast rotators. This difference is also shown in Fig.~\ref{kumulative_Verteilung_Flare_Energy_all}. In addition, the flares of the slow rotators are less energetic. The flares with the highest energies are found for early-M dwarfs, while late-M dwarfs have higher flare rates. For partly convective M dwarfs the FFD increases in energy with decreasing rotation period. This effect has also been described by \citet{2019ApJ...871..241D}, who compiled the average FFDs by combining all available quarters of \textit{Kepler} data for a sample of 347 stars with measured rotation periods. Because their sample included a wide range of SpTs, \citet{2019ApJ...871..241D} split the targets into different color and accordingly, mass bins. Their sample did not include fully convective M dwarfs, however. We find that the trend of an increasing FFD with decreasing rotation period seems to revert for fast-rotating fully convective M dwarfs (SpT group M3.5-M4) with $P_{\rm rot}<2$\,d, such that stars with shorter $P_{\rm rot}$ show lower flare energies. We do not see any trend with rotation period for the latest M dwarfs (SpT group M5-M6). Our sample of only three stars in that group is too small to draw firm conclusions, however.

The FFD is usually modeled with a power-law function. The power law predicts a very high number of low-energy flares that cannot be detected because the noise present in any observed LC imposes a detection threshold. This completeness limit is seen as the break in the power law for low energies in the FFDs. To calculate our completeness limit, we estimated the minimum energy that a flare would need to have to be detected at a given threshold. Our short-cadence flares are required to have five consecutive data points above the detection threshold of 3$S_{\mathrm{flat}}$. The minimum flare duration accordingly is $\text{approximately}$\,240\,s. To be conservative, we chose a flare duration of 360\,s in order to allow points of the flare to be below the detection threshold (as explained in Sect.~\ref{flare_validation}, our final flare duration includes points below the threshold). The duration was used as input parameter for the flare template of \citet{2014ApJ...797..122D}. Then we varied the flare amplitude until 240\,s of the total flare duration were above 3$S_{\mathrm{flat}}$ and would therefore have been detected by our flare-search algorithm. Finally, we calculated the energy by multiplying the quiescent luminosity with the ED (integral under the resulting flare template). All flares with energies above this estimated detection threshold energy are considered in the power-law fit, which is defined as
\begin{equation}
\rm log(\nu)=\beta\,\rm log(E_{\rm F})+C
\end{equation}
\citep[see, e.g.,][]{1976ApJS...30...85L}. We used Poisson uncertainties for the energy values in the fitting process to prevent the highest energy flares from receiving too much weight. 

Finally, we compared the slopes, $\beta$, of the power-law fit for the different SpT groups. Within the error bars there is no difference in the average slope, $\overline{\beta}$ (see Table~\ref{average_beta}). A T-test that resulted in T between -0.46 and 1.5 and p-values between 0.21 and 0.82 for different combinations of the SpT groups confirmed that there is no significant SpT dependence on $\beta$. 

An example of the power-law fit for three selected M dwarfs with different rotation periods is shown in Fig.~\ref{kumulative_Verteilung_Flare_Energy_selected}. The power-law slopes increase with increasing rotation period. By plotting the slopes of all fitted FFDs over $P_{\rm rot}$ , we found a slightly negative trend that is not significant ($\sim$1.2\,$\sigma$), however. This negative trend in the $P_{\rm rot}$-$\beta$ plane indicates that the FFDs become steeper for longer rotation periods. This behavior was also seen in the simulations of \citet{2019ApJ...871..241D} for a theoretical 0.5\,$M_{\odot}$ star (approximately an M0 dwarf). 

The derivative of the cumulative flare number distribution is given by 
\begin{equation}
\dfrac{\mathrm{d}\nu}{\mathrm{d}E_{\rm F}}\sim E_{\rm F}^{-\alpha},
\end{equation}
where $\alpha=1+|\,\beta\,|$ \citep{2000ApJ...535.1047A,2007A&A...468..463S}. Using the average $\beta$ for all fitted flare number distribution, we obtain $\alpha=1.84 \pm 0.14$. This is in excellent agreement with the value reported by \citet{2019ApJ...873...97L} for their sample of 548 flaring M dwarfs. Furthermore, our power-law slope is also consistent with those measured for the Sun \citep[e.g., $\alpha=1.8 \pm 0.2,$][]{2015ApJ...802...53A}. Because $\alpha$ is close to 2, we can expect a high number of nanoflares, which might play an important role in the heating of the stellar atmosphere \citep{1988ApJ...330..474P,1991SoPh..133..357H}.

From Fig.~\ref{kumulative_Verteilung_Flare_Energy_all} we can derive a superflare frequency $\nu_{\rm SF}$[$N_{\mathrm{flares}}/d$] for the fast and slow rotators. For the stars with $P_{\rm rot}>10$\,d, we detect only five flares with an energy $> 10^{33}$\,erg, while there are 86 flares with energies in the superflare regime for the fast rotators. The corresponding flare rates are 0.002\,$N_{\mathrm{flares}}/d$ and 0.094\,$N_{\mathrm{flares}}/d$ for the slow and fast rotators, respectively. This means that the superflare rate is about 50 times higher for the fast rotators. 

\citet{2019A&A...622A.133I} analyzed \textit{K2} observations of the young open clusters Pleiades (M\,45, 125\,Myr) and Praesepe (M\,44, 630\,Myr) and found 751 flare candidates in a sample of 1761 late-K to mid-M stars. A higher number of flares were detected in the Pleiades, which confirms the expected age dependence of the flaring activity.  By fitting the FFDs, they found the power-law slopes in their sample to be between $\sim$2.0-2.4, depending on the effective temperature. This is in general slightly steeper than the slopes that we find in our sample. In Fig.~\ref{kumulative_Verteilung_Flare_Energy_selected} we overplot the power-law fits of \citet[][from their Table~4]{2019A&A...622A.133I} for Praesepe and the Pleiades in the corresponding $T_{\rm eff}$ range on our three example FFDs. For Praesepe, the fits in both temperature ranges match the FFD of our slowly rotating example ($P_{\rm rot}\sim32$\,d). This is in agreement with the study by \citet{2019ApJ...879..100D}, in which the Praesepe M dwarfs were found to have rotation periods of $P_{\rm rot}<40$\,d. The power-law fits for the Pleiades are between our slow and fast rotators. According to the rotation period distribution of the Pleiades presented by \citet{2016AJ....152..113R}, the cluster M dwarfs rotate with a period of $P_{\rm rot}<10$\,d. We would therefore have expected that the power-law fits for the Pleiades agree with our fast-rotating M dwarfs. However, we found higher flare rates than those measured in the Pleiades.
%and, hence, the power-law fits of \citet{2019A&A...622A.133I} are not consistent with our FFD measurements.

\begin{table}[h]
\centering
\caption{Average slope of the power-law-fits of the FFDs for different SpT groups.}
\label{average_beta}
\begin{tabular}{lc}
\hline \hline
SpT group & $\overline{\beta}$  \\ \hline
K7...M1 &  -0.78$\pm$0.01 \\
M2...M3  & -0.81$\pm$0.10  \\
M3.5...M4 & -0.85$\pm$0.09 \\
M5...M6 & -0.83$\pm$0.05 \\\hline
\hline
\end{tabular}
\end{table}

%\begin{figure}
%  \centering
%  \includegraphics[width=0.3\textwidth,angle=270]{kumulative_Verteilung_Flare_Sigma_period.ps}
%  \caption{Cumulative distribution of flares with a given flare $\sigma$ for stars with rotations periods lower (purple) and higher (orange) than 10\,d. Stars with lower period show a higher number of high-$\sigma$ flares.}
%  \label{kumulative_Verteilung_Flare_Sigma}
%\end{figure}

\section{Summary and conclusions}
\label{summary}

Studies of the rotation-activity relation of M-type stars are essential to enhance our understanding of stellar dynamos and angular momentum evolution. The behavior at the fully convective boundary where the sample size is still small is particularly interesting. A drastic change in the level of photometric activity at a critical rotation period of $\text{about }$10\,d was identified by S16 using \textit{K2} long-cadence data of 134 M dwarfs. An abrupt drop in activity at a similar period ($\sim$10\,--\,30\,d) is also suggested from studies of $H_{\alpha}$ emission \citep[see, e.g.,][their Fig.~7]{2015ApJ...812....3W}. Furthermore, a clear bimodal behavior was also seen in the fraction of stars with detectable H$_{\alpha}$ emission (activity fraction) by \citet[][their Figs.~5 and 6]{2015ApJ...812....3W}. On the other hand, in  X-ray and $H_{\alpha}$ studies, fast-and slow-rotator regimes are often modeled with a broken power law, that is, a smooth change rather than a sharp one. We note, however, that a gap at intermediate X-ray activity levels of $L_{\rm x}/L_{\rm bol}=10^{-4}$ supporting a clear cut was identified by Magaudda et al. (2020,  A\&A subm.).

The photometric activity study by S16 (and the follow-up work by Raetz et al. 2019, AN, submitted) is based on \textit{K2} long-cadence data, which represents a bias especially in the analysis of flares. In particular, the low cadence limits the study to long flares with durations of $>$1\,h. Here, we compiled a sample of 56 M dwarfs from the Superblink proper motion catalog by \citet{2011AJ....142..138L} with simultaneous \textit{K2} long- and short-cadence data. Eight of them were observed in two \textit{K2} campaigns, which results in 64 individual LCs. The goal was to compare results of long- and short-cadence data in order to address potential biases in the study of the relation of rotation and photometric activity of M dwarfs.

For our analysis, we developed a procedure for the data reduction of short-cadence LCs. Our method is based on the method of \citet{2014PASP..126..948V} and was used to extract publicly available long-cadence LCs. We showed that our method provides reliable LCs with sufficient precision for this work. 

From our corrected LCs we extracted the rotation period and a wealth of photometric activity indicators (standard deviation of the full LC, standard deviation of the flattened and cleaned LC, amplitude of the rotational signal, flare frequency, peak flare amplitude, and flare duration). We found rotation periods for $\text{about }$82\% of all targets in our M-dwarf sample, which is in agreement with previous studies, for instance, \citet{2014ApJS..211...24M}, S16, and Raetz et al. (2019, AN, submitted). The activity diagnostics connected with the rotation cycle were derived simultaneously to the period search, while the flare parameters were determined with our flare-finding procedure. 

For all presented photometric activity diagnostics we found the change in behavior at a rotation period of $\text{about }$10\,d described above. Fast rotators show a higher activity level than slow rotators. This change is most drastic in the flare frequency determined from the short-cadence data: the average is 1.60\,$N_{\mathrm{flares}}/d$ for fast and 0.07\,$N_{\mathrm{flares}}/d$ for slow rotators. This transition period seen for the photometric activity indicators reflects the turn-off from the saturated to the correlated regime in X-rays (e.g., S16). Because our sample is small, we do not see that the critical period increases with decreasing mass, as found by \citet{2003A&A...397..147P} and Magaudda et al. (2020,  A\&A subm.). We showed for the first time that the same bimodal distribution is also present in the durations of flares detected in short-cadence LCs. Almost all flares detected on slow rotators are shorter than 1\,h, while fast rotators have flares that last up to $\text{about }$4\,h. We confirmed the bimodal behavior of the standard deviation of the flattened and cleaned LC, $S_{\mathrm{flat}}$, with the rotation period that was previously detected by S16 in long-cadence data. This trend is more prominent in long-cadence data. $S_{\mathrm{flat}}$ reflects the trend visible in the activity diagnostics related to the rotation cycle and flares, which therefore indicates that the underlying noise is most likely caused by unresolved spot and flare activity.

When we compared the parameters of long-cadence flares with the parameters of their short-cadence counterparts, we found that the peak flare amplitudes are underestimated 
and the flare durations are overestimated. This is in agreement with the study by \citet{2018ApJ...859...87Y}. While \citet{2018ApJ...859...87Y} found the flare energies in long cadence to be underestimated, we found them to be almost identical in long and short cadence. This is expected because the flare energy is proportional to the integral under the flare, and thus the opposite trends of flare amplitude and flare duration cancel each other in the flare energy. Specifically, the longer durations and lower amplitudes of flares in long-cadence data yield similar values for the integral as the shorter durations and higher amplitudes of flares in short-cadence data. 

As a consequence of the underestimated long-cadence flare amplitudes, we see a downward shift in the flare amplitude - rotation relation with respect to the short-cadence data. Despite this shift, we found the same overall behavior for the relative peak flare amplitudes in short- and long-cadence data. This means that the conclusions of flare amplitude studies based on long-cadence data, for example, S16, do not depend on the data-point cadence.

The bimodal distribution of the durations of short-cadence flares with rotation period is only barely seen for the durations of long-cadence flares (middle panel of Fig.~\ref{period_Flare_A_D_LC_SC}). This is a result of the overestimation of the duration in long-cadence data, which is stronger for shorter flares because of the strong artificial quantization caused by the data-point cadence of $\sim$30\,min.

We found that at a given $P_{\rm rot}$, the total flare frequency is significantly higher for the short-cadence data than for the long-cadence data (by a factor of 4.6), which is in agreement with the numbers given in \citet{2018ApJ...859...87Y}. The difference in the flare rate between long- and short-cadence data is larger in the fast-rotator regime. As a result of the lower flare rate, the time between two flares (flare waiting time) is longer for slower rotators. We did not find any correlation between the flare waiting time and the energy difference between consecutive flares. This means that the process of magnetic energy build-up does not depend on previous energy releases. The same result was obtained by \citet{2014ApJ...797..121H} for the very active M4 star GJ\,1243. Interestingly, the stars with the highest flare rates tend not to have the flares with the highest energies. This indicates that one star can release the same amount of energy in more smaller amplitude flares or with a smaller number of more energetic flares.

Using all 1644 detected flares on stars with measured rotation periods, we studied the relation between flare durations and amplitudes. Generally, longer duration flares have higher peak amplitudes, as previously described by \citet{2014ApJ...797..121H} for GJ\,1243, for example. We found a relation between flare durations and absolute amplitudes in terms of luminosity that only shows a dependence on the SpT. We confirm the findings of \citet{2009AJ....138..633K}, obtained from SDSS Stripe 82 observations of flaring M dwarfs, that earlier M dwarfs tend to have more luminous flares than later M dwarfs.

We compiled the cumulative FFD for all stars in our \textit{K2} M-dwarf sample. We found that all stars have similar slopes, with a slight trend that the slopes become steeper for longer rotation periods. This behavior has been observed in simulations by \citet{2019ApJ...871..241D} and in observations by \citet{2019ApJ...873...97L}. The latter found a slightly steeper slope for their whole sample of 548 flaring M dwarfs, comprising fast and slow rotators, than for fast rotators alone ($1.82\pm0.02$ and $1.78\pm0.02$ for the whole sample and for the fast rotators, respectively). This indicates an increasing slope with increasing rotation period. The power-law slope of on average $\text{about }$1.8 of our FFD is consistent with solar flares observed in the extreme ultraviolet, EUV \citep{2000ApJ...535.1047A}, soft X-rays \citep{1995PASJ...47..251S}, and hard X-rays \citep{1993SoPh..143..275C}, and with the sample of superflares on the Sun-like stars of \citet{2013ApJS..209....5S}. This shows that the power-law distribution of our \textit{K2} M dwarf sample is roughly on the same power-law line as superflares in Sun-like stars, solar flares, microflares, and nanoflares \citep[see Fig.~9 in][]{2013ApJS..209....5S}.

We confirmed the trend of a decrease in the observed flare frequency as a function of rotation period pointed out by \citet{2019ApJ...871..241D} for SpT up to the fully convective boundary. We did not see any trend with the rotation period for the very latest M dwarfs. The FFDs also revealed that the highest flare frequencies were not found around the fastest rotating stars, but in the period range between $2\,\mathrm{d}<P_{\rm rot}<10\,\mathrm{d}$. An increased flare rate for stars of intermediate rotation periods was also seen by \citet{2019ApJ...870...10M} in the MEarth data set. The highest flare rate in our sample is observed on the known eclipsing M-dwarf binary CU\,Cnc\footnote{We re-extracted the LCs of CU\,Cnc and CV\,Cnc because they were found to fall into the same aperture mask that produced an identical LC for both stars. CU\,Cnc and CV\,Cnc were then analyzed individually. For the known eclipsing M-dwarf binary CU\,Cnc, we determined the eclipse times and refined the linear ephemerides, which are ten times more precise, but in agreement with previously published values.} , and flares might therefore be induced by the presence of a close companion. Studies by \citet{2012AJ....144...93M} and \citet{2017AJ....154..118S}, for example, showed that the fraction of $H_{\alpha}$-active M dwarfs is observed to be noticeably higher for stars in close binaries than in the isolated field M dwarfs. These close binaries are also well known to show strong flares \citep[e.g., II Peg,][]{1992MNRAS.255...48M}.

By overplotting the flare distribution power-law fits for Praesepe and the Pleiades of \citet{2019A&A...622A.133I} on three selected M dwarfs from our sample, we found that the fast rotators show a higher flare rate than the M dwarfs in the Pleiades. Based on this, the two fast-rotating mid-M dwarfs that we show as examples might be young. For these two stars, \citet{2015ApJ...798...41A} found significant near-UV excess emission. The fastest rotator of our example stars ($P_{\rm rot}\sim1$\,d) falls into the saturated regime of the near-UV activity-age relation \citep[Fig.~13 of][]{2015ApJ...798...41A}, which indicates an age $<160$\,Myr, while the other fast rotator ($P_{\rm rot}\sim6$\,d) shows a near-UV excess emission consistent with a Hyades age ($\text{approximately }$the age of Praesepe). The flare data of \citet{2019A&A...622A.133I} of the Pleiades are therefore inconsistent with our FFD measurements.

One result of the analysis of the photometric data collected in the main \textit{Kepler} mission was the detection of superflares on solar-type (G-type main-sequence) stars \citep[e.g.,][]{2012Natur.485..478M,2013ApJS..209....5S,2013ApJ...771..127N}. Superflares with energy up to $10^{35}$\,erg also occur on slowly rotating stars \citep[e.g.,][]{2014MNRAS.442.3769K,2019ApJ...876...58N}. However, the frequency for superflares on these slow rotators is very low compared with rapidly rotating stars. \citet{2012Natur.485..478M} reported that superflares with energies between $10^{34}$\,erg and $10^{35}$\,erg occur on (slowly rotating) sun-like stars once in 800-5000\,yr. In our M-dwarf sample, we detected only a small number of superflares (flare energy $> 10^{33}$\,erg) for slowly rotating stars, while the superflare rate is 50 times higher for fast-rotating stars. No superflares above $10^{34}$\,erg were observed on slow rotators. This is consistent with the flare frequency of \citet{2012Natur.485..478M} if slowly rotating M dwarfs show the same superflare occurrence rate as G stars in the same period regime. Because our observation baseline of $\sim$80\,d is short, it is unlikely that we would catch such a high-energy flare. For the fast rotators, however, we found a frequency of superflares above $5\,\cdotp10^{34}$\,erg for M stars that is about twice higher than the frequency reported by \citet[][their Fig.~3 revealed an average flare frequency of $\sim$0.2\,$N_{\mathrm{flares}}/\mathrm{yr}=5.5\,\cdotp10^{-4}N_{\mathrm{flares}}/$d for stars with $P_{\rm rot} < 10$\,d]{2012Natur.485..478M} for G stars.

With our sample size of 56 targets, the number of M dwarfs from the \citet{2011AJ....142..138L} catalog with high-cadence observations is still small. NASA's \textit{TESS} (Transiting Exoplanet Survey Satellite) mission, launched in 2018 April, observes in its two-year prime mission $\sim$7000 \citet{2011AJ....142..138L} M dwarfs in a 2min cadence mode. In this way, \textit{TESS} dramatically enlarges our sample of continuous high-cadence monitored bright M-dwarfs and allows us to study the activity and in particular the morphology of flares in depth. In the future, the \textit{PLATO} mission with its unprecedented precision, short cadence, and long observational baseline will allow us to study the magnetic activity indicators in so far unrivaled detail.

\begin{acknowledgements}
We would especially like to thank A. Vanderburg for his public release of the analyzed long cadence \textit{K2} light curves and for his additional work in producing the short cadence light curves of our targets. We would like to thank amateur astronomer M. Raetz for providing observations of the eclipsing binary CU\,Cnc. AS's work is supported by the STFC grant no. ST/R000824/1. This paper includes data collected by the K2 mission. Funding for the K2 mission is provided by the NASA Science Mission directorate. This research has made use of the SVO Filter Profile Service (http://svo2.cab.inta-csic.es/theory/fps/) supported from the Spanish MINECO through grant AYA2017-84089.
\end{acknowledgements}

\bibliographystyle{aa}
\bibliography{literatur}

%\begin{thebibliography}{}

%  \bibitem[1966]{baker} Baker, N. 1966,
%      in Stellar Evolution,
%      ed.\ R. F. Stein,\& A. G. W. Cameron
%      (Plenum, New York) 333
%\end{thebibliography}

\appendix

\section{CU\,Cancri and CV\,Cancri}
\label{CUCnc}

In our \textit{K2} M dwarf sample we found two LCs that show clear eclipses. The LCs of the two stars look almost identical, and the eclipses occurred at the same times. To investigate the reason for the identical LC shapes and to determine the linear ephermeris in flare rates, we studied the LCs of the two stars, which we identified as CU\,Cnc and CV\,Cnc, in more detail.

CU\,Cnc (EPIC\,211944670, GJ\,2069A, HIP\,41824) is a known eclipsing M-dwarf binary \citep{1999A&A...341L..63D} and the A component (also often called Aab) of a multiple system of at least five M dwarfs \citep{1999A&A...344..897D,2004A&A...425..997B}. CU\,Cnc, whose components are close to the fully convective boundary, is a valuable benchmark system with which we can improve our knowledge of the physics of low-mass stars. The common proper motion companion CV\,Cnc (EPIC\,211944856, GJ 2069B) is itself a visual binary (components B and C). An additional component designated D was discovered $\sim$0.68\,arcsec distant from A. Components B, C, and D were resolved by adaptive optics imaging \citep{1999A&A...344..897D,2004A&A...425..997B}.

CU\,Cnc and CV\,Cnc are both classified as flare stars, and flare activity on CU\,Cnc in optical wavelengths was reported by \citet{2012MNRAS.423.3646Q} and \citet{2017ApJ...835..251W}. The first authors determined an $R$-band flare rate of about 0.05 flares per hour. 

\begin{figure}
  \centering
  \includegraphics[width=0.34\textwidth,angle=270]{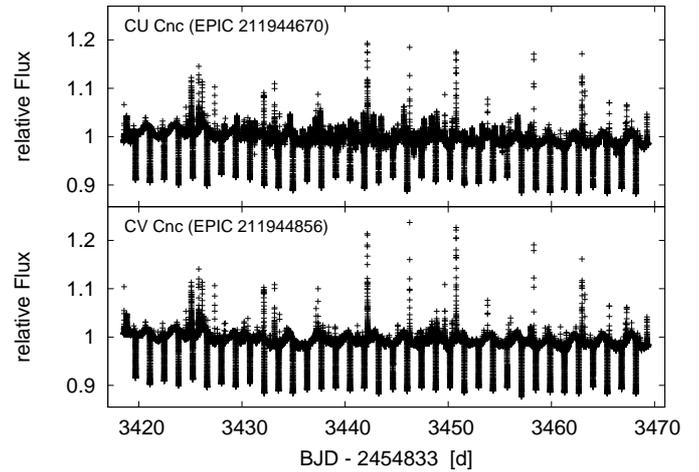}
  \caption{Corrected short-cadence LCs of CU\,Cnc (EPIC 211944670) and CV\,Cnc (EPIC 211944856).}
  \label{vdb_Epic211944670_Epic211944856_SC}
\end{figure}

The visual binary system CU\,Cnc (Aab) and CV\,Cnc (BC) is separated by $\sim$12\,arcsec. Because \textit{Kepler} has an image scale of 3.98\,arcsec per pixel, CU\,Cnc and CV\,Cnc are only separated by $\sim$3 pixels. The optimal aperture determined by \citet{2014PASP..126..948V} on the target pixel files of both targets is large and includes the light of both stars. As a result, the two LCs that were obtained in \textit{K2} campaign 18 look very similar and even show the same eclipses (see Fig.~\ref{vdb_Epic211944670_Epic211944856_SC}).

\begin{figure}
  \centering
  \includegraphics[width=0.34\textwidth,angle=270]{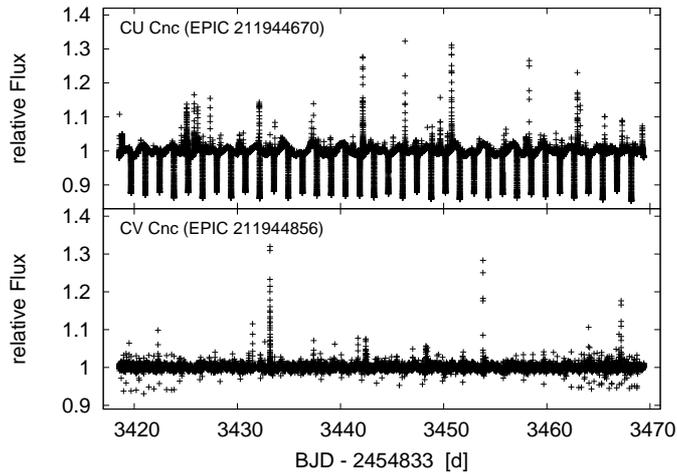}
  \caption{Short-cadence LCs of CU\,Cnc (EPIC 211944670) and CV\,Cnc (EPIC 211944856) after the re-extraction with two customized masks with smaller aperture, as described in Sect.\ref{CUCnc}.}
  \label{Epic211944670_Epic211944856_SC}
\end{figure}

To create a separate LC for CU\,Cnc and CV\,Cnc, we repeated the LC extraction with a mask with smaller aperture. Hereby, the apertures were chosen by eye-inspection, and several different apertures were tested. The aperture mask that yielded the cleanest LCs (most free from the contribution of the other star, no eclipses in the LC of CV\,Cnc visible) for each star was chosen as the final mask. The LC extraction was made in the same way as for all the other stars using the \textit{Kepler}GO/lightkurve code. The final LCs are given in Fig.~\ref{Epic211944670_Epic211944856_SC}. The LC of CV\,Cnc, the fainter component that was previously strongly contaminated by the light of CU\,Cnc, is now free from eclipses and shows a different rotation period (2.8\,d and 0.6\,d for CU\,Cnc and CV\,Cnc, respectively).

\begin{figure}
  \centering
  \includegraphics[width=0.34\textwidth,angle=270]{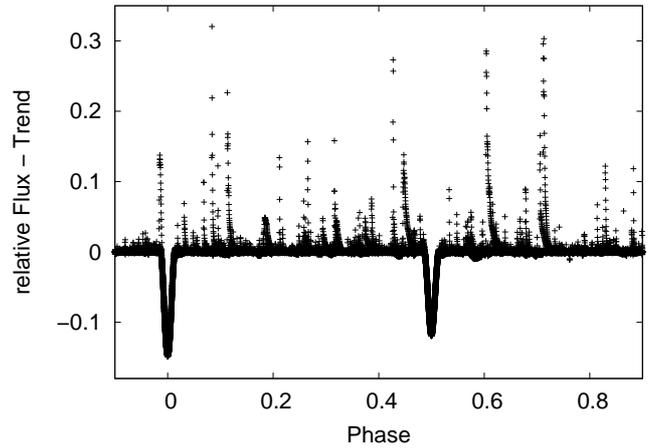}
  \caption{Phase-folded LC of CU\,Cnc (EPIC 211944670). The rotation signal was removed before the folding.}
  \label{Epic211944670_phase}
\end{figure}

In total, we found 36 eclipses of CU\,Cnc in the \textit{K2} LC. Including the time at epoch zero given in \citet{2017ApJ...835..251W} and one minimum observation of an amateur astronomer in February 2019 (M. Raetz, priv. com.) to enlarge the observational baseline, we have 38 eclipse-time measurements. Using an error-weighted linear fit to all eclipse times, we were able to improve the orbital period to 2.7714677\,$\pm$\,0.0000002\,d, which is $\text{about }$200ms lower and ten times more precise than the value determined by \citet{2017ApJ...835..251W}. Fig.~\ref{Epic211944670_phase} shows the LC of CU\,Cnc folded in phase with our updated orbital period.

Our flare detection validated 149 flares in the short-cadence LC of CU\,Cnc and 45 events for CV\,Cnc. Using the total duration of $\sim$50.8\,d of campaign 18, we calculated a flare rate of 0.12 and 0.04 $N_{\mathrm{flares}}/h$ for CU\,Cnc and CV\,Cnc, respectively. Our analysis thus yields a 2.4 times higher flare rate for CU\,Cnc than was reported by \citet{2012MNRAS.423.3646Q}, which they obtained from their $R$-band observations. One reason might be the higher flare detection sensitivity of \textit{K2} due to the passband that also covers somewhat bluer wavelengths. 

The two LCs correspond to components A and  D for CU\,Cnc, and B and C for CV\,Cnc. The individual flare activity for each star within each binary cannot be resolved. Small flares are hidden in the eclipses of CU\,Cnc. The sigma-clipping process that was used to create the smoothed LC removed the eclipses, and any flares go unnoticed in the flare-detection algorithm. Its flare rate might therefore be even higher. Flares exist in the primary and also in the secondary eclipses, which indicates that both components contribute to the total flare rate of the CU\,Cnc system.


\section{M dwarf sample: stellar parameters and rotation-activity measurements}

\begin{table*}[h]
\caption{Stellar parameters of the 56 stars in our \textit{K2} M dwarf sample. Eight stars were observed in two campaigns. SpT is given as subclass of M dwarfs (negative values correspond to SpTs earlier than M).}
\label{stellar_param_table}
\begin{small}\begin{tabular}{llccr@{\,$\pm$\,}lr@{\,$\pm$\,}lr@{\,$\pm$\,}lr@{\,$\pm$\,}lr@{\,$\pm$\,}lc}
\hline \hline
EPIC ID & Camp. & $K_{\rm p}$ & log\,$L_{\mathrm{Kp,0}}$ & \multicolumn{2}{c}{$M_{\mathrm{Ks}}$} & \multicolumn{2}{c}{$R_{\mathrm{\ast}}$  } & \multicolumn{2}{c}{ $M_{\mathrm{\ast}}$ } &  \multicolumn{2}{c}{$T_{\mathrm{eff}}$} &  \multicolumn{2}{c}{d} & SpT* \\ 
&  & [mag] & [erg/s] & \multicolumn{2}{c}{[mag]} & \multicolumn{2}{c}{[R$_{\odot}$]} & \multicolumn{2}{c}{ [M$_{\odot}$]} &  \multicolumn{2}{c}{[K]} & \multicolumn{2}{c}{[pc]} & (M subclass) \\\hline
202059204       &       c00     &       11.30   &       31.42   &       5.55    &       0.02    &       0.51    &       0.02    &       0.539   &       0.010   &       3725    &       93      &       27.46   &       0.36    &       1.23    \\
202059229       &       c00     &       10.50   &       31.56   &       5.35    &       0.02    &       0.55    &       0.02    &       0.574   &       0.011   &       3689    &       82      &       22.38   &       0.29    &       0.99    \\
201611969       &       c01     &       11.55   &       31.34   &       5.94    &       0.02    &       0.45    &       0.01    &       0.473   &       0.009   &       3552    &       81      &       28.16   &       0.37    &       2.13    \\
205467732       &       c02     &       12.93   &       30.41   &       7.38    &       0.06    &       0.27    &       0.01    &       0.266   &       0.009   &       3096    &       81      &       18.18   &       0.58    &       4.56    \\
206262336       &       c03     &       12.03   &       30.34   &       7.93    &       0.03    &       0.22    &       0.01    &       0.211   &       0.005   &       3237    &       88      &       11.13   &       0.22    &       3.78    \\
210317378       &       c04     &       12.72   &       31.06   &       6.29    &       0.02    &       0.40    &       0.01    &       0.418   &       0.008   &       3404    &       82      &       34.93   &       0.46    &       3.13    \\
210434433       &       c04     &       12.14   &       31.19   &       5.90    &       0.03    &       0.46    &       0.01    &       0.480   &       0.010   &       3448    &       82      &       30.87   &       0.61    &       2.93    \\
210460280       &       c04     &       11.20   &       31.70   &       5.12    &       0.02    &       0.59    &       0.02    &       0.613   &       0.012   &       3752    &       81      &       36.19   &       0.47    &       1.05    \\
210489654       &       c04     &       12.62   &       30.93   &       6.34    &       0.02    &       0.40    &       0.01    &       0.410   &       0.008   &       3235    &       81      &       28.78   &       0.39    &       3.97    \\
210579749       &       c04     &       9.98    &       31.54   &       5.51    &       0.02    &       0.52    &       0.02    &       0.547   &       0.010   &       3643    &       83      &       17.23   &       0.22    &       1.55    \\
210580056       &       c04     &       9.30    &       31.81   &       5.07    &       0.02    &       0.60    &       0.02    &       0.622   &       0.012   &       3817    &       89      &       17.20   &       0.22    &       0.08    \\
211480655       &       c05,c18 &       12.21   &       31.14   &       5.94    &       0.04    &       0.45    &       0.01    &       0.475   &       0.010   &       3319    &       87      &       30.42   &       0.66    &       3.58    \\
211642294       &       c05,c18 &       10.79   &       31.25   &       6.07    &       0.02    &       0.43    &       0.01    &       0.453   &       0.009   &       3434    &       90      &       17.82   &       0.20    &       2.65    \\
211892034       &       c05     &       11.58   &       31.60   &       5.21    &       0.02    &       0.57    &       0.02    &       0.598   &       0.011   &       3459    &       87      &       38.54   &       0.53    &       2.46    \\
211945363       &       c05,c18 &       12.35   &       31.02   &       6.37    &       0.05    &       0.39    &       0.01    &       0.406   &       0.011   &       3334    &       81      &       28.04   &       0.71    &       3.39    \\
212285603       &       c06     &       10.67   &       30.89   &       6.66    &       0.04    &       0.35    &       0.01    &       0.362   &       0.009   &       3291    &       87      &       11.20   &       0.29    &       3.53    \\
212518629       &       c06,c17 &       12.46   &       30.76   &       6.93    &       0.06    &       0.32    &       0.01    &       0.324   &       0.010   &       3302    &       80      &       21.99   &       0.65    &       3.58    \\
212556566       &       c06,c17 &       12.86   &       30.51   &       7.34    &       0.02    &       0.27    &       0.01    &       0.272   &       0.005   &       3204    &       82      &       19.70   &       0.26    &       4.23    \\
212560714       &       c06,c17 &       11.27   &       31.66   &       5.54    &       0.02    &       0.52    &       0.02    &       0.541   &       0.010   &       3905    &       93      &       35.96   &       0.47    &       -1.10   \\
212681564       &       c06,c17 &       11.82   &       30.74   &       7.72    &       0.02    &       0.24    &       0.01    &       0.231   &       0.005   &       3126    &       81      &       16.08   &       0.21    &       4.42    \\
212776174       &       c06,c17 &       10.03   &       31.82   &       5.14    &       0.03    &       0.59    &       0.02    &       0.609   &       0.012   &       3855    &       86      &       24.19   &       0.47    &       -0.47   \\
201659529       &       c10     &       12.13   &       30.95   &       6.58    &       0.02    &       0.36    &       0.01    &       0.374   &       0.007   &       3355    &       80      &       23.38   &       0.31    &       3.20    \\
245919787       &       c12     &       10.33   &       31.80   &       4.38    &       0.02    &       0.73    &       0.02    &       0.737   &       0.014   &       3732    &       84      &       27.35   &       0.36    &       1.17    \\
210651981       &       c13     &       11.25   &       31.90   &       4.36    &       0.04    &       0.74    &       0.02    &       0.741   &       0.015   &       3549    &       83      &       46.83   &       1.08    &       2.29    \\
246774176       &       c13     &       11.49   &       31.41   &       5.60    &       0.02    &       0.51    &       0.02    &       0.532   &       0.010   &       3394    &       83      &       29.72   &       0.39    &       2.79    \\
248601792       &       c14     &       13.34   &       30.01   &       8.22    &       0.02    &       0.19    &       0.01    &       0.188   &       0.004   &       3039    &       80      &       13.93   &       0.18    &       4.84    \\
211889233       &       c16     &       11.59   &       31.83   &       5.02    &       0.02    &       0.61    &       0.02    &       0.630   &       0.012   &       3883    &       81      &       50.39   &       0.69    &       -0.31   \\
212405060       &       c17     &       12.50   &       31.23   &       5.85    &       0.02    &       0.47    &       0.01    &       0.488   &       0.009   &       3427    &       80      &       38.16   &       0.52    &       3.05    \\
212424143       &       c17     &       12.01   &       31.02   &       6.28    &       0.03    &       0.40    &       0.01    &       0.420   &       0.009   &       3351    &       87      &       24.22   &       0.47    &       3.43    \\
212438743       &       c17     &       11.06   &       31.65   &       5.51    &       0.02    &       0.52    &       0.02    &       0.546   &       0.011   &       3775    &       82      &       32.24   &       0.46    &       0.29    \\
212518679       &       c17     &       12.16   &       31.66   &       5.30    &       0.02    &       0.56    &       0.02    &       0.582   &       0.011   &       3806    &       87      &       54.16   &       0.72    &       0.43    \\
212578572       &       c17     &       12.06   &       32.10   &       4.50    &       0.11    &       0.71    &       0.03    &       0.717   &       0.022   &       3960    &       87      &       85.73   &       4.35    &       -1.39   \\
212610229       &       c17     &       11.75   &       31.44   &       6.12    &       0.02    &       0.43    &       0.01    &       0.445   &       0.009   &       3394    &       82      &       34.84   &       0.47    &       2.96    \\
212611828       &       c17     &       12.85   &       31.07   &       6.24    &       0.07    &       0.41    &       0.02    &       0.425   &       0.014   &       3344    &       84      &       37.43   &       1.39    &       3.19    \\
212652474       &       c17     &       12.00   &       31.74   &       5.33    &       0.02    &       0.55    &       0.02    &       0.577   &       0.011   &       3907    &       87      &       54.99   &       0.74    &       -0.96   \\
212662886       &       c17     &       11.66   &       31.72   &       5.10    &       0.03    &       0.59    &       0.02    &       0.617   &       0.012   &       3716    &       89      &       46.10   &       0.91    &       1.27    \\
212663959       &       c17     &       12.23   &       30.96   &       6.69    &       0.03    &       0.35    &       0.01    &       0.357   &       0.008   &       3361    &       81      &       24.89   &       0.49    &       3.06    \\
212679181       &       c17     &       12.01   &       31.51   &       5.44    &       0.02    &       0.53    &       0.02    &       0.558   &       0.011   &       3512    &       84      &       42.55   &       0.56    &       2.33    \\
212681560       &       c17     &       11.91   &       31.37   &       5.72    &       0.02    &       0.49    &       0.01    &       0.512   &       0.010   &       3443    &       84      &       34.48   &       0.47    &       2.81    \\
212710520       &       c17     &       11.72   &       31.80   &       5.17    &       0.02    &       0.58    &       0.02    &       0.604   &       0.011   &       3915    &       87      &       51.50   &       0.68    &       -0.62   \\
212712577       &       c17     &       10.86   &       31.94   &       4.85    &       0.03    &       0.64    &       0.02    &       0.660   &       0.013   &       3934    &       88      &       40.86   &       0.80    &       -1.43   \\
212734783       &       c17     &       11.60   &       31.05   &       6.37    &       0.03    &       0.39    &       0.01    &       0.405   &       0.009   &       3414    &       87      &       20.68   &       0.40    &       2.90    \\
212807182       &       c17     &       12.23   &       31.31   &       6.10    &       0.02    &       0.43    &       0.01    &       0.448   &       0.009   &       3583    &       82      &       37.33   &       0.49    &       1.73    \\
212810669       &       c17     &       12.54   &       30.38   &       8.16    &       0.05    &       0.20    &       0.01    &       0.193   &       0.005   &       3445    &       84      &       14.79   &       0.39    &       2.76    \\
212840848       &       c17     &       8.91    &       31.85   &       5.15    &       0.02    &       0.58    &       0.02    &       0.609   &       0.011   &       3921    &       99      &       15.07   &       0.20    &       -0.99   \\
212846068       &       c17     &       11.24   &       31.70   &       5.25    &       0.04    &       0.57    &       0.02    &       0.590   &       0.013   &       3813    &       88      &       37.11   &       0.97    &       0.36    \\
212874829       &       c17     &       11.85   &       31.60   &       5.47    &       0.02    &       0.53    &       0.02    &       0.553   &       0.011   &       3743    &       87      &       43.80   &       0.59    &       0.95    \\
251512053       &       c17     &       10.65   &       31.87   &       4.86    &       0.02    &       0.64    &       0.02    &       0.658   &       0.012   &       3870    &       85      &       34.19   &       0.46    &       -0.02   \\
251540467       &       c17     &       11.49   &       31.65   &       5.30    &       0.02    &       0.56    &       0.02    &       0.582   &       0.011   &       3817    &       90      &       39.43   &       0.53    &       0.37    \\
251550724       &       c17     &       10.16   &       31.23   &       6.02    &       0.02    &       0.44    &       0.01    &       0.460   &       0.009   &       3448    &       86      &       13.12   &       0.18    &       2.79    \\
251567386       &       c17     &       12.31   &       31.48   &       5.55    &       0.02    &       0.52    &       0.02    &       0.540   &       0.010   &       3570    &       94      &       46.85   &       0.62    &       2.15    \\
251583820       &       c17     &       12.44   &       30.79   &       7.01    &       0.03    &       0.31    &       0.01    &       0.312   &       0.007   &       3400    &       81      &       22.54   &       0.44    &       3.02    \\
251584738       &       c17     &       10.91   &       31.40   &       5.76    &       0.02    &       0.48    &       0.01    &       0.503   &       0.010   &       3659    &       84      &       22.52   &       0.29    &       1.31    \\
211944670       &       c18     &       10.55   &       31.28   &       5.50    &       0.02    &       0.52    &       0.02    &       0.548   &       0.010   &       3282    &       81      &       16.59   &       0.22    &       3.82    \\
211944856       &       c18     &       11.02   &       31.09   &       6.62    &       0.02    &       0.36    &       0.01    &       0.368   &       0.007   &       2869    &       93      &       16.62   &       0.22    &       5.58    \\
211992989       &       c18     &       12.80   &       30.68   &       6.83    &       0.02    &       0.33    &       0.01    &       0.337   &       0.007   &       3411    &       79      &       23.47   &       0.31    &       2.72    \\
\hline\hline
\end{tabular}\end{small}
\\
* calculated with Eq.~\ref{SpT_calib}
\end{table*}

\begin{table*}[h]
\caption{Rotation period and activity indicators obtained in short- and long-cadence LCs for all 56 stars in our \textit{K2} M dwarf sample. $R_{\mathrm{per}}$: full amplitude of the rotation signal, $S_{\mathrm{ph}}$: standard deviation of the full LC, $S_{\mathrm{flat}}$: standard deviation of the flattened and cleaned LC, and $\nu$: flare frequency.}
\label{rot_flares_table}
\begin{small}\begin{tabular}{llccc|cccc|ccc}
\hline \hline
\multicolumn{5}{c|}{} & \multicolumn{4}{c|}{short cadence} & \multicolumn{3}{c}{long cadence}\\\hline
EPIC ID & Camp. & $P_{\rm rot}$ & $R_{\mathrm{per}}$ & log\,$E_{\mathrm{F,max}}$ & $S_{\mathrm{ph}}$ & $S_{\mathrm{flat,SC}}$ & $N_{\mathrm{flares,SC}}$ &  $\nu_{\rm SC}$ &  $S_{\mathrm{flat,LC}}$ & $N_{\mathrm{flares,LC}}$ &  $\nu_{\rm LC}$ \\ 
&  & [d] & [\%] & [erg] & [ppm] & [ppm] &  & [$N_{\mathrm{flares}}/d$] & [ppm] &  & [$N_{\mathrm{flares}}/d$]\\\hline
202059204       &       c00     &       7.76    &       2.29    &       33.41   &       8760    &       381     &       52      &       1.4347  &       426     &       8       &       0.2207  \\
202059229       &       c00     &       5.06    &       2.50    &       33.49   &       7441    &       389     &       51      &       1.4071  &       397     &       11      &       0.3035  \\
201611969       &       c01     &       25.88   &       0.50    &       31.80   &       3736    &       272     &       11      &       0.1374  &       66      &       9       &       0.1124  \\
205467732       &       c02     &       1.32    &       0.95    &       33.03   &       8580    &       1651    &       59      &       0.7593  &       524     &       18      &       0.2317  \\
206262336       &       c03     &       9.73    &       3.57    &       33.31   &       16086   &       772     &       197     &       2.8501  &       813     &       54      &       0.7812  \\
210317378       &       c04     &       27.09   &       1.80    &       32.27   &       7696    &       589     &       16      &       0.2331  &       194     &       6       &       0.0874  \\
210434433       &       c04     &       23.26   &       1.35    &       32.29   &       4510    &       332     &       1       &       0.0146  &       85      &       1       &       0.0146  \\
210460280       &       c04     &       43.72   &       0.92    &       ---     &       4090    &       188     &       0       &       ---     &       48      &       0       &       ---     \\
210489654       &       c04     &       ---     &       ---     &       32.19   &       7291    &       573     &       44      &       0.6419  &       497     &       6       &       0.0875  \\
210579749       &       c04     &       11.24   &       0.70    &       ---     &       2954    &       122     &       5       &       0.0729  &       47      &       3       &       0.0438  \\
210580056       &       c04     &       21.43   &       0.69    &       31.33   &       2667    &       98      &       0       &       ---     &       38      &       0       &       ---     \\
211480655       &       c05,c18 &       ---     &       ---     &       32.14   &       13427   &       486     &       6       &       0.0484  &       200     &       0       &       ---     \\
211642294       &       c05,c18 &       48.25   &       1.73    &       33.30   &       5798    &       157     &       6       &       0.0802  &       53      &       4       &       0.0535  \\
211892034       &       c05     &       36.75   &       0.44    &       ---     &       2036    &       279     &       0       &       ---     &       62      &       0       &       ---     \\
211945363       &       c05,c18 &       60.20   &       0.94    &       31.41   &       3030    &       360     &       13      &       0.1035  &       85      &       8       &       0.0637  \\
212285603       &       c06     &       ---     &       ---     &       31.43   &       4161    &       192     &       4       &       0.0507  &       51      &       0       &       ---     \\
212518629       &       c06,c17 &       78.53   &       1.52    &       31.58   &       13623   &       460     &       12      &       0.0832  &       106     &       6       &       0.0416  \\
212556566       &       c06,c17 &       1.02    &       1.63    &       33.14   &       19581   &       20061   &       189     &       1.3103  &       1098    &       27      &       0.1872  \\
212560714       &       c06,c17 &       27.61   &       0.33    &       32.07   &       5191    &       201     &       1       &       0.0127  &       46      &       0       &       ---     \\
212681564       &       c06,c17 &       61.23   &       0.85    &       31.43   &       35062   &       582     &       3       &       0.0208  &       140     &       0       &       ---     \\
212776174       &       c06     &       18.19   &       0.77    &       33.08   &       1990    &       209     &       5       &       0.0347  &       36      &       1       &       0.0069  \\
201659529       &       c10     &       44.25   &       2.01    &       32.44   &       16046   &       449     &       4       &       0.0579  &       134     &       3       &       0.0434  \\
245919787       &       c12     &       5.65    &       4.09    &       34.75   &       24753   &       799     &       82      &       1.0400  &       971     &       8       &       0.1015  \\
210651981       &       c13     &       2.45    &       4.16    &       34.63   &       19047   &       912     &       104     &       1.2916  &       1196    &       12      &       0.1490  \\
246774176       &       c13     &       2.54    &       1.13    &       34.12   &       11177   &       886     &       153     &       1.9225  &       891     &       30      &       0.3770  \\
248601792       &       c14     &       0.69    &       2.51    &       33.40   &       42222   &       22419   &       26      &       0.3334  &       2882    &       12      &       0.1539  \\
211889233       &       c16     &       10.76   &       2.75    &       33.46   &       13156   &       391     &       39      &       0.4905  &       529     &       2       &       0.0252  \\
212405060       &       c17     &       47.35   &       1.66    &       31.60   &       8549    &       650     &       1       &       0.0152  &       167     &       1       &       0.0152  \\
212424143       &       c17     &       6.63    &       2.07    &       34.32   &       27537   &       676     &       147     &       2.1931  &       532     &       43      &       0.6415  \\
212438743       &       c17     &       ---     &       ---     &       ---     &       2686    &       285     &       0       &       ---     &       79      &       0       &       ---     \\
212518679       &       c17     &       34.56   &       0.33    &       ---     &       1705    &       322     &       0       &       ---     &       67      &       0       &       ---     \\
212578572       &       c17     &       32.82   &       0.72    &       32.33   &       2786    &       456     &       1       &       0.0153  &       93      &       0       &       ---     \\
212610229       &       c17     &       31.27   &       0.38    &       ---     &       2266    &       446     &       0       &       ---     &       101     &       0       &       ---     \\
212611828       &       c17     &       1.09    &       2.12    &       34.27   &       22052   &       1183    &       160     &       2.4475  &       1283    &       24      &       0.3671  \\
212652474       &       c17     &       27.38   &       0.78    &       32.09   &       3054    &       301     &       3       &       0.0459  &       65      &       0       &       ---     \\
212662886       &       c17     &       37.13   &       0.50    &       32.09   &       2431    &       264     &       2       &       0.0306  &       58      &       2       &       0.0306  \\
212663959       &       c17     &       ---     &       ---     &       31.30   &       5142    &       333     &       1       &       0.0153  &       74      &       1       &       0.0153  \\
212679181       &       c17     &       32.35   &       0.78    &       32.37   &       4125    &       300     &       15      &       0.2295  &       74      &       7       &       0.1071  \\
212681560       &       c17     &       33.58   &       0.49    &       32.27   &       4178    &       338     &       5       &       0.0765  &       78      &       3       &       0.0459  \\
212710520       &       c17     &       ---     &       ---     &       31.30   &       8785    &       245     &       1       &       0.0153  &       48      &       0       &       ---     \\
212712577       &       c17     &       29.48   &       0.43    &       31.86   &       3537    &       182     &       4       &       0.0612  &       49      &       3       &       0.0459  \\
212734783       &       c17     &       ---     &       ---     &       30.92   &       2697    &       267     &       1       &       0.0153  &       61      &       0       &       ---     \\
212807182       &       c17     &       19.47   &       0.40    &       ---     &       3514    &       410     &       0       &       0.0000  &       102     &       0       &       ---     \\
212810669       &       c17     &       25.89   &       1.21    &       31.94   &       4598    &       395     &       17      &       0.2600  &       119     &       7       &       0.1071  \\
212840848       &       c17     &       ---     &       ---     &       ---     &       1049    &       80      &       0       &       ---     &       27      &       0       &       ---     \\
212846068       &       c17     &       19.90   &       0.32    &       ---     &       1694    &       200     &       0       &       ---     &       44      &       0       &       ---     \\
212874829       &       c17     &       30.26   &       0.93    &       33.70   &       4073    &       338     &       16      &       0.2447  &       89      &       5       &       0.0765  \\
251512053       &       c17     &       30.26   &       0.29    &       ---     &       1589    &       199     &       0       &       ---     &       51      &       0       &       ---     \\
251540467       &       c17     &       14.63   &       0.90    &       31.16   &       3514    &       236     &       1       &       0.0153  &       58      &       0       &       ---     \\
251550724       &       c17     &       46.42   &       1.04    &       31.58   &       3732    &       126     &       5       &       0.0765  &       40      &       4       &       0.0612  \\
251567386       &       c17     &       ---     &       ---     &       ---     &       12824   &       488     &       0       &       ---     &       124     &       0       &       ---     \\
251583820       &       c17     &       52.05   &       0.59    &       32.06   &       2585    &       499     &       4       &       0.0612  &       123     &       0       &       ---     \\
251584738       &       c17     &       41.58   &       0.56    &       31.72   &       2747    &       195     &       1       &       0.0153  &       48      &       0       &       ---     \\
211944670       &       c18     &       2.77    &       0.95    &       33.71   &       17108   &       1192    &       149     &       2.9323  &       996     &       23      &       0.4526  \\
211944856       &       c18     &       0.62    &       0.71    &       33.54   &       17141   &       894     &       45      &       0.8856  &       769     &       11      &       0.2165  \\
211992989       &       c18     &       ---     &       ---     &       ---     &       1045    &       350     &       0       &       ---     &       82      &       0       &       ---     \\
\hline\hline
\end{tabular}             \end{small}
\end{table*}

\section{FFDs for the different SpT groups}

\begin{figure*}
  \centering
  \includegraphics[width=0.3\textwidth,angle=270]{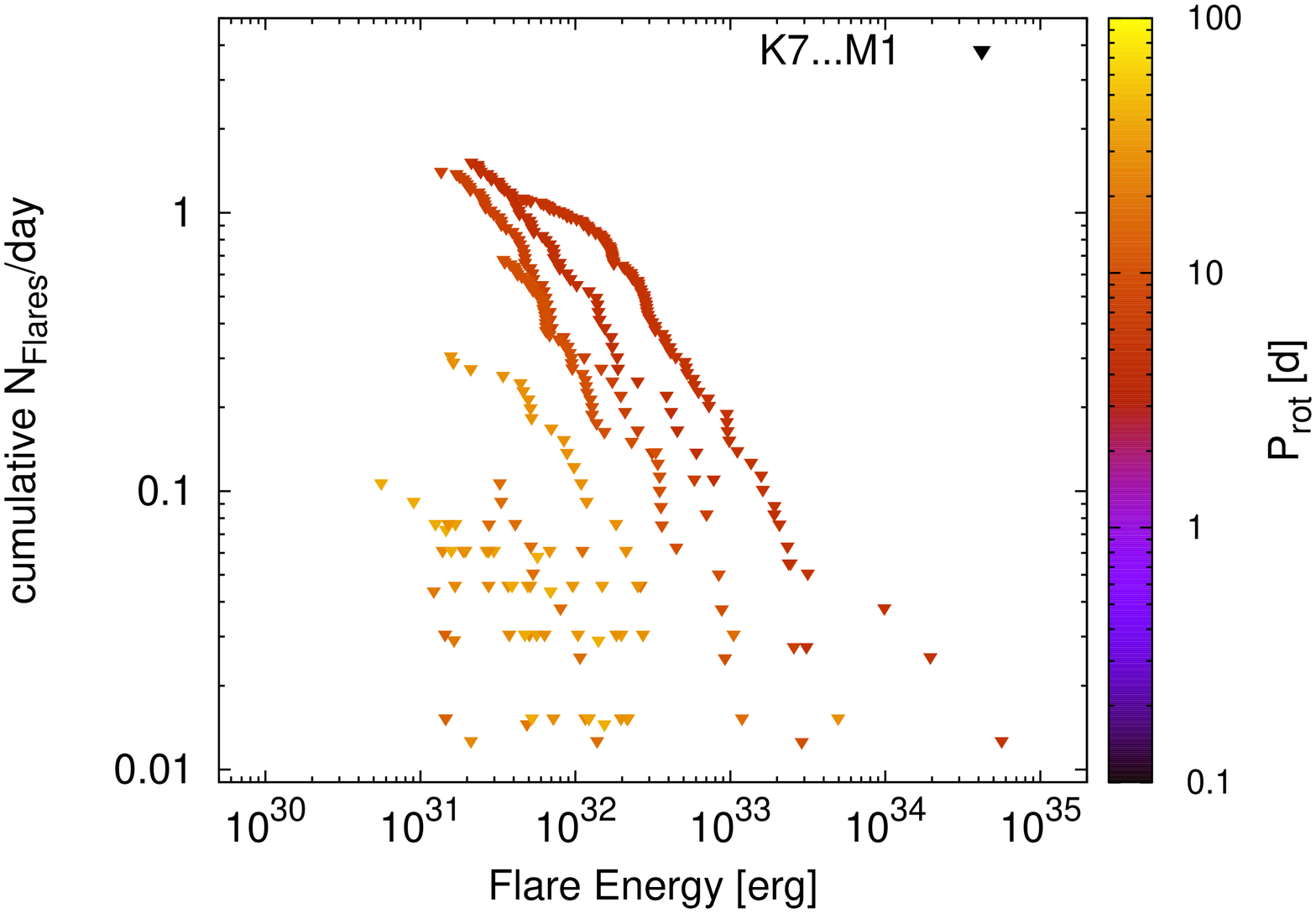} 
  \includegraphics[width=0.3\textwidth,angle=270]{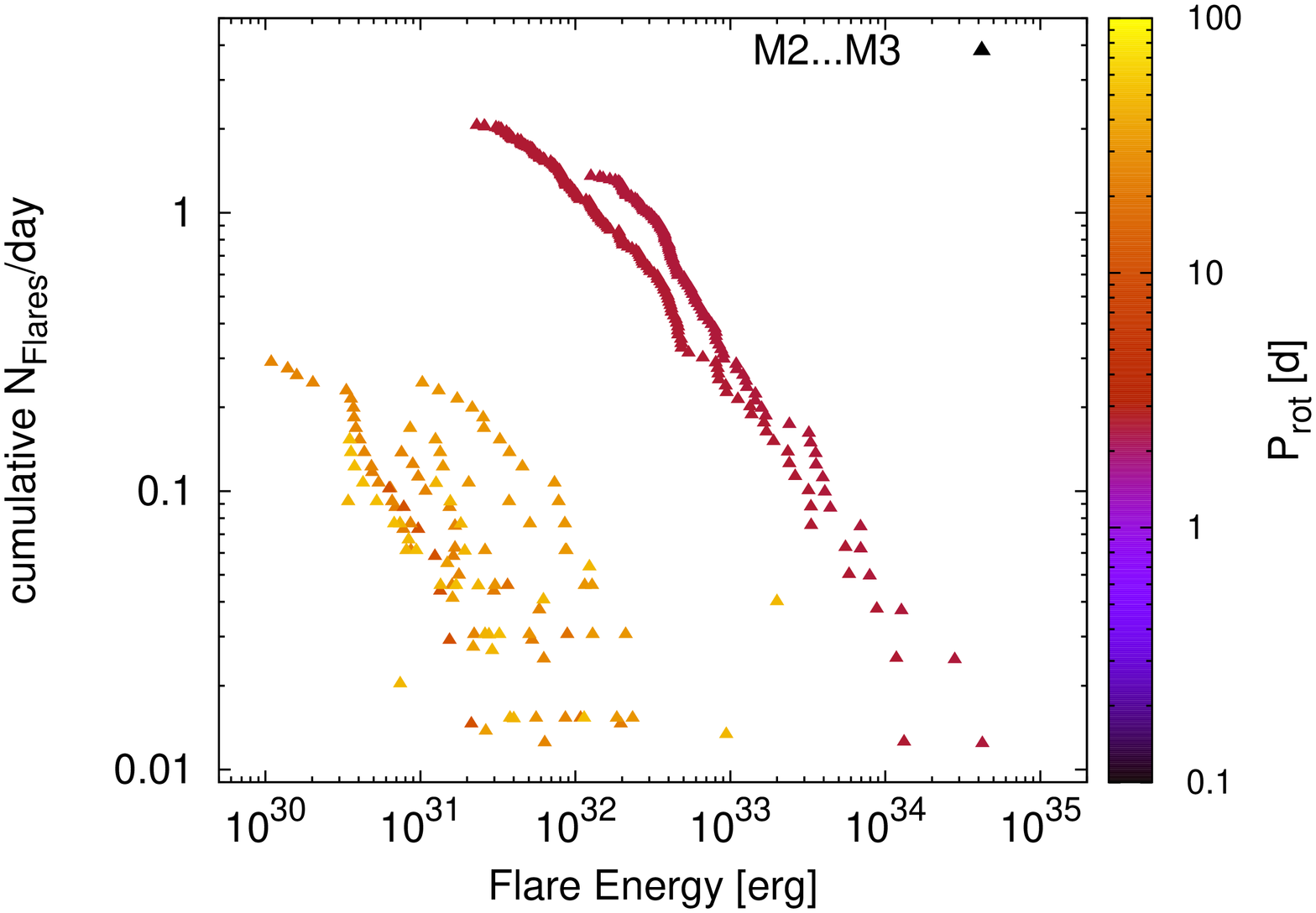}
  \includegraphics[width=0.3\textwidth,angle=270]{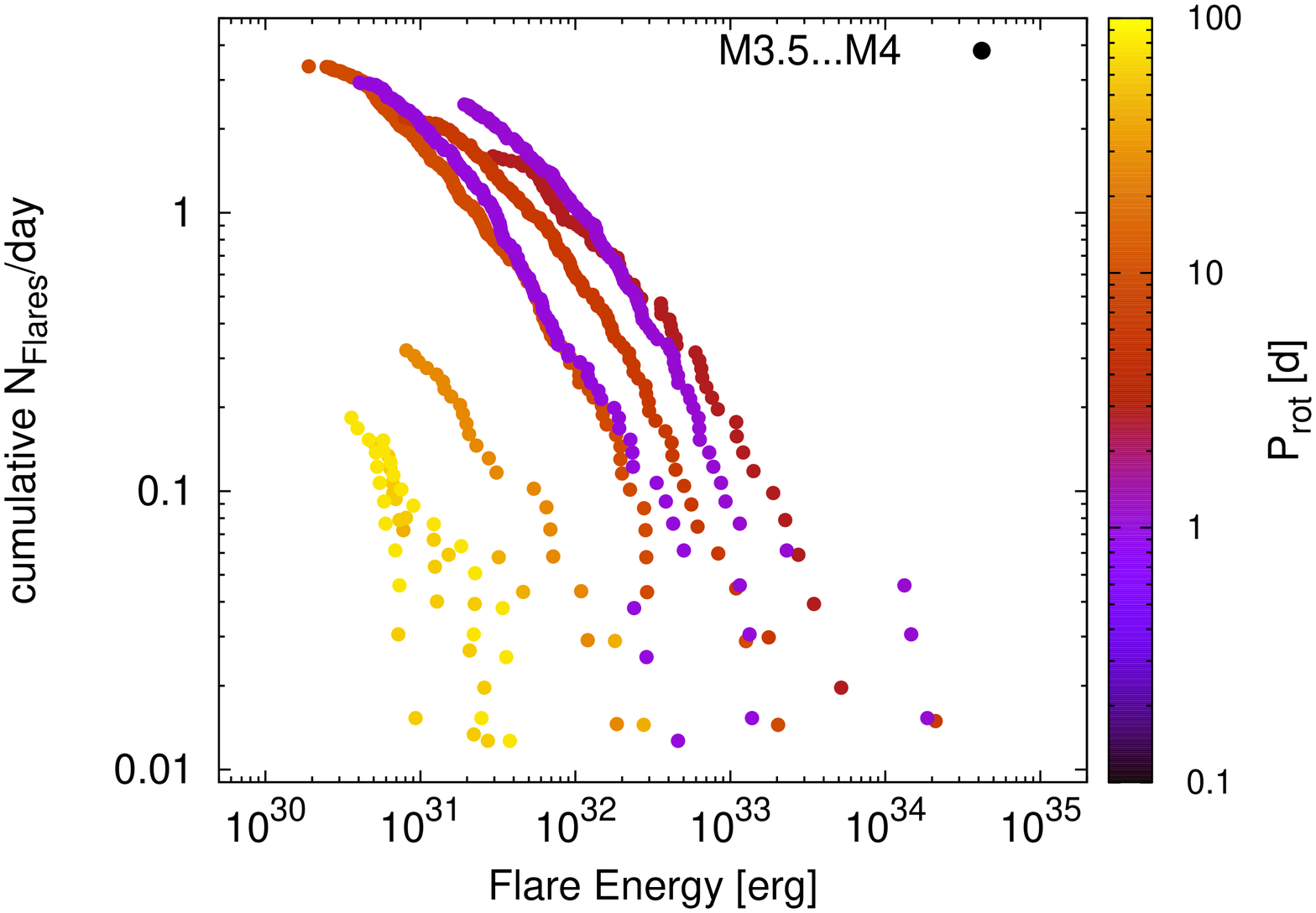}
  \includegraphics[width=0.3\textwidth,angle=270]{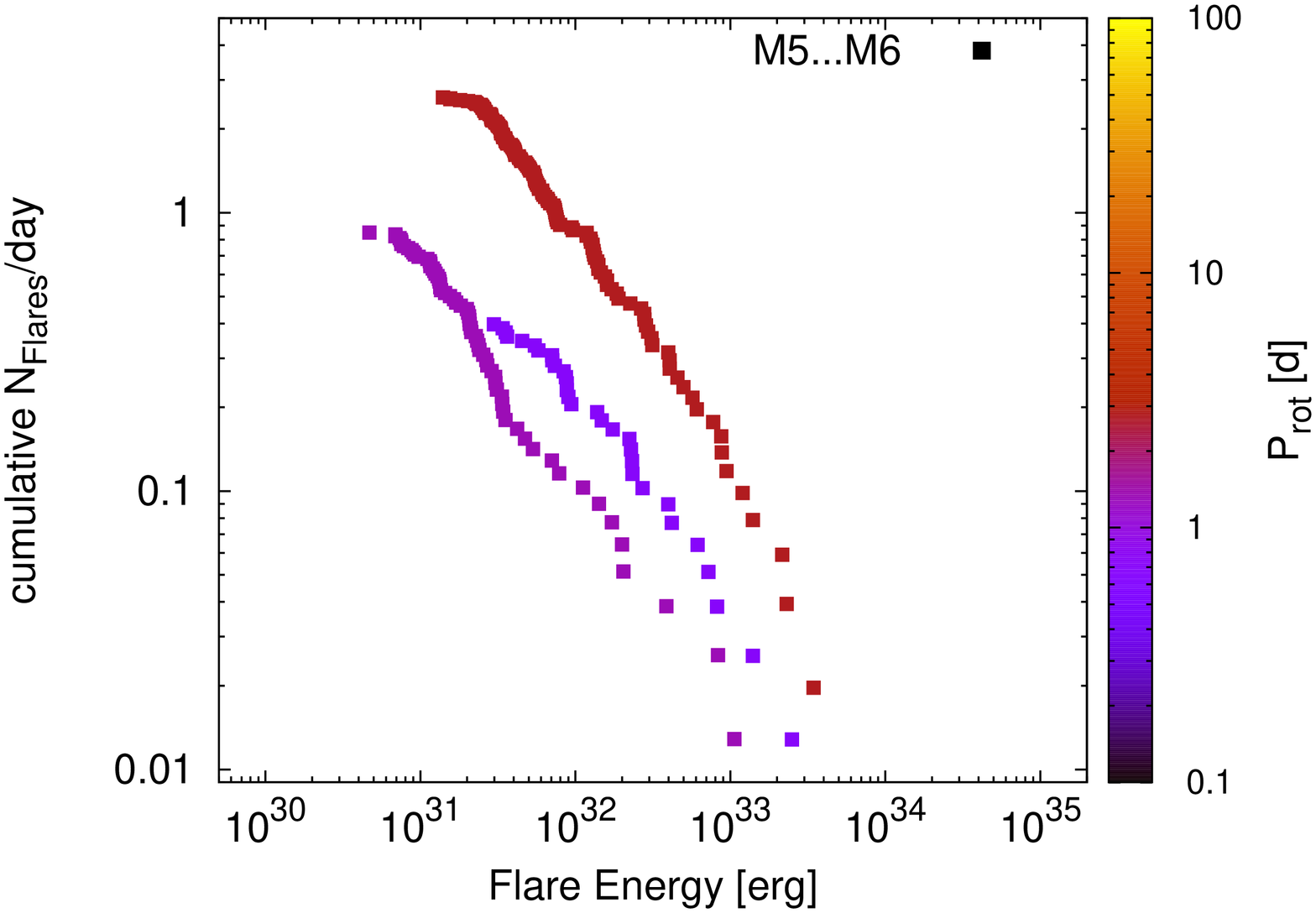}
  \caption{Cumulative frequency distribution of the  flare energy for all targets with measurable rotation periods that show flares in short-cadence LCs. The four different panels show the FFDs for different SpT groups. The rotation period of the stars is color-coded.}
  \label{FFD_SpT_groups}
\end{figure*}

%\begin{figure}
%  \centering
%  \includegraphics[width=0.3\textwidth,angle=270]{kumulative_Verteilung_Flare_Energy_all_SpT.ps} 
%  \caption{\textbf{Cumulative flare energy frequency distribution for all targets with measurable rotation period that show flares in short cadence LCs. Color-coded is the spectral type that were obtained from the $V-J$ calibration given in Eq.~\ref{SpT_calib} and shown in Fig.~\ref{Pecaut_Mamajek_VJ_SpT_fit}. Negative values denote spectral types earlier than M.}}
%  \label{FFD_all_SpT}
%\end{figure}

\end{document}